%% file: main.jhep.tex
\documentclass[a4paper,11pt]{article}
\usepackage{jheppub} 

\makeatletter
\def\@fpheader{}
\makeatother

\usepackage{units}
\usepackage{amsmath}
\usepackage{graphicx}
\usepackage{dcolumn}
\usepackage{pbox}
\usepackage{amssymb}
\usepackage{epsfig}
\usepackage{slashed}
\usepackage{amssymb}
\usepackage{ mathrsfs }
\usepackage[usenames,dvipsnames]{xcolor}

\usepackage{multirow}
\usepackage{lineno}
\usepackage{autobreak}

\usepackage{comment}

\usepackage{subcaption}
\usepackage{graphicx}

\usepackage{bbold}

\newcommand{\tr}{\mathrm{Tr}}
\newcommand{\trans}{\mathrm{T}}

\definecolor{MyLightBlue}{rgb}{0.22,0.51,0.9}
\definecolor{BrickRed}{rgb}{0.8, 0.25, 0.33}
\RequirePackage{hyperref}
\hypersetup{colorlinks, citecolor=blue,linkcolor=BrickRed, urlcolor=MyLightBlue}

\usepackage{bbding}
\usepackage{pifont}
\usepackage{wasysym}
\usepackage{amssymb}
\usepackage{tabu}

\newcommand{\oline}{\overline}

\usepackage{autobreak}

\newcommand{\la}{\langle}
\newcommand{\ra}{\rangle}

\title{\Large Predictive Dirac Neutrino Spectrum with Strong CP Solution \\[0.05in]in $\pmb{SU(5)_L \times SU(5)_R}$ Unification}

\author[a]{\bf K.S. Babu,}
\emailAdd{babu@okstate.edu}
\affiliation[a]{Department of Physics, Oklahoma State University, Stillwater, OK 74078, USA}

\author[b]{\bf Rabindra N. Mohapatra,}
\emailAdd{ rmohapat@umd.edu}
\affiliation[b]{Maryland Center for Fundamental Physics \& Department of Physics, University of Maryland, College Park, Maryland 20742, USA}

\author[c]{\bf Anil Thapa}
\emailAdd{wtd8kz@virginia.edu}
\affiliation[c]{Department of Physics, University of Virginia, Charlottesville, Virginia 22904-4714, USA}

\abstract{We develop a grand unified theory  of matter and forces based on the gauge symmetry $SU(5)_L\times SU(5)_R$ with parity interchanging the two factor groups. Our main motivation for such a construction is to realize a minimal GUT embedding of left-right symmetric models that provide  a parity solution to  the strong CP problem without the axion. We show how the gauge couplings unify with an intermediate gauge symmetry $SU(3)_{cL}\times SU(2)_{2L}\times U(1)_{L}\times SU(5)_R$, and establish its consistency with proton decay constraints. The model correctly reproduces the observed fermion masses and mixings and leads to {\it naturally light Dirac neutrinos} with their Yukawa couplings suppressed by a factor $M_I/M_G$, the ratio of the intermediate scale to the GUT scale. We call this mechanism type II-Dirac seesaw. Furthermore, the model predicts  $\delta_{CP} = \pm (130.4 \pm 1.2)^\circ $ and  $m_{\nu_1} = (4.8-8.4)$ meV for the Dirac CP phase and the lightest neutrino mass.  We demonstrate how the model solves the strong CP problem via parity symmetry.}

\begin{document}
\maketitle
\flushbottom

\section{Introduction}
Grand unification of forces and matter~\cite{Pati:1974yy,Georgi:1974sy,Georgi:1974yf} is an appealing framework for beyond the standard model physics, as it provides an understanding of the disparate forces of nature, and also provides a connection between apparently diverse building blocks of matter such as the quarks and leptons. It further promises to elucidate the nature of the neutrino, explains its tiny mass, and addresses the origin of matter-antimatter asymmetry in the universe. Concrete realizations of this idea predict new phenomena such as the decay of the proton which can be used to test this approach and has rightfully spurred dedicated decades-long experimental efforts to discover such decays. Grand unification (GUT) has thus become a widely accepted paradigm for physics beyond the standard model for nearly half a century. 

The most common frameworks for realizing GUTs are based on simple gauge symmetry groups such as $SU(5)$, $SO(10)$ and $E_6$. While these models are phenomenologically appealing and quite successful, there has been no direct experimental support for any of them so far. It is important, therefore, to pursue alternative unification ideas that are testable. In this paper, we develop and analyze in detail one such alternative based on the gauge group $SU(5)_L\times SU(5)_R$ with parity symmetry that interchanges the two   $SU(5)$ factors~\cite{Davidson:1987mi,Cho:1993jb,Mohapatra:1996fu,Lee:2016wiy,Emmanuel-Costa:2011hwa,Tavartkiladze:2016imo,Lonsdale:2014wwa}, so that there is a single unified force. The main motivation for developing such a theory is to provide a minimal GUT embedding of left-right symmetric models which solve the strong CP problem via parity symmetry without the axion.  A successful left-right symmetric model of this type was proposed  by two of us some time ago~\cite{Babu:1989rb}  which has been followed up and extended in several recent papers~\cite{Hall:2018let, Hall:2019qwx,Dunsky:2020dhn,Harigaya:2022wzt,Carrasco-Martinez:2023nit,Craig:2020bnv,deVries:2021pzl,Dcruz:2022rjg,Babu:2023srr,Bonnefoy:2023afx}. The key idea behind this solution to the strong CP problem is that the instanton-induced strong CP-violating Lagrangian term $\overline{\theta} G \widetilde{G}$ is parity violating and therefore a theory that obeys parity symmetry could lead to vanishing $\overline{\theta}$ \cite{Beg:1978mt,Mohapatra:1978fy,Mohapatra:1995xd,Kuchimanchi:1995rp,Mohapatra:1997su,Babu:2001se,Kuchimanchi:2023imj}.  Parity is of course broken in nature, which would induce nonzero $\overline{\theta}$, but in the model of Ref.~\cite{Babu:1989rb} this occurs only at the two-loop level.  The $SU(5)_L\times SU(5)_R$ model with parity symmetry thus has the potential to solve the strong CP problem without the axion, in the framework of a GUT.

Unified theories based on product groups accompanied by a discrete symmetry, such as $SU(3)_c \times SU(3)_L \times SU(3)_R$ trinification with a $Z_3$ symmetry~\cite{Rujula:1984,Babu:1985gi}, $SU(3)_q \times SU(3)_\ell \times SU(3)_L \times SU(3)_R$ quartification with a $Z_4$ symmetry~\cite{Babu:2003nw}, as well as  $SO(10)_1 \times SO(10)_2 \times SO(10)_3$~\cite{Babu:2007mb} and $SU(5)_1 \times SU(5)_2 \times SU(5)_3$~\cite{FernandezNavarro:2023hrf} with a $Z_3$ symmetry, where each gauge group acts on a separate family in the last two cases, have been developed to varying degrees of detail in the literature. The $SU(5)_L \times SU(5)_R$ model developed here is similar in spirit to these models, but with the main motivation being parity solution to the strong CP problem.

Achieving the desired symmetry breaking pattern consistent with gauge coupling unification, along with realistic fermion mass generation within $SU(5)_L\times SU(5)_R$ grand unification turns out to be somewhat challenging.  One reason for this is that the GUT scale value of the weak mixing angle in this framework is ${\rm sin}^2 ~\theta_W (M_G)= \frac{3}{16}$, as opposed to this value being ${\rm sin}^2 ~\theta_W (M_G)= \frac{3}{8}$ in conventional GUTs such as $SU(5)$.  This mixing angle should therefore {\it increase} in running down in energy.  Consistent unification can be achieved, as we show here, with an intermediate symmetry group $SU(3)_{cL}\times SU(2)_{2L}\times U(1)_{L}\times SU(5)_R$. This is the simplest scenario that we have found for symmetry breaking. With an unbroken $SU(5)_R$ surviving down to the intermediate scale, one should worry about proton decay mediated by the ($X_R^\mu,\,Y_R^\mu)$ gauge bosons of this group.  We have found a natural flavor structure that emerges within the framework which suppresses proton decay mediated by these gauge bosons completely, as these couplings always involve a heavy field.

The fermion mass matrices of the model have a general structure that resembles the universal seesaw mechanism~\cite{Davidson:1987mh,Berezhiani:1985in} wherein the usual fermions acquire their masses by mixing with vector-like fermions present in the theory.  However, such a seesaw cannot be universal with the assumed intermediate symmetry, since in that case there would be rapid proton decay.  In the absence of certain scalars that are needed for universal seesaw, we find that proton decay is naturally suppressed.  

The model developed here is very predictive in the neutrino sector.  As we shall see, neutrinos are naturally light Dirac fermions in this framework, with their Yukawa couplings suppressed by an overall factor of  $M_I/M_G$, the ratio of the intermediate scale to the GUT scale.  
We call such a suppression mechanism {\it type-II Dirac seesaw} due to its similarity with type-II seesaw in Majorana neutrino models. The model predicts normal ordering of neutrino masses. The quark-lepton connection present in the unified theory enables us to predict the Dirac CP phase appearing in neutrino oscillations to be $\delta_{CP} = (130.4 \pm 1.2)^\circ $ or $(229.6\pm1.2)^\circ$, and the lightest neutrino mass to be $m_{\nu_1} = (4.8-8.4)$ meV. These predictions will provide tests of the model. Naturally, there will be no neutrinoless double beta decay within this model.

The main results of the paper can be summarized as follows. 
{\it (i)} we have presented a successful embedding of the parity solution to strong CP problem in an $SU(5)_L\times SU(5)_R$ GUT framework and shown its consistency with gauge coupling unification; {\it (ii)} neutrinos are naturally light Dirac fermions within the model with quantitative predictions for two: the CP phase and the lightest neutrino mass; {\it (iii)} there is no issue with rapid proton decay even with the intermediate symmetry containing $SU(5)_R$ owing to an emergent flavor structure; and {\it (iv)} we have shown the vanishing of $\overline{\theta}$ through one-loop diagrams, including various contributions from the GUT symmetry breaking sector. The two-loop induced $\overline{\theta}$ is compatible with neutron electric dipole moment (EDM) limits, with a mild fine-tuning of parameters at the few percent level. Neutron EDM should be within reach of forthcoming experiments for the model to be not finely tuned.

The paper is organized as follows. We describe the salient features of the model including the symmetry breaking sector and its motivations in Sec. \ref{sec:Model}. The mechanism of fermion mass generation is discussed in Sec. \ref{sec:fermionmass}. In Sec. \ref{sec:unification}, we discuss the unification of gauge couplings where we also determine the unification scale and the intermediate scale as well as the values of the various couplings at these scales. In Sec.~\ref{sec:fermionfitting} we show how the neutrino and the down-type quark mass matrices are connected leading to  predictions for the leptonic CP phases and the lightest neutrino mass. Sec.~\ref{sec:protondecay} is devoted to a discussion of proton decay in the model where we show the consistency of the model. In Sec.~\ref{sec:thetabar},
we evaluate all the one-loop corrections to the strong CP parameter $\overline{\theta}$ and show that they are all vanishing. Here we also estimate the leading two-loop contributions and discuss its implications for neutron electric dipole moment. In Sec. \ref{sec:Conclusions} we conclude.  In several appendices we present various technical details, including decomposition under subgroups, threshold corrections,  and two-loop renormalization group equations for the Yukawa coupling matrices that are used in our analysis of gauge  coupling unification,  fermion mass fitting and $\overline{\theta}$ estimation.

\section{Model}\label{sec:Model} 
The unified theory that we develop in this paper is based on the gauge group $SU(5)_L \times SU(5)_R$ with parity symmetry which interchanges the two $SU(5)$ factors.  This theory provides a natural embedding of the left-right symmetric theory of Ref. \cite{Babu:1989rb} which provides a parity solution to the strong CP problem.
We start with a brief review of the model of Ref. \cite{Babu:1989rb}. It is based on the gauge group $SU(3)_c\times SU(2)_L\times SU(2)_R\times U(1)_{B-L}$ with the assignment of the three families of fermions as follows:
\begin{align}
Q_L\ (3,2,1,1/3) &= \begin{pmatrix}
 u_L \\
 d_L \\
\end{pmatrix} , \hspace{5mm}
Q_R\ (3,1,2,1/3) = \begin{pmatrix}
 u_R \\
 d_R \\
\end{pmatrix} , \nonumber\\[5pt]

\ell_L\ (1,2,1,-1) &= \begin{pmatrix}
 \nu_L \\
 e_L \\
\end{pmatrix}, \hspace{5mm}
\ell_R\ (1,1,2,-1) = \begin{pmatrix}
 \nu_R \\
 \ell_R \\
\end{pmatrix} .
\label{eq:fermion}
\end{align}
This is supplemented by three families of vector-like quarks and leptons denoted as
\begin{equation}
U_{L,R}(3,1,1,4/3),~~~D_{L,R}(3,1,1,-2/3),~~~E_{L,R}(1,1,1,-2)~.
    \label{eq:vecferm}
\end{equation}
Parity symmetry ($P$) can be defined within the model, under which $f_L \leftrightarrow f_R$ where $f$ stands for any of the fermions, along with $W_L^\mu \leftrightarrow W_R^\mu$ for the $SU(2)_{L,R}$ gauge bosons. 
Note the absence of an electrically neutral vector-like lepton, which is crucial for realizing Dirac neutrinos in the framework \cite{Babu:1988yq,Babu:2022ikf}.
The Higgs sector of the model is very simple, consisting of a pair of parity symmetric doublets, $\{\chi_L(1,2,1,1) +  \chi_R(1,1,2,1) \}$ with the transformation $\chi_L \leftrightarrow \chi_R$ under $P$.

We now proceed to embed this model into a grand unified $SU(5)_L \times SU(5)_R$ framework. To achieve this, we note that under the $SU(5)_L \times SU(5)_R$ gauge symmetry, all left-handed fermions of Eq. (\ref{eq:fermion}) neatly fit into ${\bf 10} + {\bf \overline{5}}$ of $SU(5)_L$, while all right-handed fermions fit into ${\bf 10} + {\bf \overline{5}}$ of $SU(5)_R$. The  vector-like fermions of Eq. (\ref{eq:vecferm}) complete these multiplets without needing any other fermions. Thus the full fermion content of this left-right symmetric model fits into  two basic anomaly-free chiral representations of $SU(5)_L \times SU(5)_R$. Their grouping under $SU(5)_L \times SU(5)_R$ is given by
\begin{align}
   \psi_{L,R} = \left[\begin{array}{c}
D_{1}^{c} \\
D_{2}^{c} \\
D_{3}^{c} \\
e \\
-\nu
\end{array}\right]_{L,R} \, , \hspace{10mm} 
\chi_{L,R} =\frac{1}{\sqrt{2}}\left[\begin{array}{ccccc}
0 & U_{3}^{c} & -U_{2}^{c} & u_{1} & d_{1} \\
-U_{3}^{c} & 0 & U_{1}^{c} & u_{2} & d_{2} \\
U_{2}^{c} & -U_{1}^{c} & 0 & u_{3} & d_{3} \\
-u_{1} & -u_{2} & -u_{3} & 0 & E^{c} \\
-d_{1} & -d_{2} & -d_{3} & -E^{c} & 0
\end{array}\right]_{L,R} \, . 
\label{eq:fermionsector}
\end{align}
We denote these fields transforming under $SU(5)_L \times SU(5)_R$ as \{$\psi_L(\overline{\bf 5},1) + \psi_R(1,\overline{\bf 5})\}$, and
$\{\chi_L({\bf 10},1) +  \chi_R(1, {\bf 10})\}$. There are three copies of them corresponding to the three generations. As in the left-right symmetric model, right-handed neutrinos, $\nu_R$, are naturally present in the $\psi_R(1,\overline{\bf 5})$ multiplet.  Under parity, the fields transform as $\psi_L \leftrightarrow \psi_R$, $\chi_L \leftrightarrow \chi_R$ along with $V^\mu_L \leftrightarrow V^\mu_R$, where $V^\mu_{L,R}$ denote the gauge bosons of $SU(5)_{L,R}$ symmetry. Parity symmetry would imply that there is a single gauge coupling in the theory, identified as $\alpha_{5L} = \alpha_{5R} = \alpha_G$. One of the main goals of this paper is to construct a realistic symmetry breaking chain that admits gauge coupling unification consistent with proton decay constraints, while reproducing the fermion masses correctly and preserving the parity solution to the strong CP problem as in the left-right symmetric model of Ref. \cite{Babu:1989rb}.

To break $SU(5)_L \times SU(5)_R$ spontaneously all the way down to $SU(3)_{c} \times U(1)_{\rm em}$ and to generate realistic fermion masses and mixings, we choose the following Higgs multiplets:\footnote{While the choice of $\{({\bf 24},1)+(1,{\bf 24})\}$ instead of $\{({\bf 75},1)+(1,{\bf 75})\}$ is more economical, it will not go well with the strong CP solution, since terms in the Higgs potential of the type $({\bf 24},1)H^\dagger_R\Phi H_L $ and $({\bf 24},1) \eta^\dagger \Phi \Phi$ with complex coefficients would be allowed, in spite of parity symmetry. These complex couplings will spoil the strong CP solution via parity symmetry, owing to scalar--pseudoscalar mixings that they generate. In presence of such mixings, the loop-induced $\overline{\theta}$ would be typically too large. It is possible to forbid such terms with an additional discrete symmetry, but not with parity alone.}
\begin{align}
   &  \{ \Sigma_L ({\bf 75},1) + \Sigma_R (1,{\bf 75}) \}, \hspace{3mm} \{ H_L ({\bf 5},1)+ H_R (1,{\bf 5}) \}, \hspace{3mm} \Phi  (\overline{\bf 5},{\bf 5}), \hspace{3mm}   \eta (\overline{\bf 15},   {\bf 15}) \, .
    \label{eq:Higgssector}
\end{align} 
Here the field $\Sigma_L$ is used to break the $SU(5)_L$ symmetry down to $SU(3)_{cL} \times SU(2)_L \times U(1)_L$ and $\Sigma_R$ appears as its parity partner. The $H_{L,R}$ fields and the $\Phi$ field are used to generate fermion masses and to break the surviving symmetry down to $SU(3)_c \times U(1)_{\rm em}$. We discuss the details of fermion mass generation in Sec. \ref{sec:fermionmass}, with the mass matrices for quarks and leptons given in Eq.~\eqref{eq:fermionmass}. 
The simplest symmetry breaking chain that we have found, consistent with realistic fermion masses and proton decay constraints, assumes a single intermediate scale, denoted as $M_I$, and proceeds as follows:
\begin{gather}
S U(5)_L \times S U(5)_R \notag\\
~~~~~~~~~~~~~~\downarrow M_G \sim \langle\Sigma_L \rangle \notag\\
S U(3)_{c L} \times S U(2)_L \times U(1)_L \times S U(5)_R \notag\\
~~~~~~~~~~~~~~~~~~\downarrow M_I \sim \langle \Phi \rangle , \langle H_R\rangle \notag\\
S U(3)_c \times S U(2)_L \times U(1)_Y \notag\\
~~~~~~~~~~~~~\downarrow M_W \sim\langle H_L\rangle \notag\\
S U(3)_c \times U(1)_{\mathrm{em}}
\label{eq:breakingchain}
\end{gather}
Here the energy scales $\langle \Phi \rangle$ and $\langle H_R \rangle$ could in principle be different, but with the simplified assumption of having a single intermediate scale, they are identified.  A hierarchy of VEVs is necessary with $M_W \ll M_I \ll M_G$ so as to be consistent with phenomenology, viz., proton decay constraints as well as collider constraint on the mass of new gauge bosons. This is achieved in the model, as we discuss towards the end of this section.

The $\eta(\overline{\bf 15}, {\bf 15})$ field of Eq. (\ref{eq:Higgssector})
plays no significant role in the symmetry breaking, which can be arranged by choosing its mass term $M^2_\eta > 0$, which we shall assume.\footnote{However, once the other fields acquire VEVs, $\eta$ will acquire an induced VEV of order $M_I$.}  Its purpose is to enable gauge coupling unification.   
In order to generate the correct value of $\sin^2\theta_W(M_Z)$, which must increase in running down in energy, it is beneficial to have new matter/scalar multiplets at scales below $M_G$ that transform nontrivially under $SU(2)_L$, but carry very small $U(1)_Y$ charges.  The fragment $(\overline{\bf 3}, {\bf 2}, -1/6,{\bf 15})$ under $[SU(3)_{cL} \times SU(2)_L \times U(1)_L] \times SU(5)_R$ from the $\eta(\overline{\bf 15}, {\bf 15})$ field  has all the desired properties, and can lead to successful unification, consistent with low energy phenomenology, if its mass is assumed to be at or around $M_I$.  Evolution of gauge couplings  is discussed in detail in Sec. \ref{sec:unification}
where we have shown in Fig. \ref{fig:unification} successful unification when this scalar fragment  $(\overline{\bf 3}, {\bf 2}, -1/6,{\bf 15})$ has a mass equal to $M_I$.  This is a simpler way of achieving gauge coupling unification in $SU(5)_L \times SU(5)_R$ theories compared to other attempts \cite{Davidson:1987mi,Cho:1993jb,Mohapatra:1996fu,Lee:2016wiy,Emmanuel-Costa:2011hwa}. 

A scalar field belonging to $(\overline{\bf 10}, {\bf 10})$ could be considered to help with gauge coupling unification 
instead of the $\eta(\overline{\bf 15}, {\bf 15})$ field
(and has been introduced in Ref. \cite{Davidson:1987mi,Cho:1993jb,Lee:2016wiy,Emmanuel-Costa:2011hwa}), which contains a fragment $(\overline{\bf 3}, {\bf 2}, -1/6, {\bf 10})$ under $[SU(3)_{cL} \times SU(2)_L \times U(1)_L] \times SU(5)_R$ with the right properties to increase $\sin^2\theta_W$ while running down in energies.   However, we find this choice to have three problems in the present context: (i) It would allow the $(2,2)$ blocks of the up-type quark and charged lepton mass matrices ${\cal M}_u$ and ${\cal M}_\ell$ (see Eq. (\ref{eq:fermionmass})) to be nonzero, which would result in rapid proton decay mediated by the $SU(5)_R$ gauge bosons having masses of order $M_I$; (ii) it would lead to complex couplings in the scalar potential of the type $(\overline{\bf 10},{\bf 10})^* \Phi \Phi \Sigma_L$ and $\Phi (\overline{\bf 10}, {\bf 10})(\overline{\bf 10}, {\bf 10}) \Sigma_L$, potentially spoiling the parity solution to the strong CP problem; and (iii) it would lead to $g_{5R}(M_I)$ becoming non-perturbative while running from $M_G$ to $M_I$, making calculations unreliable. This is the rationale behind adopting the $\eta(\overline{\bf 15}, {\bf 15})$ multiplet instead of the $(\overline{\bf 10}, {\bf 10})$ scalar field.

Under left-right parity symmetry the scalar multiplets transform as 
\begin{align}
  H_L \leftrightarrow H_R \, ,\  ~
    \Phi \leftrightarrow \Phi^\dagger \, ,\ ~ \eta \leftrightarrow \eta^\dagger \, ,\ ~ \Sigma_L \leftrightarrow \Sigma_R \,  .
    \label{eq:parity}
\end{align} 
We have constructed the full scalar potential with these fields and cross-checked it against the software package {\tt Sym2Int} \cite{Fonseca:2017lem}.  We find that there are three non-trivial couplings allowed by the gauge symmetry that can be complex: 
\begin{align}
    V \supset &\ \mu_1\ H_R^\dagger \Phi H_L + \mu_2\ \eta^\dagger \Phi \Phi + \lambda\ H_L^\dagger \eta^\dagger \Phi H_R + h.c.  .
    \label{eq:pot}
\end{align}
However, with the imposition of parity symmetry,  see Eq.~\eqref{eq:parity}, all these couplings become real. 
Thus, the full Higgs potential of the model is CP invariant, which would  admit a vacuum structure that preserves CP (for some range of parameters of the potential). In this case the scalar and pseudoscalar fields would remain unmixed, which is significant for the theory to provide parity-based solution to the strong CP problem, especially in the vanishing of the $\overline{\theta}$ parameter at the one-loop level. This is discussed in more detail in Sec. \ref{sec:thetabar}.

The {\it real} vacuum expectation values (VEVs) of these Higgs fields are denoted as
\begin{align}
    \langle H_{L,R} \rangle &= \kappa_{L,R}\ (0,0,0,0,1)^T \, , \hspace{5mm}
    \langle \Phi \rangle = v_\Phi\ {\rm diag(1,1,1,0,0)} \,
    \label{eq:VEV0}\\
& \hspace{15mm}\langle(\Sigma_{L})^{ab}_{cd} \rangle  = \langle \Sigma_L \rangle ({\bf 1_{SM}})^{ab}_{cd}\,
 \label{eq:VEV1}
 \end{align}
 where the ${\bf 1_{SM}}$ singlet from the ${\bf 75}$ is the (un-normalized) combination given by
 \begin{align}
 ({\bf 1_{SM}})^{ab}_{cd} &= \Sigma_L
 \left( {~}^{14}_{14} + {~}^{15}_{15} + {~}^{24}_{24} + {~}^{25}_{25} + {~}^{34}_{34} + {~}^{35}_{35} - {~}^{12}_{12}- {~}^{13}_{13} - {~}^{23}_{23} -3 . \, {~}^{45}_{45}\right) \,.
    \label{eq:VEVeta}
\end{align}
Here $(\Sigma_{L})^{ab}_{cd}=-(\Sigma_{L})^{ba}_{cd} = -(\Sigma_{L})^{ab}_{dc}$ and $(\Sigma_{L})^{ab}_{ad}=0$ with $(a,b,.. = 1...5)$~\cite{Abud:1984ni,Hubsch:1984pg}. Note that the SM singlet from the ${\bf 75}$, which may be obtained from the product ${\bf 10} \times {\bf \overline{10}} = {\bf 1} + {\bf 24} + {\bf 75}$ after removing the ${\bf 24}$ and ${\bf 1}$ fragments, is the combination $(q \overline{q} - U^c \overline{U^c}-  3.\, E^c \overline{E^c})$, in the notation of Eq. (\ref{eq:fermionsector}).\footnote{The SM singlets from the {\bf 1} and {\bf 24} are respectively $(q \overline{q} + U^c \overline{U^c} + E^c \overline{E^c})$ and $(q \overline{q} - 4 .\,U^c \overline{U^c} +  6.\, E^c \overline{E^c})$.}
This is the combination appearing in Eq. (\ref{eq:VEVeta}).\footnote{
This index structure is in agreement with the result given in Ref. \cite{Abud:1984ni}, after making the interchange $\{4 \leftrightarrow 1,\, 3 \leftrightarrow 5 \}$.}
The VEV for the $\Sigma_R$ field can be written in an analogous fashion by the replacement $L\to R$ in Eqs.~\eqref{eq:VEV1}-\eqref{eq:VEVeta}.
The field $\eta$ will acquire an induced VEV along its SM-singlet component, through the cubic scalar coupling proportional to $\mu_2$ in Eq. (\ref{eq:pot}). This induced VEV, which is of order $M_I$, can be parametrized as
\begin{equation}
\langle \eta_{11}^{11} \rangle =  \langle \eta_{22}^{22} \rangle =
\langle \eta_{33}^{33} \rangle =
2\langle  \eta_{12}^{12} \rangle =
2\langle \eta_{13}^{13} \rangle = 
2\langle  \eta_{23}^{23} \rangle= v_\eta~.
\end{equation}
Here $\eta_{ij}^{ab} = \eta_{ji}^{ab} = \eta_{ij}^{ba}$. 
This real-valued VEV of $\eta$ will contribute to the gauge boson and scalar masses which are  of order $M_I$, but otherwise it has no significant effect on the analysis we have carried out, such as in the one-loop calculation of the $\overline{\theta}$ parameter.

In the symmetry breaking scheme displayed in Eq. (\ref{eq:breakingchain}), parity is spontaneously broken at a scale $\langle \Sigma_L \rangle \sim M_G$. 
It can be shown that starting with a $P$-invariant potential, a minimum where $\langle\Sigma_R \rangle=0$ and $\langle \Sigma_L \rangle \neq 0$ can be realized, for a certain range of parameters of the potential~\cite{Senjanovic:1975rk}. Since at the scale $M_G$, only the $\Sigma_{L,R}$ fields acquire VEVs, we can write the relevant potential as\footnote{The full potential for the model is given in Appendix \ref{sec:Anew} and the minimazation conditions relevant for spontaneous $P$-breaking are given in Eqs. (\ref{eq:minSL})-(\ref{eq:minSR}).}
\begin{equation}
V(\la\Sigma_{L,R}\ra)=-\mu^2 \left(\la{\Sigma_L}\ra^2+\la{\Sigma_R}\ra^2\right)+\hat{\lambda}_1\left(\la{\Sigma_L}\ra^2+\la\Sigma_R\ra^2\right)^2 +\hat{\lambda}_2\left(\la{\Sigma_L}\ra^2-\la{\Sigma_R}\ra^2\right)^2~.
\end{equation}
Here $\hat{\lambda}_{1,2}$ are combinations of various quartic couplings involving the $\Sigma_{L,R}$ fields that appear in the potential.
For the choice   $\hat{\lambda}_1 >0$ and $\hat{\lambda}_2 <0$, and  $\mu^2 >0$, the minimum of the potential occurs at $\{\langle \Sigma_L \rangle \neq 0, \,\langle \Sigma_R \rangle = 0\}$, or at $\{\langle \Sigma_R \rangle \neq 0, \,\langle \Sigma_L \rangle = 0\}$~\cite{Senjanovic:1975rk}, among which we choose the former solution without any loss of generality. For subsequent symmetry breaking at the intermediate scale $M_I$, the cross couplings of $H_L$ and $H_R$ fields with the $\Sigma_{L,R}$ fields would generate unequal masses for the Higgs doublets in $H_L^d$ and $H_R^d$.  Such a setup allows for the realization of the hierarchy $\langle H_L^d \rangle \ll \langle H_R^d \rangle \sim \langle \Phi \rangle \neq 0$. We shall adopt this chain and mechanism of symmetry breaking in our analysis. While we have constructed the full Higgs potential of the model, we shall only make use of the reality of the couplings that admits real-valued VEVs, see Eq. (\ref{eq:pot}), and the mechanism of spontaneous parity breaking which leads to a hierarchical VEV structure as discussed here.

\section{Fermion Mass Generation}\label{sec:fermionmass}

In this section we develop the scheme for fermion mass generation. This includes a natural mechanism for small Dirac neutrino masses, which we term as type-II Dirac seesaw.  We also show here the predictivity of the model in the neutrino sector, while delegating a detailed numerical analysis to Sec. \ref{sec:fermionfitting}.

\subsection{Yukawa Lagrangian}
The most general gauge-invariant Yukawa interactions of the model, that are also invariant under parity symmetry, are given by the Lagrangian
\begin{eqnarray}
   - {\cal L}_{\rm Yuk} 
  &=&    \frac{(Y_u^*)_{ij}}{4} \epsilon_{\alpha \beta \gamma \delta \rho} 
    \left\{\chi_{Li}^{\alpha \beta} \chi_{Lj}^{\gamma \delta}  H_L^\rho + \chi_{Ri}^{\alpha \beta} \chi_{Rj}^{\gamma \delta}  H_R^\rho\right\} 
    +\sqrt{2}\, (Y_\ell^*)_{ij}\left\{\psi_{L i \alpha} \chi_{Lj}^{\alpha \beta } H_{L\beta}^* + \psi_{R i \alpha} \chi_{Rj}^{\alpha \beta } H_{R\beta}^*\right\} 
    \nonumber \\
     &+& (Y_D^*)_{ij}\ \overline{\psi}_{Li}^{\alpha} \, \Phi^\beta_\alpha \,\psi_{Rj\beta}  
    + h.c. 
    \label{eq:YukLag}
\end{eqnarray}
Here $(i,\,j)$ are family indices, while $(\alpha,\,\beta...)$ are $SU(5)_{L,R}$ indices. We have  $(Y_u^*)_{ij} = (Y_u^*)_{ji}$ by $SU(5)_{L,R}$ symmetry, and $(Y_D^*)_{ij} = (Y_D)_{ji}$, owing to parity. 
After spontaneous symmetry breaking, with the VEVs as shown in Eq. (\ref{eq:VEV0}), the mass matrices induced for the up-type quarks, charged leptons and down-type quarks can be written, in the notation $-{\cal L}_{\rm mass} \supset \overline{f_L} M_f f_R$, as
\begin{align}
\mathcal{M}_u=\left(\begin{array}{cc}
0 & Y_u \kappa_L \\
Y_u^{\dagger} \kappa_R & 0
\end{array}\right), ~~~~
\mathcal{M}_{\ell}=\left(\begin{array}{cc}
0 & Y_{\ell} \kappa_L \\
Y_{\ell}^{\dagger} \kappa_R & 0
\end{array}\right), 
~~~~\mathcal{M}_d=\left(\begin{array}{cc}
0 & Y_{\ell}^T \kappa_L \\
Y_{\ell}^* \kappa_R & Y_D v_\Phi
\end{array}\right)   \, .
\label{eq:fermionmass}
\end{align}
Here the $3 \times 3$ sub-matrices obey $Y_u = Y_u^T$ and $Y_D = Y_D^\dagger$, while $Y_\ell$ is a general complex matrix. 
While  these mass matrices resemble those appearing in the context of universal seesaw left-right symmetric models of Ref. \cite{Davidson:1987mh,Berezhiani:1983hm,Babu:1989rb}, there is a significant difference in that the $(2,2)$ blocks of ${\cal M }_u$ and ${\cal M}_\ell$ are zero here.  This departure from universal seesaw is crucial for the model to be compatible with proton decay constraints, with the assumed $SU(5)_R$ intermediate symmetry. Nonzero entries in the $(2,2)$ blocks of ${\cal M}_u$ and ${\cal M}_\ell$ will lead to significant mixing between $u$ and $U$ quarks as well as between  $e$ and $E$ leptons.  This in turn would result in rapid proton decay mediated by the $(X_R^\mu,\,Y_R^\mu)$ gauge bosons of $SU(5)_R$.  With these blocks being zero, as in Eq. (\ref{eq:fermionmass}), the $(X_R^\mu,\,Y_R^\mu)$ gauge bosons will have baryon and lepton number violating couplings only between light fermions and heavy vector-like fermions, preventing rapid proton decay. This issue is further discussed in Sec. \ref{sec:protondecay}.

\subsection{Type-II Dirac seesaw for small neutrino masses}

Neutrinos turn out to be naturally light Dirac particles within the model.  Although baryon $(B$) and lepton numbers ($L$) are both broken by the gauge interactions of $SU(5)_L \times SU(5)_R$,  $(B-L)$ symmetry is left unbroken, which prevents any renormalizable couplings that would allow Majorana masses for the neutrinos. There is in fact a natural mechanism for the Dirac neutrinos Yukawa couplings to be extremely small: They are proportional to the ratio $(M_I/M_G)$, the scales associated with intermediate and GUT symmetry breaking. The effective Dirac neutrino masses and Yukawa couplings are of the form
\begin{figure}[ht]
    \centering
    \includegraphics[scale=0.5]{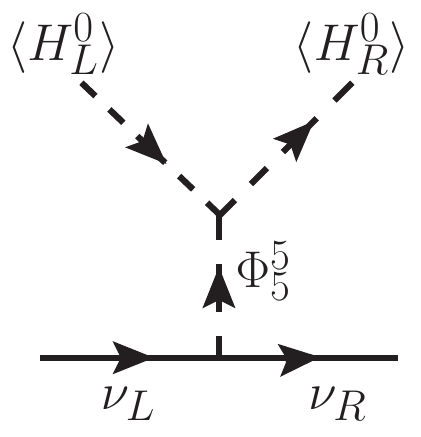}
    \caption{Diagram generating small Dirac neutrino masses via type-II Dirac seesaw. Here $\Phi_5^5$ is the neutral member of a Higgs doublet that has GUT scale mass, which acquires a small induced VEV after $SU(2)_{L,R}$ symmetry breaking triggered by the VEVs $\langle H^0_{L,R}\rangle$. }
    \label{fig:diracseesaw}
\end{figure}
\begin{equation}
{\cal L}_{\nu -{\rm mass}}^{\rm Dirac} = \frac{\overline{\nu}_L \nu_R \langle H_L^0 \rangle \langle H_R^0 \rangle}{M_G} \hspace{3mm} \Rightarrow \hspace{3mm} Y_\nu^{\rm Dirac} \sim \frac{M_I}{M_G} \approx 10^{-7}~.
\label{eq:nueff}
\end{equation}
Here we have inserted the numerical value of the ratio $(M_I/M_G)$ obtained from gauge coupling unification, see Eq. (\ref{eq:MIMGa}) of Sec. {\ref{sec:unification}}, as a reference value.  In addition to this suppression, there are flavor-dependent suppression factors in $Y_\nu^{\rm Dirac}$, leading to consistent neutrino oscillation phenomenology. 

The effective interactions shown in Eq. (\ref{eq:nueff}) arise as follows. In presence of the cubic scalar coupling $\mu_1 H_R^\dagger \Phi H_L$, as given in Eq. (\ref{eq:pot}), the neutral member ($\Phi_5^5$) of the Higgs doublet $\Phi_d(1,2,1/2)$ (the quantum numbers here refer to SM gauge symmetry) contained in $\Phi(\overline{\bf 5},{\bf 5})$ field  will acquire an induced VEV, $\langle \Phi^5_5 \rangle = v_\nu$, proportional to the product of VEVs $\kappa_{L} \kappa_R$ that break $SU(2)_L \times SU(2)_R$ symmetry, as shown in Fig. \ref{fig:diracseesaw} ~\cite{Lee:2016wiy}.  The third term of the Yukawa Lagrangian of Eq. (\ref{eq:YukLag}) will in turn lead to neutrino masses:
\begin{equation}
    v_\nu \simeq \frac{\mu_1 \kappa_L \kappa_R}{M^2_{\Phi_d}(1,2,1/2)}\, \hspace{3mm} \Rightarrow  \hspace{5 mm} M_\nu = Y_D^* \,v_\nu \, .
    \label{eq:numass}
\end{equation}
Note here that the mass of $\Phi_d(1,2,1/2)$ scalar is at the GUT scale, consistent with the extended survival hypothesis \cite{Dimopoulos:1984ha} that we adopt, while $\kappa_R = M_I$.  The original survival hypothesis of Georgi \cite{Georgi:1979md} applies to fermions in theories with multiple scales and assumes that only those fermions survive down to lower energy that cannot acquire masses consistent with symmetries. The extended survival hypothesis generalizes this idea to the scalar sector and assumes that only those scalar fields that are needed for subsequent symmetry breaking survive below the GUT scale \cite{Dimopoulos:1984ha,Mohapatra:1982aq}. We interpret the extended version of the hypothesis somewhat more broadly to also includes realism in phenomenology, with a fragment of $\eta(\overline{\bf 15}, {\bf 15})$ not involved in symmetry breaking surviving below $M_G$ to help with gauge coupling unification.  The mass parameter $\mu_1$ can be as large as $M_G$, but it could be smaller as well. Taking $\mu_1 \sim M_G$, one would obtain $Y_\nu^{\rm Dira} \sim (M_I/M_G)\sim 10^{-7}$, which is the largest allowed value for these couplings. 
This mechanism is somewhat analogous to the familiar type-II seesaw mechanism for Majorana neutrinos, where a Higgs triplet acquires an induced VEV, and leads to an effective $d=5$ operator for neutrino masses once the triplet field is integrated out. 
We therefore call this type-II Dirac seesaw~\cite{Berbig:2022hsm,Bonilla:2017ekt}. This is in contrast to type-I Dirac seesaw where local symmetries in the right-handed neutrino sector leads directly to operators of the  type $(\ell_L H_L) (\ell_R H_R)/M$~\cite{Berezhiani:1995yi,Gu:2006dc}.

\subsection{Reparametrization and predictive neutrino spectrum}

Without loss of generality one can work in a basis where the Yukawa coupling matrices $Y_D$ and $Y_u$ of Eq. (\ref{eq:fermionmass}) are chosen to be real and diagonal. This is possible by independent rotations on the three families of $\{({\bf 10},1) + (1, {\bf 10})\}$ and $\{({\bf \overline{5}}, 1) + (1, \overline{\bf 5})\}$ fermion fields.  In this basis, the $3 \times 3$ up-type quark mass matrix and the Dirac neutrino mass matrix take the form:
\begin{equation}
\hat{M}_u \equiv \hat{Y}_u \,\kappa_L =  {\rm diag.} (m_u, \, m_c, \, m_t),\hspace{5mm}
    \hat{M}_\nu \equiv Y_D^*\, v_\nu  = {\rm diag.} (m_{\nu_1}\,,m_{\nu_2}\,,m_{\nu_3} ) \, .
\end{equation}
The $3 \times 3$ charged lepton mass matrix $M_\ell$ will have a general form which can be written as
\begin{equation}
    M_\ell \equiv Y_\ell\, \kappa_L = \hat{U}^\dagger_{\rm PMNS} \hat{M}_\ell V_R^T \, , \hspace{5mm} {\rm where}\ \hat{M}_\ell = {\rm diag.} (m_e,\, m_\mu,\,m_\tau) \, .
    \label{eq:leptons}
\end{equation}
Here $\hat{U}_{\rm PMNS} = P U_{\rm PMNS} Q$, where $U_{\rm PMNS}$ is the charged lepton mixing matrix written in the canonical form with a single CP-violating phase $\delta_{CP}$, while $P$ and $Q$ are diagonal phase matrices that are unobservable in weak interactions.  (Note that there are no Majorana phases associated with charged current weak interactions in this model, since neutrinos are Dirac fermions.) The matrix $V_R$ in Eq. (\ref{eq:leptons}) is an arbitrary unitary matrix. 

We redefine the down-type quarks $(d,D)$ and the charged leptons $(e,E)$ to go from the original basis in which the mass matrices of Eq. (\ref{eq:fermionmass}) are written, to a new primed basis given by
\begin{align}
    &d_L =  V_R\, P^* d'_L , \hspace{3mm}  d_R =  V_R\, P^* d'_R , \hspace{3mm}
    D_L =  Q\, U_{\rm PMNS}^T\,  D'_L , \hspace{3mm} D_R =  Q\,U_{\rm PMNS}^T\, D'_R \notag\\
    &e_L = Q^* U_{\rm PMNS}^\dagger e_L' , \hspace{3mm}
 e_R = Q^*\, U_{\rm PMNS}^\dagger e_R' , \hspace{3mm}
    \nu_L = Q^* \nu_L', \hspace{3mm}
    \nu_R = Q^*\, \nu_R' \notag\\
    &E_L = V_R^*\, P E_L' , \hspace{5mm}
    E_R = V_R^* P E_R' \, .
    \label{eq:primedbasis}
\end{align}
These primes states are not quite the mass eigenstates; see further redefinitions given below in Eq. (\ref{eq:further}) that achieve this. 
The fermion mass matrices of Eq.~\eqref{eq:fermionmass} at the GUT scale in the new primed basis read as
\begin{eqnarray}
&{\cal M}_u =
\left(\begin{matrix}
     0 & \hat{M}_u \cr
    \hat{M}_u \frac{\kappa_R}{\kappa_L} & 0
    \end{matrix}\right),~~~~
     {\cal M}_\ell = 
    \left( \begin{matrix}
 0 & \hat{M}_\ell \\
    \hat{M}_\ell \frac{\kappa_R}{\kappa_L} & 0
    \end{matrix}\right), 
    \label{eq:Mufit}\\
& {\cal M}_d = 
 \begin{pmatrix} 0 & 
\hat{M}_\ell \\
\hat{M}_\ell \frac{\kappa_R}{\kappa_L} &~~ \frac{v_\Phi}{v_\nu} U_{\rm PMNS}^* \hat{M}_\nu U_{\rm PMNS}^T  
    \end{pmatrix}~.
    \label{eq:Mdfit}
\end{eqnarray}
The charged current interactions of the $W_L^{\pm \mu}$ gauge bosons with the  quarks and leptons in the new basis of Eq.~\eqref{eq:primedbasis} read as
\begin{equation}
    {\cal L}_{\rm cc} = \frac{g_{2L}}{\sqrt{2}}\left\{\left(\bar{u}_L \gamma_\mu  V_R P^*  d'_L\right) W^{\mu+}_L + \left(\bar{e}'_L U_{\rm PMNS}\  \gamma_\mu \nu'_L\right) W^{\mu-}_L  \right\} + h.c.
    \label{eq:CCurrent}
\end{equation}
We see that the $(e',\,\nu')$ fields correspond to physical leptons with their interactions written in the canonical form.  In the quark sector, while the up-type quark mass matrix is diagonal, ${\cal M}_d$ of Eq. (\ref{eq:Mdfit}) needs to be diagonalized to get to the physical basis.  This can be done by a bi-unitary transformation:
\begin{equation}
    \xi_L^\dagger {\cal M}_d \,\xi_R = {\rm diag.}(m_d, \,m_s, \,m_b,\, m_{D_1}, \,m_{D_2},\, m_{D_3})
    \label{eq:biunitary}
\end{equation}
where $\xi_{L,R}$ are $6 \times 6$ unitary matrices which can be parametrized in $3 \times 3$ block form as
\begin{eqnarray}
\xi_{L,R} = \left( \begin{matrix} \xi^{11} & \xi^{12} \cr \xi^{21} & \xi^{22}
\end{matrix} \right)_{L,R}~.
\end{eqnarray}
Denoting the light down-type mass eigenstates collectively as $d^0$ and the heavy ones as $D^0$, we have 
\begin{eqnarray}
d'_{L} &=& \xi^{11}_{L} \,d_{L}^0 + \xi_{L}^{12} \,D_{L}^0 \nonumber \\
D_{L}' &=& \xi_{L}^{21} \,d_{L}^0 + \xi_{L}^{22} \,D_{L}^0~
\end{eqnarray}
along with analogous relations for $(d'_R,\,D'_R)$ obtained with the replacement $L \rightarrow R$. The charged current quark interactions in this basis will involve the matrix $V_R P^* \xi_L^{11}$, which should be identified as the CKM mixing matrix, $\hat{V}_{\rm CKM}$.  To bring this (essentially) unitary matrix\footnote{The $3 \times 3$ matrix $\xi_L^{11}$, being a sub-block of the $6 \times 6$ unitary matrix $\xi_L$, is not  unitary in general.  However, departure from unitarity in $\xi_L$ is extremely small, of order $(m_{d_i}/m_{D_i}) \leq 10^{-10}$.} to the canonical form with a single phase, we write $\hat{V}_{\rm CKM} = P' V_{\rm CKM} Q'$, where $V_{\rm CKM}$ has a single phase, and where $P',\,Q'$ are diagonal phase matrices. We further redefine the fields so that the charged current quark interactions have the canonical form with a single phase in $V_{\rm CKM}$ while also enuring that the mass eigenvalues remain real:
\begin{eqnarray}
&u_L = P' \hat{u}_L, \hspace{3mm} U_R = P' \hat{U}_R \nonumber \\
&d_L^0=  Q^{\prime *} \hat{d}_L,\hspace{3mm}  d_R^0 = Q^{\prime *} \hat{d}_R
\label{eq:further}
\end{eqnarray}
These hatted fields $(\hat{u},\,\hat{d})$ are the mass eigenstates, and their charged current weak interactions involve the canonical CKM matrix.  These unitary transformations are relevant for proton decay discussions, which will be addressed in Sec. \ref{sec:protondecay}.

The CKM matrix, which is given by
\begin{equation}
V_{\rm CKM} = P^{\prime *} \,V_R\, P^*\, \xi_L^{11}\,Q^{\prime *}~,
\label{eq:CKM}
\end{equation}
is unconstrained in this scenario, as it contains the unspecified unitary matrix $V_R$, along with certain unspecified diagonal phase matrices. Nevertheless, the mass matrices of Eqs. (\ref{eq:Mufit})-(\ref{eq:Mdfit}) are constrained, as they have less parameters than observables.  To see this, notice that ${\cal M}_d$ in Eq. (\ref{eq:Mdfit}) contains only a single unknown parameter, the VEV ratio factor $(\kappa_R/\kappa_L)(v_\nu/v_\Phi)$, if we assume complete knowledge of the neutrino masses and mixings.  With this single parameter all three light quark masses $(m_d,\,m_s,\,m_b)$ should be fitted, which leads to two quantitative predictions.  In practice, in the fit that we carry out in Sec. \ref{sec:fermionfitting}, we use the known quark masses to predict two of the currently unknown parameters in neutrino oscillations, viz., $\delta_{CP}$ and $m_{\nu_1}$, the Dirac CP phase and the lightest neutrino mass.  It may be noted that in spite of two VEV ratios $(\kappa_R/\kappa_L)$ and $(v_\Phi/v_\nu)$ that appear in the mass matrix ${\cal M}_d$ of Eq. (\ref{eq:Mdfit}), only a single combination appears in the light down-type quark mass matrix.  This can be seen by considering a general block matrix of the form
\begin{align}
    {\cal M} =
    \begin{pmatrix}
        0 & m \\
        M' & M 
    \end{pmatrix} \, 
\end{align}
which obeys the hierarchy $m \ll M' \sim M$. Integrating out the heavy states will lead to the light-sector mass matrix  ${\cal M}_{\rm Light}  = m \left[ \mathbb{1} - ( \mathbb{1} + x\ x^\dagger)^{-1}\right]^{1/2}$, or equivalently  
\begin{align}
    {\cal M}_{\rm Light} {\cal M}_{\rm Light}^\dagger = m \left[ \mathbb{1} - ( \mathbb{1} + x\ x^\dagger)^{-1}\right] m^\dagger \, , \hspace{5mm}\ {\rm where}\ \quad x = M^{-1} M' \, . 
    \label{eq:Mdformula}
\end{align}
In the limit of $x \ll 1$, this matrix reproduces the usual seesaw formula, but Eq. (\ref{eq:Mdformula}) is valid even when $x \sim 1$~\cite{Babu:1995uu}. It becomes clear then that only a single parameter, the ratio of $(\kappa_R/\kappa_L)$ and $(v_\Phi/v_\nu)$, will appear in the light down-type quark mass matrix. 
In Sec. \ref{sec:fermionfitting} where we carry out numerical fits to fermion masses and mixings we have used the full $6\times 6$ matrix given in Eq.~\eqref{eq:Mdfit}, but we have cross-checked with the analytic expression of Eq. (\ref{eq:Mdformula}) for consistency.   We defer a discussion of fermion fits to Sec. \ref{sec:fermionfitting}, since that requires the values of the various Yukawa and gauge couplings at the GUT scale and at the intermediate scale, to which we now turn. 

\section{Gauge Coupling Unification}\label{sec:unification}
Under parity symmetry the gauge bosons of the two $SU(5)_{L,R}$ groups transform as $V_L^\mu \leftrightarrow V_R^\mu$, which requires equality of the two coupling, $\alpha_{5L} = \alpha_{5R} \equiv \alpha_G$.  This, in turn, implies that the three gauge couplings of the SM obey, at the unification scale $M_G$, the relations
\begin{equation}
    2\ \alpha_3= \alpha_2= \frac{13}{3}\ \alpha_Y  = \alpha_G\, .
    \label{eq:GUT-norm}
\end{equation}
The factor $2$ in front of $\alpha_3$ arises because the $SU(3)_c$ symmetry is embedded as a diagonal subgroup of $SU(5)_L \times SU(5)_R$, unlike $SU(2)_L$ which is entirely inside $SU(5)_L$. The factor $13/3$ in front of $\alpha_Y$ arises since hypercharge is contained in both $SU(5)$ factors, with the relations 
\begin{equation}
(B-L) = Y_L + Y_R,\hspace{6mm} \frac{Y}{2} = T_{3R} + \frac{(B-L)}{2}~.
\end{equation}
where the $(B-L)$ charges are those shown in Eqs. (\ref{eq:fermion})-(\ref{eq:vecferm}), and $T_{3R}$ is the third component of right-handed isospin.  The GUT-normalized hypercharge is therefore $\sqrt{3/13}\, (Y/2)$, leading to the factor shown in Eq. (\ref{eq:GUT-norm}).\footnote{$\sum_i(Y_i/2)^2 = 26/3$, where the sum goes over a family of fermions, including vector-like fermions shown in Eqs. (\ref{eq:fermion})-(\ref{eq:vecferm}). With the normalization factor in ($Y/2)$ of $\sqrt{3/13}$ included, this factor becomes 2, which is the same normalization as for the generators of $SU(2)_L$ -- a family contains four $SU(2)_L$ doublets.}
This yields  $\sin ^2 \theta_W(M_G) = 3/16$ at the unification scale, distinct from its value of $3/8$ in conventional GUTs such as $SU(5)$. Here $\sin^2\theta_W$ should increase while running down to the weak scale, as opposed to a decreasing value in conventional GUTs. As discussed in Sec. \ref{sec:Model}, this can be achieved in a phenomenologically consistent manner by lowering the mass of the scalar sub-multiplet $({\bf \bar{3}}, {\bf 2}, -1/6, {\bf 15}) \subset ({\bf \overline{15}, 15})$ 
(under $SU(3)_{cL} \times SU(2)_L \times U(1)_L \times SU(5)_R \subset SU(5)_L \times SU(5)_R$) below the GUT scale.  Above its mass, in going to higher momenta, $\alpha_{2L}$ will decrease faster than $\alpha_Y$, leading to a smaller value of $\sin^2\theta_W(M_G)$, while being consistent with phenomenology.\footnote{In the absence of the $({\bf \bar{3}}, {\bf 2}, -1/6, {\bf 15})$ sub-multiplet at the intermediate scale we found that $\alpha_{5R}$ grows in running down from $M_G$ to $M_I$ and becomes non-perturbative before reaching $M_I$.}  Inclusion of the $\eta({\bf \overline{15}, 15})$ preserves the strong CP solution via parity, since it induces no complex couplings in the Higgs potential, nor does it have any Yukawa interactions.

\begin{figure}
    \centering    \includegraphics[width=0.65\textwidth]{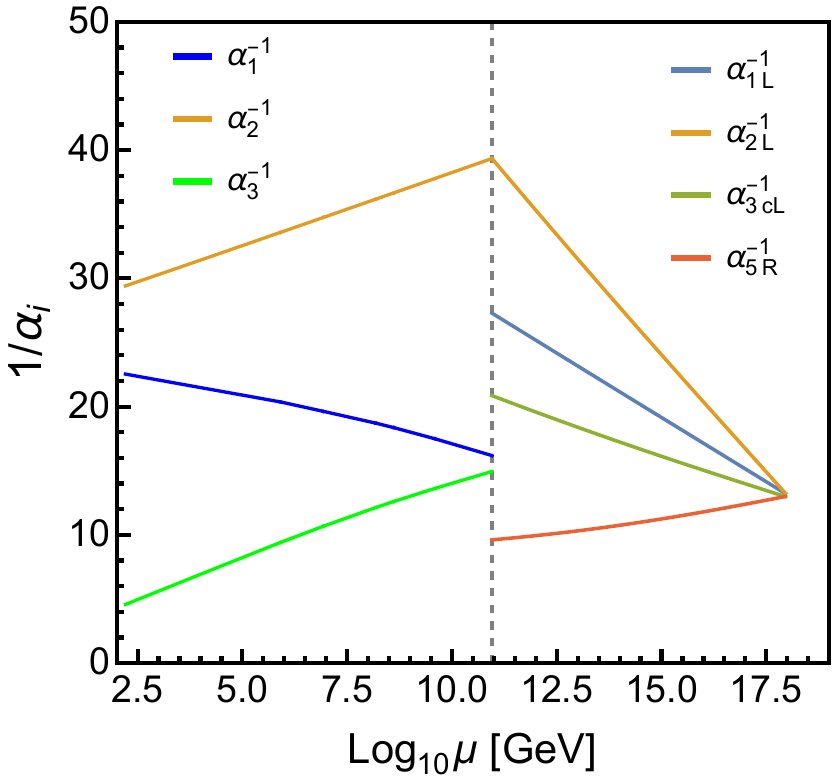}
    \caption{Evolution of the gauge coupling leading to unification in the model with the breaking chain given in Eq.~\eqref{eq:breakingchain}. We have included full three-loop RGE effects in obtaining this result. See text for a description of the benchmark point used here.   }
    \label{fig:unification}
\end{figure}

We have carried out a detailed analysis of the evolution of gauge couplings to see the prospects for unification of all gauge couplings with the symmetry breaking chain shown in Eq. (\ref{eq:breakingchain}) having a single intermediate scale $M_I$. As shown in Eq. (\ref{eq:breakingchain}), above $M_I$, the gauge symmetry is $SU(3)_{cL} \times SU(2)_L \times U(1)_L \times SU(5)_R$, while it is the Standard Model below $M_I$.
The two-loop renormalization group equations (RGE) for the gauge coupling evolution can be written in the form
\begin{align}
16 \pi^2 \frac{d g_i}{d t} =g_i^3 b_i+\frac{g_i^3}{16 \pi^2}\left[\sum_{j} b_{i j} g_j^2-\sum_{k} C_{ik} {\rm Tr}\left(Y_{k}^\dagger Y_k\right)\right] . 
\label{eq:gaugerun}
\end{align}
Here $b_i$ are the one-loop beta-function ($\beta$) coefficients, while $b_{ij}$ and $C_{ik}$ are the two-loop coefficients arising from gauge and Yukawa interactions respectively. For $m_t<\mu<M_I$, the indices take the values $i,j = (1_Y,2_L,3_c)$ and $k = (u,d,\ell)$. For $m_I<\mu<M_G$, we replace $\{b_i,b_{ij},C_{ik}\}$ by $\{b'_i,b'_{ij},C'_{ik}\}$ where $i,j = (1_L,2_L,3_{cL},5_R)$ and $k = (u_L,u_R,d_L,d_R,\ell_L,D)$.
The Yukawa coupling matrices in this momentum range are defined through Eq. (\ref{eq:YukI}) of Appendix \ref{sec:RGE}. 
 
In the momentum range $m_t \leq \mu \leq M_I$ we 
evolve the normalized gauge couplings $\alpha_{1Y} = (13/3) \alpha_Y$, as well as $\alpha_{2L}$ and $\alpha_{3cL}$.  In the range $M_I \leq \mu \leq M_G$ we run the normalized couplings $(\alpha_{1L},\, \alpha_{2L},\,\alpha_{3L}$) where $\alpha_{1L} = (5/3) \alpha_{L}$ with the charges under $U(1)_L$ of the various fields given in Eq. (\ref{eq:A1}) of Appendix. \ref{sec:A1}.  The factor (5/3) is the familiar $SU(5)$ normalization factor, but now associated with $U(1)_L$ rather than $U(1)_Y$. 
The unification conditions at $\mu = M_G$ are then
\begin{equation}
\alpha_{1L}^{-1}(M_G) = \alpha_{2L}^{-1}(M_G) = \alpha_{3cL}^{-1}(M_G) = \alpha_{5R}^{-1}(M_G) \equiv \alpha_G^{-1}~.
\label{unif}
\end{equation}
At $M_I$ the following boundary conditions hold:
\begin{align}
\alpha_{1 Y}^{-1}\left(M_I\right) & =\frac{8}{13} \alpha_{5 R}^{-1}\left(M_I\right)+\frac{5}{13} \alpha_{1 L}^{-1}\left(M_I\right) \, ,\\
\alpha_{3 c}^{-1}\left(M_I\right) & =\alpha_{3cL}^{-1}\left(M_I\right)+\alpha_{5 R}^{-1}\left(M_I\right) \, .
\label{eq:matchingcond}
\end{align}

The entire set of fermion fields will survive down to the intermediate scale $M_I$, since their masses arise only after the intermediate symmetry breaking. These fields, and the minimal set of scalar fields at $M_I$ needed for symmetry breaking, fermion mass generation, and to achieve gauge coupling unification, under the gauge group $[SU(3)_{cL}\times SU(2)_L \times U(1)_L]\times SU(5)_R$,  are:
\begin{eqnarray}
{\rm Fermions:}&& Q_L({\bf 3}, {\bf 2}, 1/6, 1) + \chi_R(1, 1, 0, {\bf 10})+ L_L(1, {\bf 2}, -1/2, 1)+ D_R({\bf 3},1, -1/3, 1)
      \notag \\
     && +~ U_R({\bf 3}, 1, 2/3, 1) + E_R(1, 1, -1, 1) + \psi_R(1, 1, 0, {\bf \oline{5}}) 
     \label{eq:MIfermion} \\
{\rm Scalars:}&&  H_R(1,1,0,{\bf 5}) + \Phi_D({\bf 3},1,-\frac{1}{3},{\bf \overline{5}}) + H_L^d(1,{\bf 2},\frac{1}{2},1) + \eta_K({\bf \bar{3}}, {\bf 2}, -1/6, {\bf 15}) \label{eq:MIscalar}
\end{eqnarray}
The scalar fields $(H_R,\,\Phi_D,\,H_L^d)$ of Eq.~\eqref{eq:MIscalar} are required for symmetry breaking and fermion mass generation, while the scalar field $\eta_K({\bf \bar{3}}, {\bf 2}, -1/6, {\bf 15})$ is used to achieve gauge coupling unification.  This choice of scalar spectrum is consistent with the extended survival hypothesis \cite{Dimopoulos:1984ha,Mohapatra:1982aq} that we have adopted.

The one-loop and two-loop $\beta$-function coefficients below $M_I$  are those of the SM (with the normalization $\alpha_{1Y} = (13/3) \alpha_Y$) and are given by
\begin{align}
&\left(b_{1Y}, b_{2L}, b_{3c}\right) = \left(\frac{41}{26},-\frac{19}{6},-7 \right), \label{eq:betaSM1}
\\[4pt]  
&   b_{ij} = 
\begin{pmatrix}
199/338 & 27/26 & 44/13 \\
9/26 & 35/6 & 12 \\
11/26 & 9/2 & -26
\end{pmatrix}, 
~~~~~~~  C_{ik} = 
\begin{pmatrix}
   17/26 & 5/26 & 15/26\\
   3/2 & 3/2 & 1/2 \\
   2 & 2 & 0
\end{pmatrix}~.
 \label{eq:betaSM}
\end{align} 
The gauge symmetry above $M_I$  is enhanced to $[SU(3)_{cL}\times SU(2)_L \times U(1)_L]\times SU(5)_R$.
With the fermions and scalar fields shown in Eqs. (\ref{eq:MIfermion})-(\ref{eq:MIscalar}) contributing to the RGE above $M_I$, the beta function coefficients are:
\begin{align}
& \left(b_{1 L}^{\prime}, b_{2 L}^{\prime}, b_{3 c L}^{\prime}, b_{5 R}^{\prime}\right) =\left(\frac{74}{15},\frac{13}{3},-\frac{7}{6},-\frac{20}{3}\right) \, , \label{eq:betaMI1} \\[4pt]
& b'_{ij} = 
\begin{pmatrix}
326/75 & 36/5 & 332/15 & 216/5\\
12/5 & 310/3 & 132 & 504 \\
83/30 & 99/2 & 307/3 & 360 \\
9/5 & 63 & 120  & 6338/15
\end{pmatrix}, 
& C'_{ik} = 
\begin{pmatrix}
   17/10 & 0 & 1/2 & 0 & 3/2 & 1\\
   3/2 & 0 & 3/2 & 0 & 1/2 & 0 \\
   2 & 0 & 2 & 0 & 0 & 5/2 \\
   0 & 9/2 & 0 & 5 & 0 & 3/2
\end{pmatrix} \, .
\label{eq:betaMI}
\end{align}
These $\beta$-functions coefficient are obtained using {\tt PyR@TE} package \cite{Sartore:2020gou} and cross-checked against known results. 

Before exploring gauge coupling unification numerically, it is worthwhile to analyze the analytic solutions to the one-loop RGE, which are given in Eq.~(\ref{eq:soln})  in Appendix \ref{app:analytical}.  There we have also included the calculable threshold effects from the vector-like fermion sector, which arise owing to the relations $M_{U_1}: M_{U_2}: M_{U_3} = m_u:m_c:m_t$,  $M_{E_1}: M_{E_2}: M_{E_3} = m_e:m_\mu:m_\tau$ and a less trivial relation for the vector-like down-quark mass ratios.  If we ignore these threshold effects from vector-like fermions for simplicity we would obtain the following expression for $\sin^2\theta_W(m_t)$ to one-loop accuracy:
\begin{align}
    \sin^2\theta_W(m_t) = \frac{3}{16}\left[ 1 + \frac{ \alpha}{6 \pi} \left\{ 13 (b_{1Y}- b_{2L}) \log\frac{m_t}{M_I} + (5 b'_{1L} - 13 b'_{2L} + 8 b'_{5R}) \log\frac{M_I}{M_G} \right\} \right] \, .
    \label{eq:weak}
\end{align}
Substituting the one-loop $\beta$-function coefficients from Eqs.~\eqref{eq:betaSM1} and \eqref{eq:betaMI1}, we obtain
\begin{equation}
   \sin^2\theta_W(m_t)  = \frac{3}{16} \left[ 1 + \frac{\alpha}{6 \pi} \left\{ -\frac{185}{3} \log\frac{M_I}{m_t} + (46 + 39) \log\frac{M_G}{M_I}  \right\} \right] \, .
   \label{eq:sin2tw}
\end{equation}
It is clear from the above equation that without the intermediate scale symmetry, which may be realized by setting $M_I \to M_G$, $\sin^2\theta_W(m_t)$ will be smaller than its value at $M_G$, which is inconsistent.  The factor 39 multiplying log$(M_G/M_I)$ in Eq. (\ref{eq:sin2tw}) arises from the $\eta_K({\bf \bar{3}}, {\bf 2}, -1/6, {\bf 15})$ scalar fragment, which helps in realizing the right value of $\sin^2\theta_W(m_t) \simeq 0.23$, consistent with other phenomenological constraints.  Without this contribution, $M_I$ would be too low to be consistent with LHC limits on vector-like quarks, see Eq. (\ref{eq:wrong}) and discussions in Appendix \ref{app:analytical}.
 
We have carried out the full two-loop RGE analysis numerically, which improves the one-loop results.  The two-loop coefficient of $g_{5R}$ is relatively large, $b'_{5_R 5_R}=6338/15$, and has an opposite sign compared to  the one loop coefficient of $b'_{5_R}=-20/3$. Since $\alpha_{5R}(M_I) \sim 0.1$, the two-loop terms correct the one-loop result significantly. We also investigated the three-loop beta-function coefficient of the $g_{5R}^7$ term, and found it to be  \cite{Poole:2019kcm,Sartore:2020gou}  $b''_{5_R 5_R 5_R}=2035937/108$, which shows slow convergence.
For better accuracy we therefore used the full three-loop beta funcations for our analysis. The following input values at $m_t (m_t) = 162.8$ GeV were used for the gauge couplings.   
\begin{align}
    \alpha_1 (m_t) = 0.01024 \, , \quad \alpha_2 (m_t) = 0.0340 \, , \quad \alpha_3 (m_t) = 0.1094 \, .
\end{align}
We first run these gauge couplings using Eq.~\eqref{eq:gaugerun} from $\mu=m_t$ to to the intermediate scale $M_I$, which is to be determined self-consistently. We have also included the two-loop threshold effects from the vector-like fermions by evaluating RGE at different vector-like fermion mass scales. These masses are obtained self-consistently from the fermion mass fits. For the up-type vector-like quarks and charged leptons we use the ratios $M_{U_1}: M_{U_2}: M_{U_3} = m_u:m_c:m_t$,  and $M_{E_1}: M_{E_2}: M_{E_3} = m_e:m_\mu:m_\tau$, and set $M_{U_3} = M_I$, along with $M_{E_3} = (m_\tau/m_t) M_I$. The running masses of the light quarks and leptons are listed at a scale $\mu = M_I$ in Table \ref{tab:runningmass}, which are used to determine the masses of the vector-like fermions.  For the down-type vector-like quarks we use the values obtained from our fermions mass fits  given for our benchmark point in Eq.~\eqref{eq:massesdown}. We ignore all GUT scale threshold effects.  We also ignore such threshold effects from scalar fields at $M_I$.  Furthermore, the scalar field $\eta_K({\bf \bar{3}}, {\bf 2}, -1/6, {\bf 15})$ is kept at $M_I$ for this numerical scan.  Since this field is not involved in symmetry breaking, keeping its mass at $M_I$ is not necessary, but is assumed for simplicity.  We have also explored two other scenarios where its mass is kept away from $M_I$, which will be discussed later in this section. Above $M_I$, the gauge couplings $\{\alpha_{1L},\alpha_{2L},\alpha_{3cL},\alpha_{5R}\}$ were run all the way to the unification scale $M_G$ identified from the matching conditions given in Eq.~\eqref{eq:matchingcond}. This numerical procedure involved scanning the space of three unknown parameters $\{\alpha_{5R}, M_I, M_{G} \}$ numerically such that all four gauge couplings are unified at a single value at $\mu=M_G$. We found successful unification when the gauge couplings at the intermediate scale take values given by
\begin{align}
    \alpha_{1L} (M_I)= 0.0367 , \hspace{3mm} \alpha_{2L} (M_I) = 0.0255 , \hspace{3mm} \alpha_{3L} (M_I) = 0.049, \hspace{3mm} \alpha_{5R} (M_I) =0.106 \, .
    \label{eq:couplings}
\end{align}
The intermediate scale ($M_I$), unification scale ($M_G$) and the unified gauge coupling ($\alpha_G$) were found for the benchmark point to be
\begin{equation}
    M_I= 9.02 \times 10^{10}\  \mathrm{GeV}, \quad M_G= 8.0 \times 10^{17}\ \mathrm{GeV}  ,
\quad \alpha_G^{-1} = 13.18   ~. \label{eq:MIMGa}
\end{equation}
The evolution of the gauge couplings in the various momentum regimes that lead to successful unification is depicted in Fig. \ref{fig:unification} where we have used the full three-loop RGE. 
We note that without the inclusion of the three-loop effects, the values of $M_I$ and $M_G$ would be $M_I = 3.38\times10^{10}$ GeV,\, $M_G = 3.09\times10^{17}$ GeV and the gauge coupling $\alpha_{5R}$ would be $\alpha_{5R}(M_I) = 0.122$.  The slow convergence of the loop-expansion can be understood partly because $\alpha_{5R}(M_I) = 0.122$ is not so small, and partly because the actual expansion parameter is not $\alpha/(4\pi)$, but rather $N \alpha/(4 \pi)$ where $N$ is an effective number of degrees of freedom, which is also not so small in the present scenario.

Note that the unification scale shown in Eq. (\ref{eq:MIMGa}) is below the Planck scale, and the unified gauge coupling $\alpha_G$ has a perturbative value. These aspects are important for the consistency of the theory and the reliability of calculations. The value of $M_G$ is also quite compatible with proton decay limits.  In Sec. \ref{sec:protondecay} we shall establish the consistency of the framework with proton decay mediated by the $SU(5)_R$ gauge bosons which have masses of order $M_I \simeq 10^{11}$ GeV.
\begin{figure}
\centering{
    \includegraphics[width=0.49\textwidth]{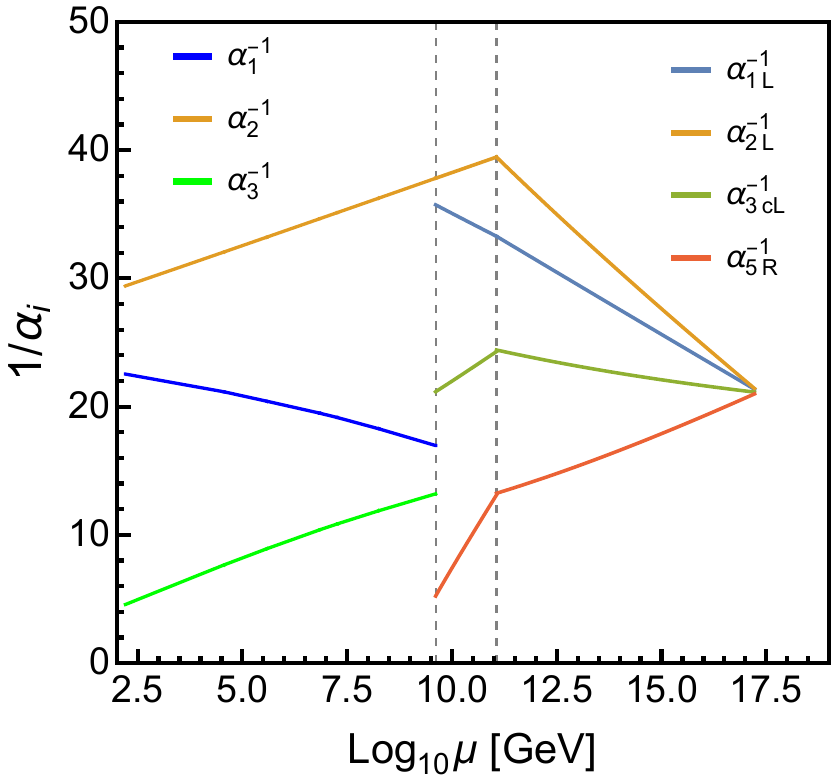}~
    \includegraphics[width=0.49\textwidth]{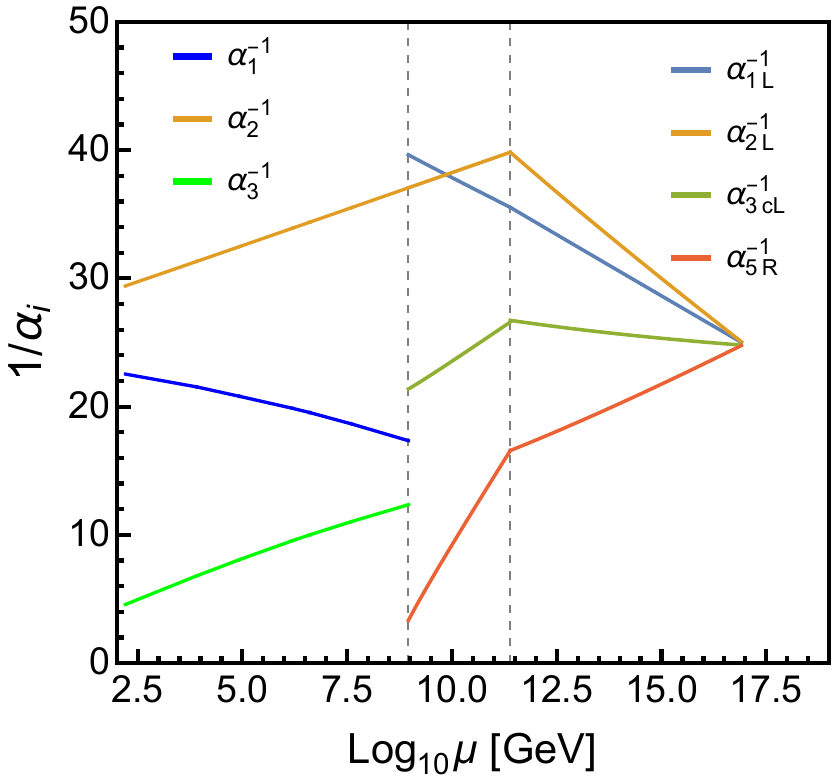}\\~~~~~~~(a)~~~~~~~~~~~~~~~~~~~~~~~~~~~~~~~~~~~~~~~~~~~~~~~~~~~~~(b)
}
    \caption{Gauge coupling evolution leading to unification with the mass of $\eta_K({\bf \bar{3}}, {\bf 2}, -1/6, {\bf 15}) \subset ({\bf \overline{15}, 15})$ above the intermediate scale. Left (Right) panel corresponds to keeping its mass at 28 (270) times the intermediate scale $M_I$.     }
    \label{fig:unificationp}
\end{figure}

The mass of the scalar field  $\eta_K$ $({\bf \bar{3}}, {\bf 2}, -1/6, {\bf 15}) \subset ({\bf \overline{15}, 15})$ is not required to be at $M_I$, as it has no role in symmetry breaking. It is therefore of interest to see how the unification picture changes if its mass is taken to be different from $M_I$.  If its mass is below $M_I$, we found the unification scale became higher, which may be not desirable. However, if its mass is above $M_I$, the scale of unification can be somewhat lowered.  This is depicted in Fig.~\ref{fig:unificationp} for two benchmark scenarios. The figure on the left (right) panel corresponds to keeping  the mass $\eta_K$ $({\bf \bar{3}}, {\bf 2}, -1/6)$ at $28 \times M_I$ ($270\times  M_I$). The intermediate scale $M_I$, the unification scale $M_G$, and the various gauge couplings for these two scenarios of Fig.~\ref{fig:unificationp} {\it (a)} and {\it (b)} are found to be:
\begin{align}
  (a)\ m_{\eta_K} &= 28\ M_I : \notag\\  
&\alpha_{1L} (M_I)=  0.0280, \hspace{3mm} \alpha_{2L} (M_I) = 0.0264 , \hspace{3mm} \alpha_{3L} (M_I) = 0.0472, \hspace{3mm} \alpha_{5R} (M_I) = 0.191 \, , \notag \\
& M_I= 4.08 \times 10^{9}\  \mathrm{GeV}, \quad M_G= 1.65 \times 10^{17}\ \mathrm{GeV} ,  \quad \alpha_G^{-1}=21.4 \, . \\
 (b)\ m_{\eta_K} &= 270\ M_I : \notag\\
& \alpha_{1L} (M_I)=  0.0252, \hspace{3mm} \alpha_{2L} (M_I) = 0.0270 , \hspace{3mm} \alpha_{3L} (M_I) = 0.0468, \hspace{3mm} \alpha_{5R} (M_I) = 0.30 \, , \notag \\
& M_I= 9.14 \times 10^{8}\  \mathrm{GeV}, \quad M_G= 7.20 \times 10^{16}\ \mathrm{GeV} , \quad \alpha_G^{-1} = 25.10 \, .
\end{align}
It is important to note that further increasing the mass of $\eta_K$ will decrease the GUT scale. However, such an increase would also result in the increase of the gauge coupling $\alpha_{5R}$, making it non-perturbative.

We conclude this section by noting that successful unification of gauge couplings has been achieved with perturbative values of all gauge couplings and the unification scale lying below the Planck scale, in the range $M_G = (7 \times 10^{16} - 8 \times 10^{17})$ GeV.

\section{Fermion Mass Fitting}\label{sec:fermionfitting}
In this section, we show that the model can successfully reproduce all fermion masses and mixings. In fact, the model is quite predictive in the neutrino sector, which can serve as a test of the model.  Having determined the gauge couplings at various momentum scales, we are now ready to analyze these predictions quantitatively.    
As discussed in Sec. \ref{sec:fermionmass}, a successful fit to all fermion masses would require finding acceptable values of the light down-type quark masses ($m_d,\,m_s,\,m_b)$ from the eigenvalues of the $6 \times 6$ matrix ${\cal M}_d$ of
Eq. (\ref{eq:Mdfit}), while being consistent with neutrino oscillation data. In practice, we use the down-type quark masses as inputs and predict the currently unknown parameters in the neutrino sector, viz., ($\delta_{CP}$,\, $m_{\nu_1}$) -- the CP-violating oscillation parameter and the lightest neutrino mass.

\begin{figure}
    \centering    \includegraphics[width=0.7\textwidth]{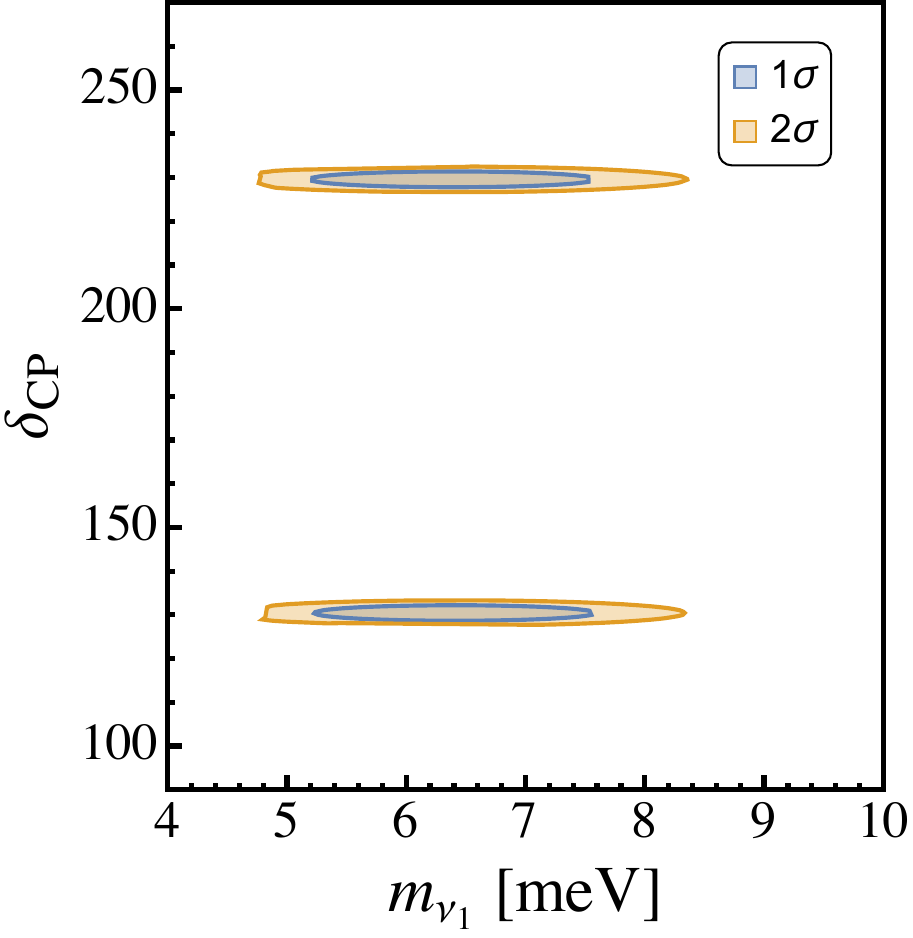}
    \caption{$1\sigma$ (blue) and $2\sigma$ (orange) normal hierarchy prediction for $\delta_{CP}$ and the lightest neutrino mass $m_{\nu_1}$ with best fit for $\delta_{CP} = (130.4 \pm 1.2)^\circ$ or $(229.6\pm1.2)^\circ$. We use $3\sigma$ uncertainties for the rest of the oscillation parameters.     }
    \label{fig:nuprediction}
\end{figure}

Neutrino Dirac masses arise from the Yukawa couplings $Y_D$  and the induced VEV $v_\nu$, given in Eq.~\eqref{eq:numass}. To obtain fits to the down-type quark masses arising from ${\cal M}_d$ of Eq.~\eqref{eq:Mdfit}, we use the neutrino oscillation data (within $3\sigma$ uncertainties \cite{Esteban:2020cvm}) as input and find the eigenvalues of $\mathcal{M}_d^\dagger \mathcal{M}_d$  from which we derive the light mass eigenvalues $(m_d,m_s,m_b)$. The eigenvectors of ${\cal M}_d$ of Eq.~\eqref{eq:Mdfit} do not play any role here, since the CKM matrix is arbitrary containing an unknown unitary matrix $V_R$ (see Eq. (\ref{eq:CKM})) which may be adjusted freely. 

We find it convenient to perform the fermion mass fits at the intermediate scale $M_I$, rather than at the unification scale $M_G$, where Eq. (\ref{eq:Mdfit}) is valid.  This is because we don't have simple expressions for the eigenvalues of ${\cal M}_d$ that can be applied at $M_G$.  The form of ${\cal M}_d$ will be modified at $M_I$ due to the renormalization group evolution of its various elements. We express ${\cal M}_d(M_I)$ as
\begin{equation}
    {\cal M}_d (M_I)=\left(\begin{array}{cc}
0 & m_\ell (m_t) \frac{\eta_\ell \eta'_{\ell_L}}{\eta'_{d_L}} \\
m_\ell(m_t)   \frac{\kappa_R}{\kappa_L} \frac{\eta_\ell \eta'_{\ell_L}} {\eta'_{d_R}} &~~~ \frac{v_\Phi \eta_\nu \eta_U^2}{v_\nu \eta'_D} U_{\rm PMNS}^* \hat{M}_\nu U_{\rm PMNS}^T
\end{array}\right)   \, .
\label{eq:FitMD}
\end{equation}
Here we have introduced various scaling factors $\eta_i$ and $\eta'_\alpha$.   The $\eta_i$  factors 
correspond to the running of the Standard Model Yukawa couplings from $\mu=m_t$ to $\mu=M_I$:
\begin{equation}
    \eta_{u_i,d_i,e_i,\nu_i} = \frac{Y_{u_i,d_i,e_i,\nu_i}(M_I)}{Y_{u_i,d_i,e_i,\nu_i}(m_t)}~.
\end{equation}
These factors are the same as the ratios of the running masses at the two energy scales.  Since the effect of the lighter quark and lepton Yukawa couplings on the RG evolution is negligible, we have $\eta_u = \eta_c$, $\eta_e = \eta_\mu = \eta_\tau \equiv \eta_\ell$, and $\eta_d = \eta_s$. The factor $\eta_U$ is the RGE factor for the running of $U_{\rm PMNS}$, which is in principle flavor dependent, but all these factors are essentially one, with the largest deviation from one being  proportional to $Y_\tau^2 \sim 10^{-4}$.  It is also worth noting that the $\eta$ factors appearing in the (2,2) block of Eq. (\ref{eq:FitMD}) may be absorbed into $v_\Phi$, and therefore are not really needed for the fit.
The $\eta'_\alpha$ factors are the running factors of the various Yukawa couplings of Eq. (\ref{eq:YukI}) between $M_I$ and $M_G$ defined as
\begin{equation}
    \eta_\alpha' = \frac{Y_\alpha(M_G)}{Y_\alpha(M_I)}~~~~{\rm for}~\alpha = (u_L, \,u_R,\, d_L,\, d_R,\, D)~.
\end{equation}
Here again there is a flavor dependence, but the lighter fermions of the same type have the same RGE factor.  

The $\eta_i$ factors are obtained by solving the SM renormalization group equations to two-loop accuracy.  We choose the benchmark values of $M_I= 9.02 \times 10^{10}$ GeV,  corresponding to the unification picture shown in Fig. \ref{fig:unification}.  We include the threshold effects from the vector-like fermions which have masses below $M_I$ (except for $U_3$ which has a mass  $M_{U_3} = M_I$ for this fit). The $\eta_i$ values are found to be
\begin{align}
  \{ \eta_{u,c},\  \eta_{t},\ \eta_{d,s},\ \eta_b,\ \eta_{\ell} \} = \{ 0.491 ,   0.537 , 0.50 ,  0.458 , 1.00 \} \, ,
    \label{eq:etaMI}
\end{align}
where $\eta_{\ell} \equiv \eta_{e,\mu,\tau}$.

To obtain the $\eta'_\alpha$ factors to go from $M_I$ to $M_G$, we numerically solve the two-loop beta functions for the Yukawa couplings $(Y_{uL}, Y_{uR},Y_{dL}, Y_{\ell L}, Y_D)$. These matrices relevant for $M_I \leq \mu \leq M_G$ are defined in Eq. (\ref{eq:YukI}) in Appendix \ref{sec:RGE} along with their boundary conditions at $M_G$ given in Eq. (\ref{eq:boundary}).  The full set of two-loop RGE is presented in Appendix \ref{sec:RGE}. Corresponding to the unification picture of Fig. \ref{fig:unification} and the values of the gauge couplings given in Eq. (\ref{eq:couplings}) and the scales given in Eq. (\ref{eq:MIMGa}) we find the $\eta_\alpha'$ factors to be
\begin{align}
   & \{ \eta'_{u_L,c_L},\ \eta'_{t_L},\ \eta'_{u_R,c_R},\ \eta'_{t_R},\ \eta'_{d_L,s_L} \}= \{ 0.468,   0.48 ,  0.0951 , 0.134  , 0.468 \} , \notag \\
    & \{ \eta'_{b_L},\ \eta'_{\ell_L},\ \eta'_{d_R,s_R},\ \eta'_{b_R},\ \eta'_{D} \} = \{ 0.475 , 0.805 , 0.150 ,  0.092,   0.30 \} \, . 
    \label{eq:etaMG}
    \end{align}
These values are obtained by setting the boundary conditions  $Y_{\ell}= Y_{\ell L}$ and $Y_u=Y_{uL}$ at $M_I$ and then randomly choosing the rest of the Yukawa couplings and accepting only those that match the boundary conditions at the GUT scale,  $Y_{uL}=Y_{uR}$ and $Y_{dL} = Y_{dR} = Y_\ell^T$, see Eq. (\ref{eq:boundary}). It turns out that due to the smallness of all Yukawa couplings, except for $Y_t$, in this running no off-diagonal entries in $Y_{u_L}$, $Y_{u_R}$ and $Y_{\ell_L}$ are induced by the RGE flow. This makes it relatively easy to select the random Yukawa couplings that match the GUT scale boundary conditions. 

We choose as input the quark and lepton masses at a scale $\mu=m_t$ the values tabulated in Ref. \cite{Huang:2020hdv}.  These values are summarized in Table \ref{tab:runningmass}.  Here we also list the running masses at $\mu = M_I$, using the $\eta_i$ factors given in Eq. (\ref{eq:etaMI}). The goal for the fermion mass fit is then to reproduce correctly the masses of the down-type quarks arising from Eq.~\eqref{eq:FitMD} with the charged lepton masses at $M_I$ taken as input. For this analysis we set $M_I = 9.02 \times 10^{10}$ GeV, see Eq. (\ref{eq:MIMGa}), and vary the known neutrino oscillation parameters in their 3-sigma ranges.  We then numerically scan the parameters of Eq.~\eqref{eq:FitMD} to obtain fits to the down-type quarks masses. Correct masses are reproduced, but only for a narrow range of the currently unknown parameters $m_{\nu_1}$ and $\delta_{CP}$.  The 1-sigma and 2-sigma allowed regions of these parameters are shown in the $(m_{\nu_1}-\delta_{CP})$ plane in Fig. \ref{fig:nuprediction}.  At the 2-sigma level this analysis shows that the neutrino parameters should lie in the range
\begin{eqnarray}
\delta_{CP} &=&  (130.4 \pm 1.2)^\circ~{\rm  or} ~  (229.6\pm1.2)^\circ \nonumber \\
m_{\nu_1} &=& (4.8 - 8.4)~{\rm meV}~.
\end{eqnarray}
The two solutions for $\delta_{CP}$, which differ by a change of its sign, lead to the same eigenvalues of ${\cal M}_d^\dagger {\cal M}_d$, and cannot be distinguished by the down-type quark masses.

\begin{table}
    \centering
    \begin{tabular}{|c|c|c|}
   \hline
        $\mu/{\rm GeV}$  & $m_t$ \cite{Huang:2020hdv} & $M_I$ \\ \hline
       $m_d/{\rm MeV}$  & $2.56 \pm 0.18$ & $1.28 \pm 0.090$\\ 
       $m_s/{\rm MeV}$ & $50.9 \pm 4.41$ & $25.45 \pm 2.205$\\
       $m_b/{\rm GeV}$ & $2.702 \pm 0.025$ & $1.237\pm 0.0114$\\
       \hline
       $m_u/{\rm MeV}$ & $1.18 \pm 0.20$ & $0.579 \pm 0.098$ \\
       $m_c/{\rm GeV}$ & $0.594 \pm 0.017$ & $0.2916 \pm 0.0083$\\
       $m_t/{\rm GeV}$ & $161.98 \pm 0.75$ & $86.98 \pm 0.40$\\
       \hline
       $m_e/{\rm MeV}$ & $0.48583 \pm 0.00045$ & $0.48568 \pm 0.00045$ \\
       $m_\mu/{\rm GeV}$ & $0.102347 \pm 0.000019$ & $0.10231 \pm 0.000019$\\
       $m_\tau/{\rm GeV}$ & $1.73850 \pm 0.00014$ & $1.73815 \pm 0.00014$ \\
       \hline
    \end{tabular}
    \caption{Running masses of the quarks and charged lepton at $\mu =m_t$ and  at  $\mu = M_I$. These are obtained with $M_I = 9.02 \times 10^{10}$ GeV, and by including the threshold corrections arising from vector-like fermions $(U,D,E)$ which have masses below $M_I$. 
    }
    \label{tab:runningmass}
\end{table}
We present a benchmark fit for the masses at the intermediate scale by setting the parameter $v_\Phi \eta_\nu \eta_u^2/(v_\nu \eta'_D) = 1.0 \times 10^7 M_I$:
\begin{align}
   {\cal M}_d (M_I) =& 
    \begin{pmatrix}
    0& 0& 0& -0.008 & 0 & 0 \\
    0& 0 & 0 & 0 & 0.17 & 0 \\
    0& 0& 0 & 0 & 0 & -2.95 \\
    -7.53\cdot10^{5}& 0 & 0 & 1.19\cdot10^{4} & 4.42 \cdot 10^6\, e^{ 1.06\, i}& 9.60 \cdot 10^6\, e^{ 2.76\, i} \\
    0 & 1.58 \cdot 10^{8} & 0 &  \bullet & 2.75 \cdot 10^7 & 2.03 \cdot 10^7\, e^{ -0.044\, i} \\
    0 & 0 & -4.38 \cdot 10^{9} &  \bullet & \bullet & 2.26 \cdot 10^7 \\
    \end{pmatrix}
    \label{eq:BMfit}
\end{align}
Here the $``\bullet"$ represents the complex conjugated entry such that the lower $(2,2)$ block of Eq.~\eqref{eq:BMfit} is hermitian. The matrix ${\cal M}_d$ can be block-diagonalized with a biunitary transformation given in Eq. (\ref{eq:biunitary}). While the $d_L-D_L$ mixing angles, parametrized by the off-diagonal entries of $\xi_L$ are extremely small, of order $m_{d_i}/M_{D_i}$, this is not the case for $d_R-D_R$ mixing, which can be as large as ${\cal O}(1)$. As a result, the seesaw approximation is not very good when applied to Eq. (\ref{eq:BMfit}).  Numerical diagonalization of Eq. (\ref{eq:BMfit}) yields the light and heavy eigenvalues to be
\begin{align}
   & \{ m_{d},m_{s},m_{b} \} (M_I) = \{ 8.06 \times 10^{-4}\ {\rm GeV},\ 2.83 \times 10^{-2}\ {\rm GeV},\ 1.24 \ {\rm GeV}  \} ,\  \label{eq:masslightd} \\
     & \{ m_{D_1}, m_{D_2}, m_{D_3} \}  (M_I) = \{ 1.05 \times 10^7\ {\rm GeV},\ 1.62 \times 10^8\ {\rm GeV},\ 4.38 \times 10^9\ {\rm GeV} \} .
     \label{eq:massesdown}
\end{align}
This fit corresponds to  $m_d\ (2 \, {\rm GeV}) = 2.92\ {\rm MeV}$, $m_s (2\ {\rm GeV}) = 0.102$ GeV, and  $m_b (m_b) = 4.16$ GeV. These values are in good agreement with observations, although the value of $m_d$ is on the lower side.  However, this is consistent with one lattice evaluation which finds  $m_d(2\ {\rm GeV})=3.68\pm0.29\pm0.10$ of Ref.~\cite{Aoki:2012st}. 

The model only admits normal ordering of neutrino masses. The predictions of the model for the CP-violating phase and the lightest neutrino mass are shown in Fig.~\ref{fig:nuprediction} where the allowed region is the shaded one. For values that lie outside of the shaded region we find that $m_d(2\,{\rm GeV})$ becomes way smaller than acceptable.   For instance, if $\delta_{CP} = 200^\circ$, which is outside the shaded region, the maximum value of the down-quark mass is $m_d\, (2 \, {\rm GeV}) \simeq 24.9$ keV, well below the experimental value. Similarly, for $m_{\rm Lightest} = 1$ meV, which is outside the shaded region, the maximum down-quark mass is found to be  $m_d\, (2 \, {\rm GeV}) \simeq 23.5$ keV.  

With the neutrino fit, we can evaluate the VEV $v_\nu$ and from there the value of the cubic scalar coupling $\mu_1$ (see Eq. (\ref{eq:numass})). For this benchmark fit we find $v_\nu = 3.3 \times 10^{-7}$ GeV, leading to $\mu_1 = 1.36 \times 10^{16}$ GeV. This shows self-consistency of the analysis and confirms the naturalness of the small Dirac neutrino masses within the framework.

\section{Proton Decay}\label{sec:protondecay}
In this section, we show the consistency of the model with proton decay constraints. Since the gauge bosons of $SU(5)_R$ have masses of order $M_I \sim 10^{11}$ GeV, it is imperative to establish that they do not mediate rapid proton decay.  The $H_R(1,5)$ Higgs field, which also has a mass of order $M_I$, contains a color-triplet scalar that could potentially mediate proton decay which should be consistent with experimental limits.  Finally, the $(X_L^\mu,\,Y_L^\mu)$ gauge bosons of $SU(5)_L$ would mediate proton decay for which we estimate the lifetime.

The couplings of $X_R^\mu$ and $Y_R^\mu$ gauge bosons of $SU(5)_R$ to fermions are given, in the original basis, as \cite{Langacker:1980js}
\begin{align}
{\cal L}_{X_R^\mu,Y_R^\mu} &=\frac{g_{5R}}{\sqrt{2}} \left[\epsilon_{\alpha \beta \gamma} {(U^{T \gamma}_L}) C \gamma_\mu u_R^\beta X_R^{\mu \alpha} + (E_L^T) C \gamma_\mu d_R^\alpha X^{\mu *}_{R \alpha} \right. 
+ (D_L^{T \alpha})C \gamma_\mu  e_R X^{\mu *}_{R \alpha}   \nonumber \\
   &~~~+ \epsilon_{\alpha \beta \gamma} (U_L^{T \gamma}) C \gamma_\mu d_R^\beta Y_R^{\mu \alpha}- \left. (D_L^{T \alpha})C \gamma_\mu \nu_R Y^{\mu *}_{R \alpha} - (E_L^T) C \gamma_\mu u_R^\alpha Y^{\mu *}_{R \alpha} \right] + h.c.
\label{eq:int}
\end{align}
It turns out that these interactions do not lead to proton decay, owing to the structure of the mass matrices of the model, as given in Eq. (\ref{eq:fermionmass}). These matrices   imply that in order to identify the light up-type quark and charged lepton fields as ($u_R,\,e_R)$, one must interchange the fields as $u_R \leftrightarrow U_R$ and $e_R \leftrightarrow E_R$.  Furthermore, the fields $U_L$ and $E_L$ are heavy with no mixing with the light $u_L$ and $e_L$ fields. 
In the down-type quark sector there exists $d_R-D_R$ mixing, which could be of order one, as well as $d_L-D_L$ mixing which is suppressed by factors of the type $(Y_{d_i}/Y_{D_i})(M_W/M_I)$, which turns out to be less than $10^{-10}$ for the first two families of quarks relevant for proton decay. 
It follows then that in  Eq. (\ref{eq:int}) all terms involve at least one heavy field, except for the second last term with the $\nu_R$ field, with a suppressed coupling when converted to light $d_L$ quark field.  But this single term does not lead to proton decay since it conserves baryon and lepton numbers, as can be seen by assigning $B$ and $L$ numbers  of 1/3 and 1 to the $Y_R^\mu$ gauge boson. We thus conclude that the proposed form of the mass matrices, given in Eq. ((\ref{eq:fermionmass})), is consistent with proton lifetime limits, even when the $SU(5)_R$ gauge bosons have masses of order $M_I \sim 10^{11}$ GeV.

The $B$-violating interactions of $(X_L^\mu,\,Y_L^\mu)$ gauge bosons of $SU(5)_L$, which have masses of order $M_G$, are given in the original basis of fermion fields by
\begin{align}
{\cal L}_{X_L^\mu,Y_L^\mu} &=\frac{g_{5}}{\sqrt{2}} \left[\epsilon_{\alpha \beta \gamma} {(U^{T \gamma}_R}) C \gamma_\mu u_L^\beta X_L^{\mu \alpha} + (E_R^T) C \gamma_\mu d_L^\alpha X^{\mu *}_{L \alpha} \right. 
+ (D_R^{T \alpha})C \gamma_\mu  e_L X^{\mu *}_{L \alpha}   \nonumber \\
   &~~~+ \epsilon_{\alpha \beta \gamma} (U_R^{T \gamma}) C \gamma_\mu d_L^\beta Y_L^{\mu \alpha}- \left. (D_R^{T \alpha})C \gamma_\mu \nu_L Y^{\mu *}_{L \alpha} - (E_R^T) \gamma_\mu u_L^\alpha Y^{\mu *}_{L \alpha} \right] + h.c.
\label{intpp}
\end{align}
We now write down this Lagrangian  in terms of the physical mass eigenstates of quark and lepton fields.
We adopt the convention where the CKM matrix in the quark sector and the PMNS matrix in the lepton sector have their canonical forms each with a single phase. The relevant transformation is given in Eq. (\ref{eq:primedbasis}) and Eq. (\ref{eq:further}), along with the interchanges $u_R \leftrightarrow U_R$ and $e_R \leftrightarrow E_R$ so that the light fields are denoted as $(u,d,e)$. 
The transformed Lagrangian is then given by (with the hats and superscript $^0$ dropped from the fields):
\begin{align}
    {\cal L}_{X_L^\mu,Y_L^\mu} &=\frac{g_{5}}{\sqrt{2}} \left[\epsilon_{\alpha \beta \gamma} {(u^{T \gamma}_R}) C \gamma_\mu (P')^2\,u_L^{\beta} X_L^{\mu \alpha} + (e_R^T)  C  \gamma_\mu  \xi_L^{11} Q'^*d_L^{\alpha} X^{\mu *}_{L \alpha} \right. 
   \nonumber \\
   &~~~+ (d_R^{T \alpha}) Q'^*(\xi_R^{21})^T   C \gamma_\mu  e_L X^{\mu *}_{L \alpha}  + \epsilon_{\alpha \beta \gamma} (u_R^{T \gamma}) C \gamma_\mu (P')^2 V_{\rm CKM} d_L^{\beta} Y_L^{\mu \alpha}
     \nonumber \\
   &~~~- \left. (d_R^{T \alpha}) Q'^* (\xi_R^{21})^T U_{\rm PMNS}  C \gamma_\mu    \nu_L Y^{\mu *}_{L \alpha} - (e_R^T)  C \gamma_\mu (\xi_L^{11} Q'^* V_{\rm CKM}^\dagger) u_L^{\alpha} Y^{\mu *}_{L \alpha} \right] + h.c.
\label{eq:intp}
\end{align}

It is straightforward to derive the baryon number violating effective four-fermion operator by integrating out the $(X_L^\mu,\,Y_L^\mu)$ gauge bosons.  It may appear that Eq. (\ref{eq:intp}) differs significantly, owing to the appearance of the matrices $(\xi_R^{21})^T$ from its analogues in standard $SU(5)$ GUT, but this is not the case. The numerical structure of the matrices $|\xi_L^{11}|$ and $|\xi_R^{21}|$ corresponding to the benchmark fit for fermion masses, Eq. (\ref{eq:BMfit}), are found to be:
\begin{align}
    |\xi_L^{11}| =  
   \left(
\begin{array}{ccc}
 0.9967 & 0.0808 & 0.00029 \\
 0.0802 & 0.989 & 0.1239 \\
 0.00974 & 0.1235 & 0.9922 \\
\end{array}
\right) \, , \hspace{5mm}
    |\xi_R^{21}| =  
  \left(
\begin{array}{ccc}
 0.9997 & 0.00545 & 0.00298 \\
 0.00193 & 0.1948 & 0.8686 \\
 0 & 0 & 0.4370 \\
\end{array}
\right) \, .
\end{align} 
This structure implies that the light $d_R$ field is contained almost entirely in the original $D_R$ field.  For the leading decay mode of the proton, $p \rightarrow e^+ \pi^0$, is a function of $(\xi_R^{21})_{11}$, which is close to unity.  Thus the model prediction for the lifetime of the proton is very similar to that in standard $SU(5)$, except that in the present case it is much longer, owing to the higher unification scale $M_G = (7 \times 10^{16} - 8 \times 10^{17})$ GeV.  We estimate the corresponding lifetime of the proton to be of order $(10^{38} - 10^{42})$ yrs, which is well beyond the reach of forthcoming experiments JUNO, HyperKamiokande and DUNE (for reviews on current status and future prospects see Refs. \cite{Babu:2013jba,Dev:2022jbf}.) We note, however, that if the GUT scale is lowered in some variations of the present model by a factor of 10, the lifetime of the proton will be shortened by a factor of $10^4$, in which case the forthcoming experiments will be sensitive to its decay.

Color-triplet scalars arising from $H_R (1,{\bf 5})$  and $H_L(5,1)$, denoted as $H_{R,L}^c$, can mediate proton decay. Since the mass of $H_R^c$ is at $M_I$, one should verify that this field does not mediate rapid proton decay.  We shall see that with the flavor structure of the mass matrices given in Eq. (\ref{eq:fermionmass}), $H_R^c$ field having a mass of order $M_I$ is indeed safe from proton decay constraints. The interactions of the $H_{R}^c$ field with quarks and leptons are given by the Lagrangian (in the original basis)
\begin{eqnarray}
{\cal L}_{\rm Yuk}^{H_R^c} &\supset& (Y_u)_{ij}^* \left[\frac{1}{2}(u_{Ri} d_{Rj} -  d_{Ri} u_{Rj}) H_R^c + U_{Li} E_{Lj} (H_R^c)^*\right]  \nonumber \\
&+& (Y_\ell^\dagger)_{ij} \left[ 
(\nu_{Ri} d_{Rj} - e_{Ri} u_{Rj})(H_R^c)^* + D_{Li} U_{Lj}H_R^c\right]  + h.c.
\label{eq:int-triplet}
\end{eqnarray}
Analogous interactions of the $H_L^c$ field are obtained by the interchange $R \rightarrow L$ in Eq. (\ref{eq:int-triplet}), with identical Yukawa couplings owing to parity symmetry. Now, with the fermion mass matrix structure of Eq. (\ref{eq:fermionmass}), one must make the interchanges \{$u_R \leftrightarrow U_R, e_R \leftrightarrow E_R\}$ so that the light states are labeled as $(u,\,e)$.  Furthermore, the $(U_L,\,E_L)$ fields are entirely in the heavy states.  These features imply that the $H_R^c$ color-triplet scalar has couplings which involve at least one heavy state, except for the term $\nu_R d_R H_R^c$, which has terms with only light fermion fields.  However, with this single term, there is no baryon number violation, as can be seen by assigning $H_R^c$ a $B$ charge of $-1/3$. The field $H_R^c$ behaves as a leptoquark in this case, not mediating proton decay.

Let us comment briefly about proton decay mediated by higher dimensional operators suppressed by the Planck scale.  One might worry, with the $SU(5)_R$ surviving down to $M_I \sim 10^{11}$ GeV, that the gauge bosons could mediate proton decay in presence of such Planck suppressed operators.  One such operator is the $\bar{\chi}_R \chi_L  \Phi \Phi/M_{\rm Pl}$.  Such a term, if present, would induce a bare mass of order $M_I^2/M_P = 10$ GeV in the (2,2) block of ${\cal M}_u$ of  (\ref{eq:fermionmass}). Such a mass term would induce $u-U$ mixing, and thus lead to proton decay mediated by the $(X_R^\mu,\,Y_R^\mu)$ gauge bosons. Keeping in mind that the mass $U_1$ is $10^{-5} M_I = 10^5$ GeV, the $u_R-U_R$ mixing angle arising from here is about $10^{-4}$. The $u_L-U_L$ mixing is estimated to be much smaller, of order $(m_u \times 10\ {\rm GeV})/(10^5\ {\rm GeV})^2 = 10^{-12}$. These operators do not induce mass terms in the (2,2) block of ${\cal M}_\ell$ of  Eq. (\ref{eq:fermionmass}).  Consequently, the leading proton decay diagram from Eq. (\ref{eq:int}) would arise by combining the fourth and fifth term of this equation. The amplitude of this diagram has a suppression factor of $10^{-22}$, arising from the $u_L-U_L$ mixing suppression of $10^{-12}$ and the $d_L-D_L$ mixing suppression of order $10^{-10}$.  The lifetime of the proton arising from these diagrams is of order $10^{58}$ yrs, which is well within limits.

\section{Strong CP Solution via Parity}
\label{sec:thetabar}

Parity solution to the strong CP problem~\cite{Beg:1978mt,Mohapatra:1978fy,Mohapatra:1995xd,Kuchimanchi:1995rp,Mohapatra:1997su,Babu:2001se}
is an alternative to the popular axion solution~\cite{Peccei:1977hh,Weinberg:1977ma,Wilczek:1977pj}. The parity solution has an advantage over the axion solution in that unlike the currently viable invisible axion models~~\cite{Kim:1979if,Shifman:1979if,Dine:1981rt,Zhitnitsky:1980tq}, which are subject to destabilization by quantum gravitational effects which violate all global symmetries ~\cite{Kamionkowski:1992mf,Barr:1992qq,Holman:1992us}, these models are stable as long as the parity breaking scale is less than about 100 TeV~\cite{Berezhiani:1992pq}. The latter property makes these low scale parity-symmetric models experimentally testable. 

A realistic UV-complete implementation of the parity solution to the strong CP problem was provided in Ref.~\cite{Babu:1989rb}. The fermion content of that left-right symmetric model is the same as in the present $SU(5)_L \times SU(5)_R$ model. There is therefore a good chance that the GUT embedding of the model can also solve the strong CP problem via parity symmetry. Indeed, we find this to be the case, as will be detailed in this section.

\subsection{\texorpdfstring{\boldmath{$\overline{\theta}$}}{theta}  at tree-level}

The strong interactions admit a CP-violating $\theta$-parameter in the Lagrangian denoted as
\begin{equation}
{\cal L}^\theta_{\rm QCD} = \frac{\theta_{\rm QCD}}{32 \pi^2} G_{\mu \nu}^a \tilde{G}^a_{\mu \nu} ~
\end{equation}
where $G_{\mu\nu}^a$ stand for the gluon fields and where $\tilde{G}^a_{\mu\nu} = \frac{1}{2}\epsilon_{\mu \nu\alpha \beta}\, G^{a \,\alpha\beta}$. The parameter $\theta_{\rm QCD}$ itself is unphysical, as its value can be altered by chiral rotations on the quark fields.  A physical observable remain invariant and is given by
\begin{equation}
\overline{\theta} = \theta_{\rm QCD} + {\rm arg}\left\{{\rm det}\left(M_q\right)\right\}
\end{equation}
where $M_q$ is the quark mass matrix. $\theta_{\rm QCD}$ is odd under parity, and therefore in a parity-symmetric theory it would vanish.  The quark mass matrix is in general not parity symmetric, since its entries arise from parity breaking VEVs.  However, if the determinant of the quark mass matrix is real in a parity-symmetric theory, then $\overline{\theta} = 0$ at tree-level.  Quantum corrections would in general induce nonzero but finite value for $\overline{\theta}$, but this may be within the experimentally allowed range of $\overline{\theta} \leq 1.19 \times 10^{-10}$, arising from neutron EDM limits \cite{Dragos:2019oxn,ParticleDataGroup:2022pth}. 

In the  $SU(5)_L \times SU(5)_R$ with parity, the gauge fields transform under $P$ as 
\begin{eqnarray}
G_{L \mu}^a(t,x) \rightarrow  G_{R \mu}^a(t,-x) \times s(\mu) \nonumber \\
G_{R \mu}^a(t,x) \rightarrow  G_{L \mu}^a(t,-x) \times s(\mu) 
\end{eqnarray}
\begin{equation}
\text{where}~~  s(\mu) =
    \begin{cases}
      + 1 & \text{for}~~ \mu = 0\\
      - 1 & \text{for}~~ \mu = i = 1, 2, 3
    \end{cases}       
\end{equation}
Here $G_{L\mu}^a$ are the $SU(5)_L$ gauge bosons, while $G_{R\mu}^a$ stand for their $SU(5)_R$ counterparts.  Consistent with this transformation, a certain $\theta'$ term is allowed in the $SU(5)_L \times SU(5)_R$ theory, which is given by 
\begin{equation}
{\cal L}_\theta' = \frac{\theta'}{32 \pi^2} \left(G_{L\mu \nu}^a \tilde{G}^a_{L \mu \nu} - G_{R\mu \nu}^a \tilde{G}_{R \mu \nu}^a\right)~.
\label{eq:thetap}
\end{equation}
$SU(3)_c$ is embedded as a diagonal subgroup of $SU(3)_{cL} \times SU(3)_{cR}$ in the full theory.  
The axigluons of the broken $SU(3)$ are $G_{A\mu}^a = (G_{L\mu}^a + G_{R\mu}^a)/\sqrt{2}$, while the massless QCD gluons are $G_\mu^a = (G_{L\mu}^a-G_{R\mu}^a)/\sqrt{2}$.  The terms from Eq. (\ref{eq:thetap}) involve  only the axigluons, and therefore  $\theta_{\rm QCD}$ involving the usual gluon fields is zero.

With the quark mass matrices given in Eq. (\ref{eq:fermionmass}), det[$M_q] = {\rm det}[{\cal M}_u {\cal M}_d$] is real in the present GUT framework, implying that $\overline{\theta} = 0$ at the tree-level. We now proceed to show that all one-loop corrections to $\overline{\theta}$ are also vanishing, owing to the structure of the theory.

\subsection{Vanishing of one-loop \texorpdfstring{$\overline{\theta}$}{theta} contributions }
\label{sec:thetabar1}

Here we show that all one-loop diagrams which could potentially generate nonzero contributions to $\overline{\theta}$ are vanishing within the model.  We follow the procedure adopted in Ref. \cite{Babu:1989rb} to perform this calculation, which we find very convenient.  First of all, we evaluate the one-loop diagrams at the GUT scale, so that the fermion mass matrices have the form given in Eq. ({\ref{eq:fermionmass}}).  We work in the original basis of fermion fields, before any field rotations are performed.  For ease of writing we shall use the notation $Y_d$ which is to be identified as $Y_\ell^T$ for the Yukawa coupling matrices appearing in ${\cal M}_d$ of Eq. (\ref{eq:fermionmass}).  

The one-loop contribution to $\overline{\theta}$ are computed from corrections to the quark mass matrices ${\cal M}_u$ and ${\cal M}_d$ of Eq. (\ref{eq:fermionmass}).  These corrections arise via the  exchange of scalar fields as well as gauge fields present in the theory. We summarize these interactions here. The Yukawa interactions of scalar fields with fermions are contained in the left-right symmetric Higgs doublet couplings that appear in ${\cal M}_u$ and ${\cal M}_d$ of Eq. (\ref{eq:fermionmass}), the color-triplet couplings given in Eq. (\ref{eq:int-triplet}) and their left-handed counterparts, and the Yukawa couplings of the $\Phi(\overline{5},5)$ field given in Eq.~\eqref{eq:YukLag}, which can be expanded to the form
\begin{eqnarray}
-{\cal L}_{\rm Yuk} &\supset& (Y_D^*)_{ij} \left[  
\left(\overline{D}_{R j \beta} D_{Li}^\alpha\right) (\phi_{DD}^o)^\beta_\alpha + \left(\overline{D}_{R j \alpha} D_{Li}^\alpha \right) \phi_{DD}^s 
 + \left( D_{Ri}^T C \nu_{Rj}\right)\phi_{D\nu}
 \right. \nonumber \\
&+& \left. \left( D_{Ri}^T C e_{Rj}\right)\phi_{De} + \left(\overline{L}_{Li} D^c_{Rj} \right) \phi_{LD} + \left(\overline{L}_{Li} \nu_{Rj}\right) \phi_{L \nu} + \left(\overline{L}_{Li} e_{Rj}\right) \phi_{Le}
\right]~.
\end{eqnarray}
Here the notation used for the various scalar fields is self-explanatory, and their quantum numbers  under the SM gauge symmetry are listed in Eq. (\ref{eq:qn}) of Appendix \ref{sec:A2}.  Other scalar fields, such as the $\Sigma_L(75,1)$ and $\eta(\overline{15},15)$ used in the model play no role in this calculation, since they have no Yukawa couplings to the fermions.

The gauge bosons of the full $SU(5)_L \times SU(5)_R$ theory have various interactions involving fermions.  In particular, the $(X_R^\mu,\, Y_R^\mu)$ gauge interactions are shown in Eq. (\ref{eq:int}), while the   $(X_L^\mu,\, Y_L^\mu)$ interactions are given in Eq. (\ref{intpp}).  Additionally, the gauge interactions of the axi-gluon fields $G_A^\mu$ and the axi-$Z'$ field $Z_A^\mu$, which have masses of order $M_I$, are given by
\begin{align}
    {\cal L}^{G^\mu_A,Z^\mu_A} = -  \sqrt{\frac{3}{10}} g_5 (Y_L - Y_R) \left( \bar{f}_{L,R}^i \gamma_\mu f_{L,R}^i \right) Z_A^\mu\ - \frac{g_5}{\sqrt{2}} (T^a G_{A}^{\mu a})_{\alpha \beta} \left( \oline{f}_{L\alpha}^i \gamma_\mu f_{L\beta}^i - \oline{f}_{R\alpha}^i \gamma_\mu f_{R\beta}^i \right) .
\end{align}
Here $Y_L$ and $Y_R$ are the $U(1)_L$ and $U(1)_R$ charges of various fields, listed in Eq. (\ref{eq:A1}) of Appendix \ref{sec:A1} (The $Y_R$ charges are identical to $Y_L$ charges, but when referred to fields transforming under $SU(5)_R$). In addition, the theory has $W_R^\pm$ and $Z_R^0$ gauge bosons, as well as the SM gauge bosons.  We shall collectively denote the ($G_A^\mu,\, G^\mu,\, Z^\mu, A^\mu, Z_A^\mu$) as $V^\mu$.  It is noteworthy that all these $V^\mu$ fields have flavor diagonal couplings to fermions, unlike the $(X^\mu_{L,R},\,Y^\mu_{L,R})$ gauge bosons.  It is also worth noting that the $W_{L,R}^\pm$ gauge bosons, having couplings only to fermions of one specific chirality, will not participate in the quark mass matrix corrections, when evaluated in the Feynman gauge, which we adopt.  

\begin{figure}
    \centering
    \includegraphics[scale=0.40]{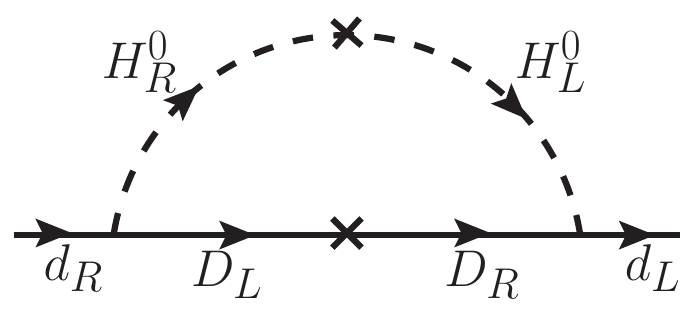}
    \includegraphics[scale=0.4]{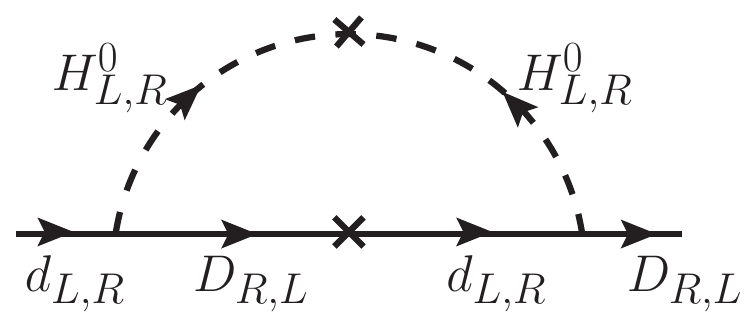}
    \includegraphics[scale=0.4]{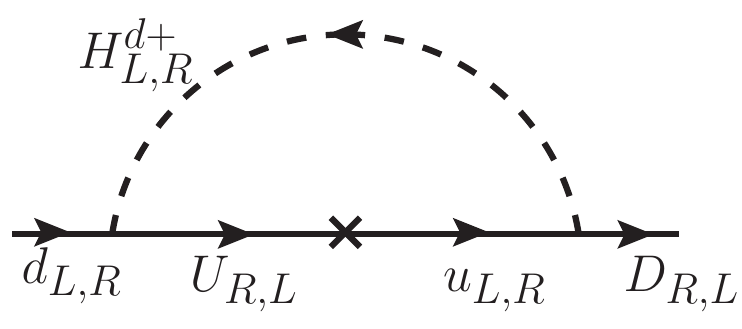} \\
  (a)~~~~~~~~~~~~~~~~~~~~~~~~~~~~~~~~~~~~(b)~~~~~~~~~~~~~~~~~~~~~~~~~~~~~~~~~~~~~(c)~~~ \\[10pt]
     \includegraphics[scale=0.4]{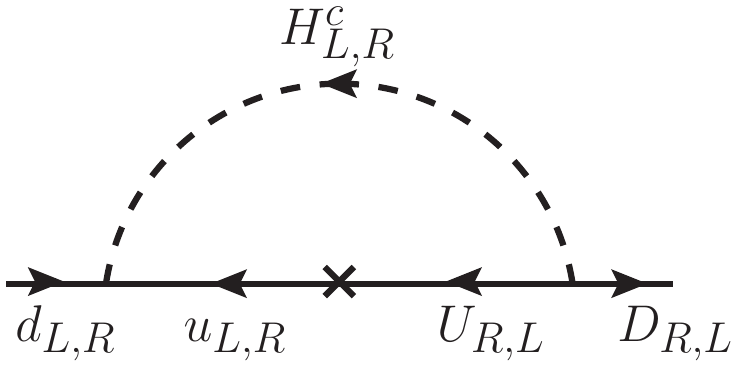}
     \includegraphics[scale=0.4]{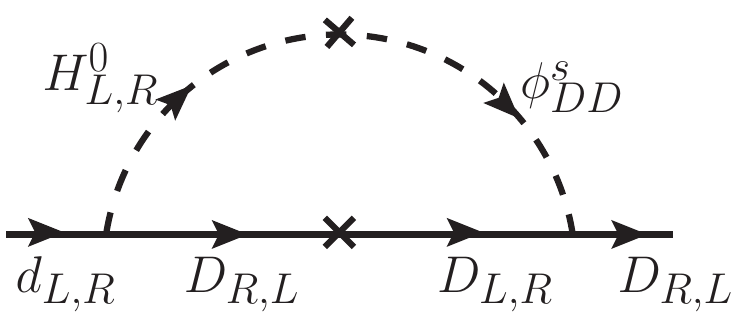}
     \includegraphics[scale=0.4]{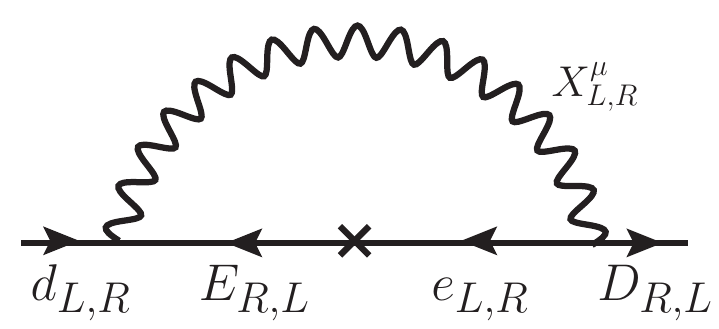}\\
     (d)~~~~~~~~~~~~~~~~~~~~~~~~~~~~~~~~~~~~(e)~~~~~~~~~~~~~~~~~~~~~~~~~~~~~~~~~~~~~(f)~~~ \\[10pt]
      \includegraphics[scale=0.4]{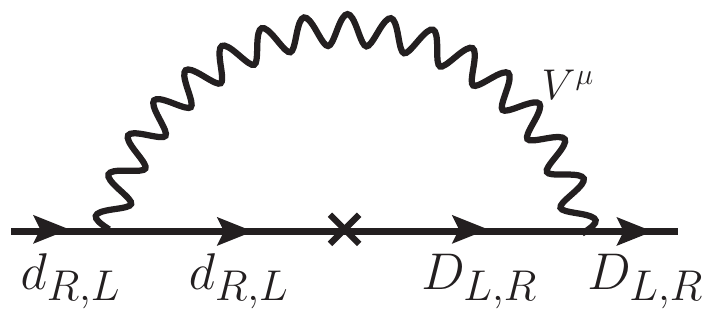}
      \includegraphics[scale=0.4]{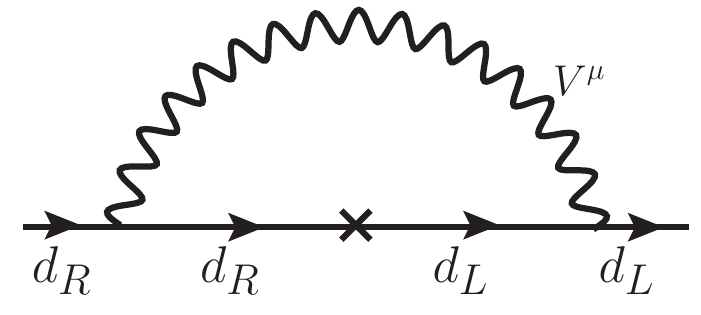}\\(g)~~~~~~~~~~~~~~~~~~~~~~~~~~~~~~~~~~~~(h)   
    \caption{Diagrams leading to one-loop radiative corrections to the down-type quark mass matrix. Here $V^\mu$ stands collectively for the gauge bosons ($G_A^\mu,\, G^\mu,\, Z^\mu, A^\mu, Z_A^\mu$) which all have flavor-consering interactions.}
    \label{fig:down_loop}
\end{figure}

\begin{figure}
    \centering
    \includegraphics[scale=0.4]{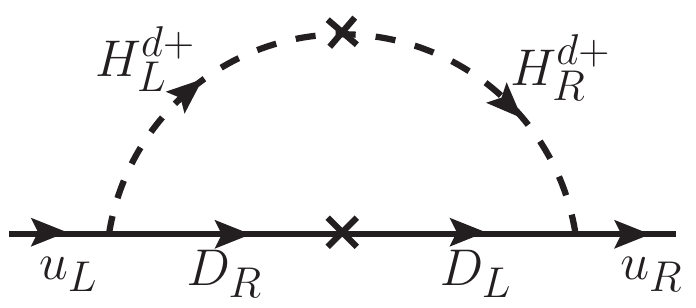}
    \includegraphics[scale=0.4]{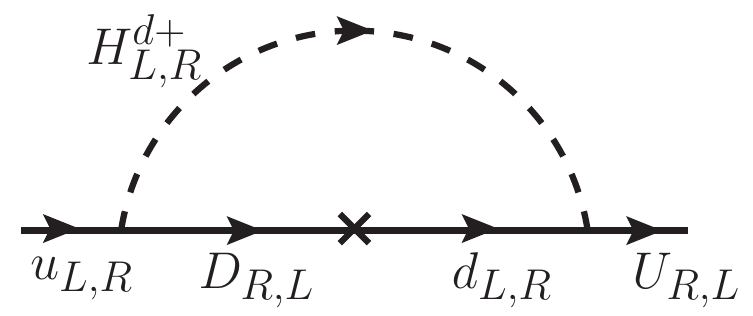}
    \includegraphics[scale=0.4]{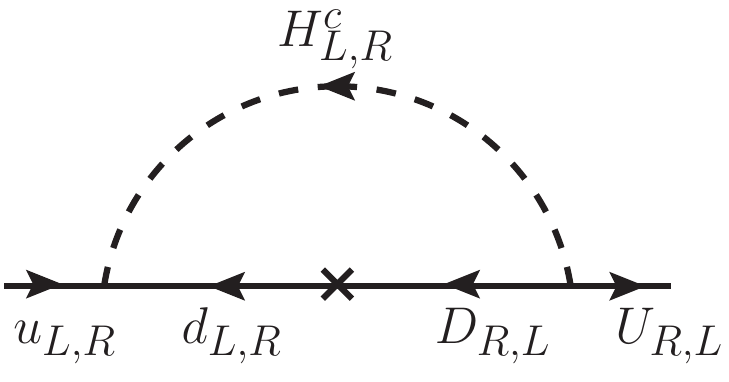} \\
(a)~~~~~~~~~~~~~~~~~~~~~~~~~~~~~~~~~~~~(b)~~~~~~~~~~~~~~~~~~~~~~~~~~~~~~~~~~~~~(c)~~~ \\[10pt]
     \includegraphics[scale=0.4]{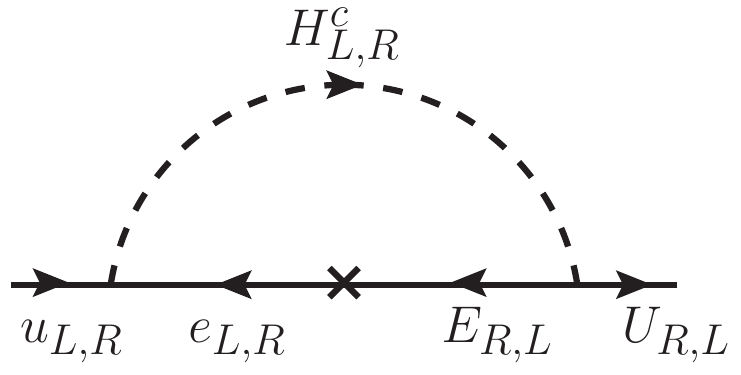}
     \includegraphics[scale=0.4]{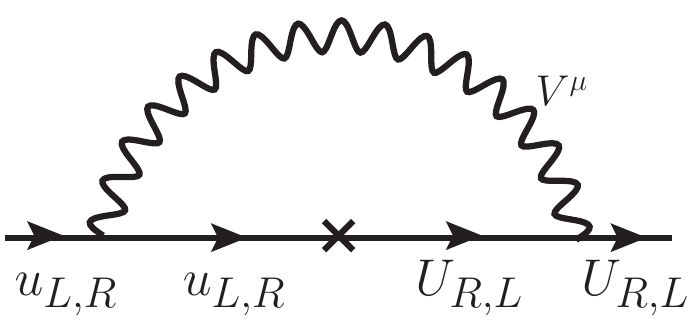}
     \includegraphics[scale=0.4]{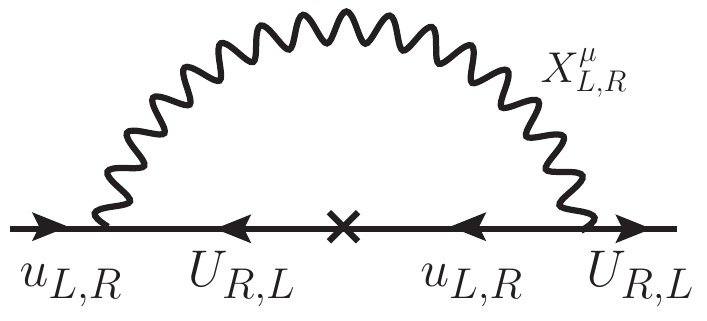}\\
(d)~~~~~~~~~~~~~~~~~~~~~~~~~~~~~~~~~~~~(e)~~~~~~~~~~~~~~~~~~~~~~~~~~~~~~~~~~~~~(f)~~~ \\[10pt]
     \includegraphics[scale=0.4]{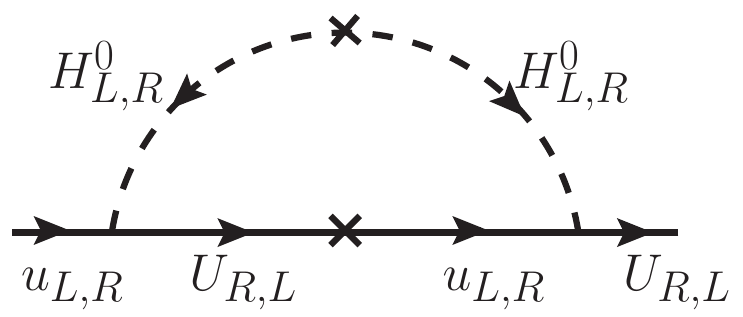}
      \includegraphics[scale=0.4]{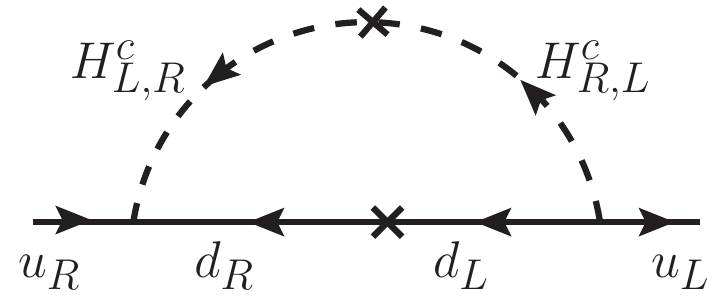}\\    (g)~~~~~~~~~~~~~~~~~~~~~~~~~~~~~~~~~~~~(h) \\[10pt]
    \caption{Diagrams leading to one-loop radiative corrections to the up-type quark mass matrix.}
    \label{fig:up_loop}
\end{figure}

Having identified all interactions that can correct the quark mass matrices, we proceed to evaluate the relevant one-loop diagrams. We adopt standard perturbation theory techniques to evaluate $\overline{\theta}$, following  the procedure outlined in Ref. \cite{Babu:1989rb}.\footnote{More recently, Fock-Schwinger method has been applied to evaluate the loop contributions to $\bar{\theta}$ in Ref.~\cite{Hisano:2023izx}.} We work in the flavor basis, where Eq. (\ref{eq:fermionmass}) is written.  The mass matrices for ${\cal M}_u$ and ${\cal M}_d$ are treated as part of the interaction Lagrangian. All the one-loop diagrams that can potentially contribute to $\overline{\theta}$ in this basis are shown in Fig. \ref{fig:down_loop} for ${\cal M}_d$ and in Fig. \ref{fig:up_loop} for ${\cal M}_u$.  The crosses in the internal fermion lines in these diagrams represent chirality flipping interactions.  Since we work in the flavor basis, there could be multiple such chirality flips which should be summed.  This can be easily done, which leads to the full tree-level propagator with all possible mass insertion, in the down-type quark sector, given by
\begin{equation}
\overline{f}_R\left(\mathcal{M}_d^{ \dagger} \frac{k^2}{k^2-\mathcal{M}_d \mathcal{M}_d^{ \dagger} }\right) f_L
\label{eq:proj}
\end{equation}
where $f_{R,L} \equiv (d_{R,L}, \,D_{R,L})^T$ are 6-component column vectors.  Analogous expressions hold for ${\cal M}_u$ and ${\cal M}_\ell$ as well.  

We can write down the loop-corrected quark mass matrix as
 \begin{eqnarray}
 M_q~=M^{(0)}_q + \delta M_q =~M^{(0)}_q (1+C)
 \end{eqnarray}
where $M^{(0)}_q$ is the tree-level quark mass matrix for $q=u,d$ given by Eq.~\eqref{eq:fermionmass} and $C=C_1+C_2+...$ is the loop contribution with subscripts denoting 1-loop, 2-loop, etc. Since Det($M^{(0)}_q$) is  real for $q=u,d$, one can write $\overline{\theta}$ as
 \begin{eqnarray}
 \overline{\theta}~=~{\rm Im~Tr}C_1  +~{\rm Im~Tr}(C_2 -\frac{1}{2}C_1^2) + ... 
\end{eqnarray}
We denote the one-loop correction to the quark mass matrix as
\begin{eqnarray}
 \delta M_q=\left(\begin{array}{cc} \delta M^q_{LL} & \delta M^q_{ LH} \\ \delta M^q_{ HL} & \delta M^q_{ HH} \end{array}\right) \, ,
\label{eq:correction}
 \end{eqnarray} 
where $H,L$ stand for the heavy and light sector respectively. 
The induced $\overline{\theta}$ from $q=d$ sector is then given by
\begin{align}
\overline{\theta}~=~{\rm Im~Tr}\left[-\frac{1}{\kappa_L\kappa_R}\delta M^d_{LL}(Y_d^\dagger)^{-1}M_D Y_d^{-1} \right. 
\left. +\frac{1}{\kappa_L}\delta M_{LH}^d Y_d^{-1} +\frac{1}{\kappa_R} \delta M_{HL}^d (Y_d^\dagger)^{-1}  \right] .
\label{eq:theta}
\end{align}
Here we have defined $M_D = Y_D v_\Phi$, which is the (2,2) block of ${\cal M}_d$. The contribution for for $q=u$ can be obtained similarly by replacing $Y_d \to Y_u$, but without the first term in Eq.~\eqref{eq:theta} since $M_U=0$. 
Note that $\delta M_{HH}^{q}$ does not contribute to $\overline{\theta}$ at one-loop order. As a result we do not include diagrams where both external legs are  the heavy $D$-quarks or $U$-quarks in Fig. \ref{fig:down_loop} and in Fig. \ref{fig:up_loop}.  It is also worth noting that the $\overline{\theta}$ corrections arising from $\delta M_{LL}^u$ will be automatically zero since $M_U = 0$, even though the correction to the mass matrix itself is nonzero. We have included such diagrams in Fig. \ref{fig:up_loop} (a) and (h), but their contribution to $\overline{\theta}$ vanishes.  

The propagators relevant to the evaluation of Figs. \ref{fig:down_loop} and \ref{fig:up_loop} can be obtained from the full tree-level propagator of Eq. (\ref{eq:proj}) by appropriate projection operators.  It is helpful to define the inverse of the propagator matrix as
\begin{equation}
\left(\mathcal{M}_d \mathcal{M}_d^{ \dagger}-k^2\right)^{-1} \equiv\left(\begin{array}{ll}
x(k^2) & ~~y(k^2) \\
y^{\dagger}(k^2) & ~~z(k^2)
\end{array}\right)~.
\end{equation}
Here $x=x^\dagger$, $z=z^\dagger$ and $y$ are $3 \times 3$ block matrices, which obey the following relations from matrix multiplication: 
\begin{align}
& \left(\kappa_R^2 Y_d^\dagger Y_d + M_D M_D^{\dagger}-k^2\right) y^{\dagger}=-\kappa_L M_D Y_d^\dagger x, \\
& \kappa_L Y_d Y_d^\dagger X+Y_d M_D^{\dagger} y^{\dagger}=\frac{1}{\kappa_L}\left(I+k^2 x\right), \\
& y=-\kappa_L H Y_d M_D^{\dagger} z, ~~
{\rm where}~~
H=H^\dagger = \left(\kappa_L^2 Y_d Y_d^\dagger-k^2\right)^{-1}~,\\
&\kappa_L y^\dagger Y_d  M_D^\dagger + z(M_D M_D^\dagger + \kappa_R^2 Y_d^\dagger Y_d  - k^2) = I~.
\end{align}
Although we have denoted $M_D^\dagger$ to be different from $M_D$ for generality, in our case owing to parity symmetry we have $M_D = M_D^\dagger$, which we shall adopt.  
The tree-level interactions corresponding to the crosses in Fig. (\ref{fig:down_loop}) can now be read off from the effective Lagrangian given by
\begin{align}
-\mathcal{L}_{\mathrm{eff}}^{\mathrm{tree}} = ~& \bar{D}_R\left[\frac{k^4}{\kappa_L} Y_d^{-1} y(k^2)\right] D_L+\bar{d}_R\left[k^2 Y_d\, \kappa_R\ \,z(k^2)\right] D_L  \notag  \\
&+\bar{D}_R\left[\frac{k^2}{\kappa_L} Y_d^{-1}\left[I+k^2\ x(k^2)\right]\right] d_L +\bar{d}_R\left[k^2 Y_d\, \kappa_R \,y^{\dagger}(k^2)\right] d_L +\text { h.c. }
\label{eq:tree1}
\end{align}
Similar expressions are valid for the up-type quark sector as well as the charged leptons sector, except that since $M_U = M_E = 0$ for these sectors, $y(k^2) = 0$, and consequently the analogs of the first and last terms of Eq. (\ref{eq:tree1}) will be vanishing.

We are now ready to evaluate the contributions of each graph in Fig. \ref{fig:down_loop} and Fig. \ref{fig:up_loop} to $\overline{\theta}$.  We begin with Fig. \ref{fig:down_loop} (a).  The correction to the down quark mass matrix from here is given by
\begin{equation}
\begin{aligned}
\delta M_{L L}^d= & \int \frac{d^4 k}{(2 \pi)^4} Y_d \frac{1}{\kappa_L} (Y_d)^{-1} \frac{k^2 y\left(k^2\right) Y_d^\dagger \lambda\, \kappa_L \kappa_R}{\left[(p-k)^2-M_{H_L^0}^2\right]\left[(p-k)^2-M_{H_R^0}^2\right]}~.
\end{aligned}
\end{equation}
Here $\lambda$ is a certain quartic coupling in the Higgs potential. It is important to recall that all couplings in the Higgs potential are real-valued in the model.  While fields with identical quantum numbers under $SU(3)_c \times U(1)_{\rm em}$ will mix, there is no mixing between scalar fields and pseudo-scalar fields owing to the reality of the Higgs potential couplings.  Using Eq.~\eqref{eq:theta} one can write the $\oline{\theta}$ contribution as
\begin{align}
& \oline{\theta}= -\operatorname{Im} \operatorname{Tr}\left(\frac{\lambda}{\kappa_L} \int \frac{d^4 k}{(2 \pi)^4} \frac{k^2 y\left(k^2\right) M_D (Y_d)^{-1}}{\left[(p-k)^2-M_{H_L^0}^2\right]\left[(p-k)^2-M_{H_R^0}^2\right]}\right)
\end{align}
The trace can be performed before doing the momentum integral, and we find
\begin{align}
& \operatorname{Tr}\left[y(k^2) M_D (Y_d)^{-1}\right]  = -\kappa_L \operatorname{Tr}\left[\left(Y_d^\dagger Y_d \kappa_L^2-k^2\right)^{-1} M_D^{\dagger}\ z(k^2) M_D\right] ~.
\end{align}
The matrix that is traced over here is a product of two hermitian matrices, and thus its trace is real.  Consequently the contribution to $\overline{\theta}$ from Fig. (\ref{fig:down_loop}) (a) is zero.

For the other diagrams we follow the same technique and summarize our results here.  Fig. \ref{fig:down_loop} (b) contains two diagrams.  While the diagram with $d_R$ incoming and $D_L$ outgoing  contributes to $\delta M_{HL}^d$, it is the hermitian conjugate of the diagram where $d_L$ is incoming and $D_R$ outgoing that contributes to $\delta M_{LH}^d$ of Eq. (\ref{eq:correction}).  This remark applies to other diagrams as well. We summarize the flavor structure of these diagrams and show that each contribution to $\overline{\theta}$ is vanishing. 
\begin{itemize}
   \item 5-(b): Fig.~\ref{fig:down_loop} (b) has the following flavor structure:
\begin{align}
    & d_R-D_L\ (\delta M_{HL}^d) : {\rm Im \, Tr}\  [Y_d^\dagger Y_d\ z(k^2) Y_d^\dagger (Y_d^\dagger)^{-1}] = 0 \\
    & D_R-d_L\ (\delta M_{LH}^d) : {\rm Im \, Tr}\  [Y_d Y_d^{-1}\ [I + k^2 x(k^2) ] Y_d Y_d^{-1}] = 0
\end{align}
The first trace is over the product of two hermitian matrices, while the second one is over a hermitian matrix, both of which are real, leading to zero contribution to $\overline{\theta}$. 
\item 5-(c): \vspace{-5mm}
\begin{align}
     & d_R - D_L\ (\delta M_{HL}^d) : {\rm Im \, Tr}\  [Y_d^\dagger Y_u\  z(k^2) Y_u^\dagger (Y_d^\dagger)^{-1}] = 0  \\
    & D_R-d_L\ (\delta M_{LH}^d) : {\rm Im \, Tr}\  [Y_u Y_u^{-1}\ [I + k^2 x(k^2) ] Y_d Y_d^{-1}] = 0
\end{align}
\item 5-(d): \vspace{-5mm}
\begin{align}
    & d_R - D_L\ (\delta M_{HL}^d) : {\rm Im \, Tr}\  [Y_d^\dagger\ z^T(k^2) Y_u^T Y_u^* (Y_d^\dagger)^{-1}] = 0  \\
    & D_R-d_L\ (\delta M_{LH}^d) : {\rm Im \, Tr}\  [Y_u^T  [I + k^2 x(k^2) ] (Y_u^T)^{-1}\ Y_d Y_d^{-1}] = 0
\end{align}
\item 5-(e): \vspace{-5mm}
\begin{align}
    & d_R - D_L\ (\delta M_{HL}^d) : {\rm Im \, Tr}\  [(\kappa_L^2 Y_d^\dagger Y_d - k^2)^{-1} M_D^\dagger z(k^2) Y_D] = 0  \\
    & D_R-d_L\ (\delta M_{LH}^d) : {\rm Im \, Tr}\  [(Y_d^\dagger Y_d \kappa_L^2 -k^2)^{-1} M_D^\dagger z(k^2) Y_D] = 0
\end{align}
\item 5-(f): \vspace{-5mm}
\begin{align}
    & d_R - D_L\ (\delta M_{HL}^d) : {\rm Im \, Tr}\  [z^*(k^2)  (Y_d^\dagger)^{-1}Y_d^\dagger] = 0  \\
    & D_R-d_L\ (\delta M_{LH}^d) : {\rm Im \, Tr}\  [(I+ k^2 x(k^2))^* ( Y_d^\dagger Y_d)^{-1}] = 0
\end{align}
\item 5-(g): \vspace{-5mm}
\begin{align}
    & d_R - D_L\ (\delta M_{HL}^d) : {\rm Im \, Tr}\  [z^\dagger(k^2) Y_d^\dagger (Y_d^\dagger)^{-1}] = 0  \\
    & D_R-d_L\ (\delta M_{LH}^d) : {\rm Im \, Tr}\  [(I+ k^2 x(k^2))^\dagger (Y_d Y_d^\dagger)^{-1}] = 0
\end{align}
\item 5-(h): \vspace{-5mm}
\begin{align}
    & d_R - d_L\ (\delta M_{HL}^d) : {\rm Im \, Tr}\  [( Y_d^\dagger Y_d \kappa_L^2 - k^2)^{-1} M_D^\dagger z(k^2) M_D] = 0 
\end{align}
\end{itemize}

Now we turn to the diagrams correcting the up-type quark mass matrix shown in Fig. \ref{fig:up_loop} and summarize our results here. As noted earlier, even though the contributions from Fig. \ref{fig:up_loop} (a) and (h) to $\delta M_{LL}^u$ are nonzero, they do not contribute to $\overline{\theta}$ owing to the condition $M_U = 0$, see Eq. (\ref{eq:theta}).

\begin{itemize}
\item 6-(b): \vspace{-5mm}
\begin{align}
    & u_R - U_L\ (\delta M_{HL}^u):  {\rm Im \, Tr}\ [Y_u^\dagger Y_d z(k^2) Y_d^\dagger (Y_u^\dagger)^{-1}] =0 \\
    & U_R - u_L (\delta M_{LH}^u):  {\rm Im \, Tr}\ [Y_d Y_d^{-1} [I + k^2 x(k^2)] Y_u Y_u^{-1} ] =0 
\end{align}
\item 6-(c): \vspace{-5mm}
\begin{align}
    & u_R - U_L\ (\delta M_{HL}^u):  {\rm Im \, Tr}\ [Y_d^* z^T(k^2) Y_d^T Y_u^\dagger (Y_u^\dagger)^{-1}] =0 \\
    & U_R - u_L (\delta M_{LH}^u):  {\rm Im \, Tr}\ [Y_u [I+ k^2 x(k^2)]^T (Y_d^T)^{-1} Y_d^T Y_u^{-1}] =0
\end{align}
\item 6-(d): \vspace{-5mm}
\begin{align}
    & u_R - U_L\ (\delta M_{HL}^u):  {\rm Im \, Tr}\ [Y_u^\dagger z^T(k^2)  Y_d Y_d^\dagger(Y_u^\dagger)^{-1}] =0 \\
    & U_R - u_L (\delta M_{LH}^u):  {\rm Im \, Tr}\ [Y_d [I + k^2 x(k^2)]^T Y_d^{-1} Y_u Y_u^{-1}] =0 
\end{align}
\item 6-(e): \vspace{-5mm}
\begin{align}
    & u_R - U_L\ (\delta M_{HL}^u):  {\rm Im \, Tr}\ [z^\dagger (k^2) Y_u^\dagger (Y_u^\dagger)^{-1}] =0 \\
    & U_R - u_L (\delta M_{LH}^u):  {\rm Im \, Tr}\ [(I+ k^2 x(k^2))^\dagger (Y_u Y_u^\dagger)^{-1}] = 0 
\end{align}
\item 6-(f): \vspace{-5mm}
\begin{align}
    & u_R - U_L\ (\delta M_{HL}^u):  {\rm Im \, Tr}\ [Y_u^* z^*(k^2) (Y_u^\dagger)^{-1}] =0 \\
    & U_R - u_L (\delta M_{LH}^u):  {\rm Im \, Tr}\ [(Y_u^*)^{-1} (I + k^2 x(k^2))^* Y_u^{-1}] = 0 
\end{align}
\item 6-(g): \vspace{-5mm}
\begin{align}
    & u_R - U_L\ (\delta M_{HL}^u):  {\rm Im \, Tr}\ [Y_u^\dagger Y_u z(k^2) Y_u^\dagger (Y_u^\dagger)^{-1}] =0 \\
    & U_R - u_L (\delta M_{LH}^u):  {\rm Im \, Tr}\ [Y_u Y_u^{-1} (I+k^2 x(k^2)) Y_u Y_u^{-1}] = 0 
\end{align}
\end{itemize}


This proves that there is no induced $\overline{\theta}$ in the model at the one-loop level.

\subsection{Renormalization group evolution of \texorpdfstring{\boldmath${\overline{\theta}}$}{thetaX}  and neutron EDM}

We have seen that the one-loop induced $\overline{\theta}$ at the scale $M_G$ is vanishing.  There is the possibility that extrapolation of the Yukawa couplings by the renormalization group equations (RGE) from the GUT scale to the weak scale could generate a nonzero $\bar{\theta}$. To study this question, we analyze the RG equations for the Yukawa matrices of the model relevant for the momentum range $M_I \leq \mu \leq M_G$ given in Appendix \ref{sec:RGE}. First we note that the evolution of $\overline{\theta}$ will involve  determinant of the matrix $Y_{uL}  Y^\dagger_{uR}$ and those with $u$ replaced by $d$. From the RG equations given in Appendix \ref{sec:RGE}, we see that the one loop expression for $\frac{d}{dt}{(Y_{uL}  Y^\dagger_{uR}})$ is a hermitian matrix, and therefore does not generate $\bar{\theta}$ if the initial $\bar{\theta}$ is zero. 

\begin{figure}
    \centering
    \includegraphics[scale=0.6]{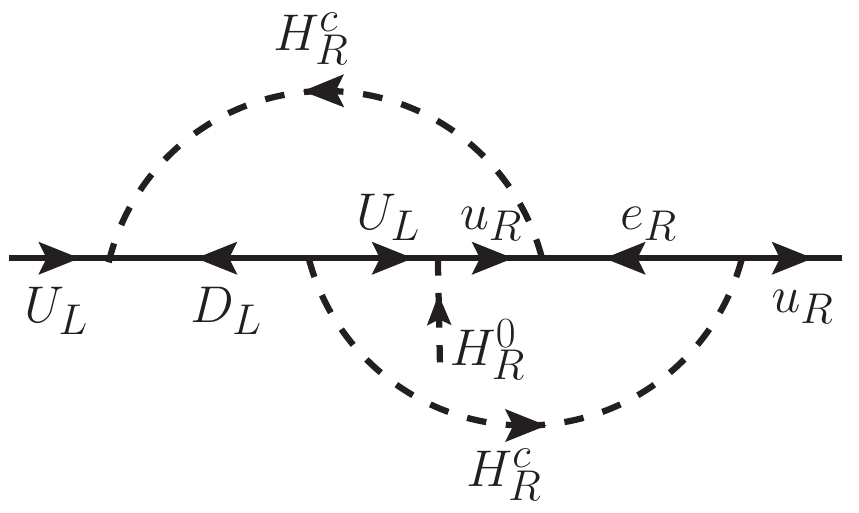}
    \caption{Dominant two-loop radiative correction contributing to $\bar{\theta}$. }
    \label{fig:twoloop_theta}
\end{figure}

To show this in more detail, we take infinitesimal extrapolations of the $\bar{\theta}$ parameters in steps of $dt$. These are given by $\frac{d}{dt} (Y_{uL}Y^\dagger_{uR}) dt$. We see that the induced value of $\bar{\theta}$ depends on how the integrand behaves.  If the integrand is real, no $\bar{\theta}$ is induced in this infinitesimal interval. Successive iterations can lead to finite shifts in $t$.  The induced $\bar{\theta}$ via RGE from the up-quark sector can be written as
\begin{eqnarray}
\delta (\bar{\theta})= {\rm Im \, Tr}\left[\frac{d}{dt}(Y_{uL}  Y^\dagger_{uR})(Y_{uL}  Y^\dagger_{uL})^{-1}\right] ~.  
\end{eqnarray}
Using the one-loop expressions for the RGE from the Appendix \ref{sec:RGE} and setting $Y_{uL}=Y_{uR}$ as the initial value, we see that the expression within the bracket is hermitian and therefore induced $\delta (\bar{\theta})=0$.
They would therefore keep the GUT scale $\bar{\theta}$ value unchanged to one loop. Several  of the two-loop corrections can also similarly be seen to give  zero contributions. However at the two-loop level there are nonzero contributions to $\overline{\theta}$. In particular, the 8th term of Eq. (\ref{eq:RGEYuR}) in the RGE for $Y_{u_R}$ generates a nonzero $\overline{\theta}$ which can be estimated to be
\begin{equation}
\overline{\theta} \approx \frac{8}{(16 \pi^2)^2}{\rm Im \,}{\rm Tr}\left[Y_{dR}^TY_{d_R}^*Y_{u_R} Y_{d_R}^\dagger Y_{d_R}(Y_{u_L} Y_{u_L}^\dagger)^{-1} \right]\rm{ln}\left(\frac{M_{H_L^c}}  {M_{H_R^c}}\right)~.
\label{eq:2theta}
\end{equation}
This term can be seen to be originating from the two-loop diagram shown in Fig. \ref{fig:twoloop_theta}.  Note that this diagram is log-divergent.  There is an analogous diagram where the color-triplet field $H_R^c$ is replaced by $H_L^c$ and the quark helicities are flipped.  Since in the computation of the RGE beta functions, it was assumed that $H_L^c$ has a mass of order $M_G$, while $H_R^c$ is at $M_I$, only the diagram of Fig. \ref{fig:twoloop_theta} contributes below $M_G$.  Above the mass scale of $H_L^c$, the combined contributions to $\overline{\theta}$ from $H_R^c$ and $H_L^c$ would nearly vanish, since this is the parity symmetric limit.  

To estimate the induced $\overline{\theta}$ from Eq. (\ref{eq:2theta}), we set the GUT-scale values of the Yukawa coupling matrices, namely, $Y_{d_R} = Y_{d_L} = Y_\ell^T$ and $Y_{u_R} = Y_{u_L} = Y_u$.  Then we use the transformed basis where the fermion fit was given, with the mass matrices given as in Eq. (\ref{eq:Mdfit}).  We can estimate $\overline{\theta}$ to be
\begin{equation}
\overline{\theta} \approx \frac{8}{(4 \pi)^4}{\rm Im \,}{\rm Tr} \left[Q^* U_{\rm PMND}^\dagger \,\hat{Y}_\ell^2U_{\rm PMNS} Q^2 \,\hat{Y}_u \,U_{\rm PMNS}^T \hat{Y}_\ell^2\, U_{\rm PMNS}^* (\hat{Y}_u)^{-1} \right]\rm{ln}\left(\frac{M_{H_L^c}}  {M_{H_R^c}}\right) ~.
\label{eq:theta22}
\end{equation}
Here all the parameters are known, except for the two phases in the diagonal matrix $Q = {\rm diag}.(e^{i\alpha_1},\,e^{i \alpha_2},\,1)$. These two phases are unobservable in low energy experiments, except through their contributions to $\overline{\theta}$ and thus to neutron EDM.

\begin{figure}
    \centering
    \includegraphics[scale=0.42]
{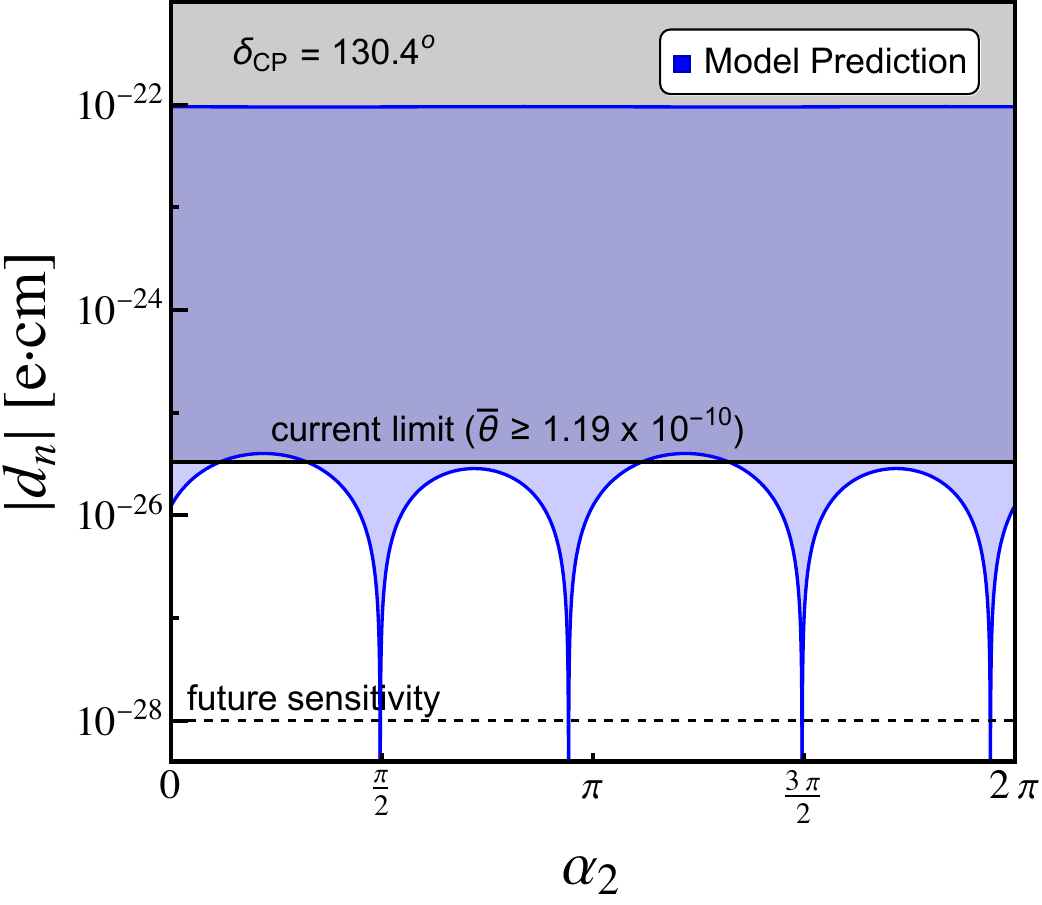}
\includegraphics[scale=0.42]
{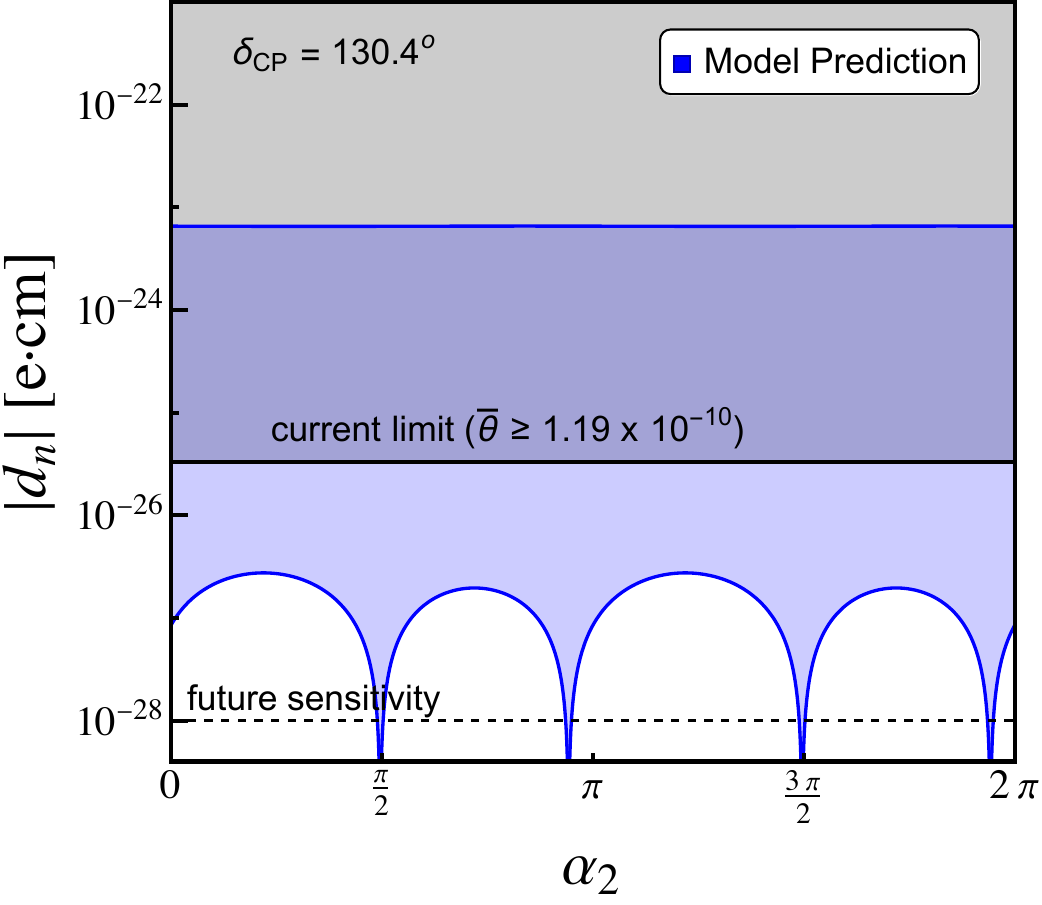}
    \caption{Prediction of the model for neutron EDM $d_n$ as a function of the phase $\alpha_2$ marginalized over the phase $\alpha_1$. The allowed values  consistent with the current EDM limit for the phase $\alpha_1$ are $0.128 + n \pi/2$ $(n=0,1,2,3)$. Here $Q \equiv {\rm diag}.(e^{i\alpha_1}, e^{i \alpha_2}, 1)$. Gray shaded region is the exclusion from the current limit \cite{Dragos:2019oxn,ParticleDataGroup:2022pth} and black dashed line represent the future sensitivity reach of ${\cal O}(10^{-28})$ e-cm \cite{Alarcon:2022ero}. }
    \label{fig:nEDM}
\end{figure}

In Fig. \ref{fig:nEDM} we have presented the induced value of $\overline{\theta}$ arising from this dominant two-loop diagram, as a function of one of the phase parameters, $\alpha_2$.  We have fixed the phase $\alpha_1 = 0.128$ modulo integer multiples of $\pi/2$.
This is the preferred value of this angle to be consistent with neutron EDM limits.
We have also shown the correlations with neutron EDM as well as its current limit and future sensitivity.  In the left panel of Fig.  \ref{fig:nEDM}, we kept the mass of $H_L^c$ equal to $M_G$ while $H_R^c$ mass is $M_I$. In the right panel, $H_L^c$ mass is kept at 3 times $H_R^c$ mass.  Such a lowering of the $H_L^c$ mass to a scale of order $10^{12}$ GeV is compatible with proton decay constraints, and has very little effect on gauge coupling unification.

A mild fine-tuning of parameters of the model  is needed for the induced $\overline{\theta}$ to be within allowed range from neutron EDM.  This can be seen by expanding the imaginary part of the trace appearing in Eq. (\ref{eq:theta22}).  Making an expansion in small quark masses, we find the leading term in the ImTr[] to be
\begin{equation}
{\rm Im \,}{\rm Tr}[~] \simeq Y_b^4 \frac{Y_t}{Y_u} {\rm Im} \left[e^{-2 i \alpha_1} (U_{31}^* U_{33})^2\right]~.
\end{equation}
For typical value of the phase $\alpha_1$, one would get $\overline{\theta}\sim 10^{-9}$ corresponding to the right panel of Fig. \ref{fig:nEDM}, and $\overline{\theta}\sim 10^{-7}$ for the left panel.  The best fit to fermion masses provides the PMNS matrix element to be $U_{31} = 0.483 e^{-0.127i}$, and the choice of $\alpha_1$ nearly cancels this phase. The tuning needed is at the level of $10^{-3}$ for the left panel, while it is only a few percent for the right panel.  One could turn this observation around and state that neutron EDM should not be too much smaller than the current experimental limit within the model; otherwise the model will be more finely tuned. It should be noted that other contributions to neutron EDM arising from heavy particle exchange are highly suppressed since the new particles have masses of order $M_I \simeq 10^{11}$ GeV.

If parity symmetry is broken at low energy scales, Planck suppressed operators will not destabilize the parity solution to the strong CP problem \cite{Berezhiani:1995yi}.  In the GUT embedding one necessarily has a high scale, and there are operators such as $\Sigma_L V^{\mu\nu}_L \widetilde{V}_{L,\mu\nu}/M_P$ contributing to $\overline{\theta}$ at the level of $10^{-3}$, which would need to be tuned.  In comparison to the tuning needed in popular axion model, which is at the level of $10^{-44}$ \cite{Kamionkowski:1992mf,Barr:1992qq,Holman:1992us}, here the needed turning is at the level of $10^{-7}$.

\section{Conclusions}
\label{sec:Conclusions}

We have developed in this paper a grand unified framework based on the gauge symmetry $SU(5)_L\times SU(5)_R$ that embeds a class of left-right symmetric models which solves the strong CP problem by parity symmetry.  Of the many possible symmetry breaking chains, we have found one that allows successful gauge coupling unification with a single intermediate scale  with $M_I \sim 10^{11}$ GeV.  The intermediate scale gauge symmetry is $SU(3)_{cL}\times SU(2)_{L} \times U(1)_L \times SU(5)_R$.  We have further shown that the gauge bosons of $SU(5)_R$ that survive down to $M_I$ do not cause any problem with rapid proton decay.  This is achieved with an interesting flavor structure for the fermion mass matrices which disallows interactions of the $(X_R^\mu,\, Y_R^\mu)$ gauge bosons of $SU(5)_R$ with light fermion fields alone.  This flavor structure also explains the observed masses and mixing in the  quark and lepton sectors.

A novel feature of the model is that neutrinos are naturally light Dirac fermions.  Their Yukawa couplings are suppressed by a factor $M_I/M_G \sim 10^{-7}$, since the Higgs doublet that induces its mass has a GUT scale mass and acquires only an induced VEV, $v_\nu \sim v\, (M_I/M_G)$, where $v \sim 174$ GeV.  We call this mechanism of obtaining suppressed Dirac mass as the type-II Dirac seesaw.  Furthermore, the model makes several predictions in the neutrino sector, which can serve as its tests.  The CP-violating phase in neutrino oscillations is found to lie in the range  $\delta_{CP} = \pm (130.4 \pm 1.2)^\circ $. The lightest neutrino mass is constrained to lie in the range $m_{\nu_1} = (4.8-8.4)$ meV. The model predicts normal ordering of neutrino masses, and obviously neutrinoless double beta decay is forbidden within the model.


We note that our model is a full GUT embedding of the left-right symmetric model that preserves the strong CP solution via parity symmetry. The fermion sector of the non-unified model fits nicely into the GUT model. We also find it remarkable that despite the presence of many new particles at the GUT and intermediate scale as well as the stringent constraints of unification, most features of the strong CP solution survive. For instance, the one loop contributions to $\bar{\theta}$- parameter vanishes as in the non-unified model. The two-loop contribution, in order to be compatible with the current neutron EDM limits, however, does require a mild fine tuning at the few percent level. The GUT model will therefore be under stress if the neutron EDM limit goes down further. We also note that due to high scale of parity breaking, the Planck scale induced P-violating operators can induce a large $\bar{\theta}$ unless their strengths are suppressed. Leaving aside the question of how to estimate the strengths of Planck scale operators in general, it is worth pointing out that the degree of suppression we need is much milder than that required in the axion models, with or without grand unification.

Before closing we wish to comment on two topics that are relevant to any GUT.  These relate to the question of dark matter and baryogenesis.  Additional particles may be added to the model which may be identified as the dark matter candidate with an appropriate symmetry that guarantees its cosmological stability. The model we have presented already contains such a particle in the scalar field $\eta$, which plays no role in symmetry breaking.  If a reflection symmetry $\eta \rightarrow -\eta$ is imposed, which survives spontaneous symmetry breaking, the lightest fragment of $\eta$ will be stable.  If this component is chosen as the SM Higgs doublet, it can serve as the inert doublet dark matter, provided that its mass is of order TeV \cite{Barbieri:2006dq}. As for baryogenesis, one possible avenue is to rely on Dirac leptogenesis \cite{Dick:1999je}, which is currently under investigation.

\section*{Acknowledgement} 
The work of KSB was supported in part by the United States Department of Energy Grant No. DE-SC0016013.
The work of AT was supported in part by the National Science Foundation under Grant PHY-2210428. KSB and AT wish to acknowledge the Center for Theoretical Underground Physics and Related Areas (CETUP*) and the Institute for Underground Science at SURF, where part of this work was done, for hospitality and for providing a stimulating environment. We thank Ravi Kuchimanchi for a useful comment on the manuscript.

\appendix

\section{Appendix}

\subsection{Decomposition of fields under intermediate symmetry}
\label{sec:A1}

Here we list the decomposition of various fields under the intermediate symmetry $SU(3)_{c L} \times SU(2)_L \times U(1)_L \times SU(5)_R$. This list is helpful in deciding which fragments of the Higgs fields survive down to $M_I$.
\begin{align}
    ({\bf 5}, \overline{\bf 5}) &= ({\bf 3}, 1 , -1/3, {\bf \overline{5}}) + (1,{\bf 2}, 1/2, {\bf \overline{5}} ) \notag \\
     ({\bf 75}, 1) &= ({\bf \overline{3}}, 1 , -5/3, 1) + ({\bf 6}, {\bf 2} , -5/6, 1) + ({\bf 3},{\bf 2}, -5/6, 1) + ({\bf 8},{\bf 3},0,1) \notag \\
     &~~+ ({\bf 8}, 1 , 0, 1) + (1,1,0,1) + ({\bf 6}, {\bf 2}, 5/6, 1) + ({\bf 3}, {\bf 2}, 5/6, 1) + ({\bf 3}, 1, 5/3, 1) \notag \\ 
     ({\bf 5},1) &= ({\bf 3}, 1, -1/3, 1) + (1, {\bf 2}, 1/2, 1) \notag\\
     ({\bf 10},1) &= ({\bf \overline{3}} ,1, -2/3, 1) + ({\bf 3}, {\bf 2},1/6,1) + (1,1,1,1) \notag \\
     ({\bf 15},{\bf \overline{15}}) &= ({\bf 3},{\bf 2},1/6,{\bf \overline{15}}) + (1,{\bf 3}, 1 , {\bf \overline{15}})  + ({\bf \overline{6}},1,-2/3,{\bf \overline{15}})
     \label{eq:A1}
\end{align}

\subsection{Higgs potential analysis}\label{sec:Anew}
Here we construct the Higgs potential involving the fields $\{H_L({\bf 5},1)+H_R(1,{\bf 5})\}$, $\Phi({\bf \oline{5}}, {\bf 5})$, and $\{\Sigma_L({\bf \oline{75}},1)+\Sigma_R(1,{\bf \oline{75}})$. While the full theory contains $\eta({\bf 15}, {\bf \oline{15}})$ as well, however for simplicity, we impose a discrete symmetry under which $\eta \to -\eta$, in presence of which the VEV of $\eta$ would be vanishing.  We also note that the standard model Higgs doublet contained in $\eta$ may be identified as the inert doublet dark matter, provided that the doublet field has a mass of order 1 TeV. With the $\eta \rightarrow -\eta$ symmetry, this field does not play any role in symmetry breaking. The most general renormalizable potential for the $(H_L,H_R,\Phi,\Sigma_L,\Sigma_R)$ fields can be written as 
\begin{eqnarray}
    V = V(H_{L,R}) + V(\Phi) + V(\Sigma_{L,R}) + V(H_{L,R},\Phi,\Sigma_{LR}) \, ,
    \label{eq:fullpot}
\end{eqnarray}
where
\begin{align}
    &V(H_{L,R}) =   - m_H^2 (H_L^\dagger H_L+H_R^\dagger H_R )    + \lambda_1  (H_L^\dagger H_L H_L^\dagger H_L+ H_R^\dagger H_R H_R^\dagger H_R)   +  \lambda_2 H_L^\dagger H_L  H_R^\dagger H_R  \, , \\[5pt]
    &V(\Phi) = -m_\Phi^2 {\rm Tr} [\Phi^\dagger \Phi] + \lambda_3 {\rm Tr} [\Phi^\dagger \Phi] {\rm Tr} [\Phi^\dagger \Phi] + \lambda_4 {\rm Tr} [\Phi^\dagger \Phi\Phi^\dagger \Phi]  \, , \\[5pt]
   & V(\Sigma_{L,R}) = - m_\Sigma^2 (\Sigma_L)^{ab}_{cd} (\Sigma_L)^{cd}_{ab}   + \mu_2  (\Sigma_L)^{ab}_{cd} (\Sigma_L)^{ef}_{ab} (\Sigma_L)^{cd}_{ef} + \lambda_{5} (\Sigma_L)^{ab}_{gh} (\Sigma_L)^{cd}_{ab} (\Sigma_L)^{ef}_{cd} (\Sigma_L)^{gh}_{ef} \notag\\
    &~~~~~~~~~~~~~~~~  + \lambda_{6} (\Sigma_L)^{ab}_{cg} (\Sigma_L)^{cd}_{ab} (\Sigma_L)^{ef}_{dh} (\Sigma_L)^{gh}_{ef} + \lambda_{7} (\Sigma_L)^{ab}_{fg} (\Sigma_L)^{cd}_{ab} (\Sigma_L)^{ef}_{ch} (\Sigma_L)^{gh}_{de} + (L \to R)  \notag \\
    &~~~~~~~~~~~~~~~~  + \lambda_{8} (\Sigma_L)^{ab}_{cd} (\Sigma_L)^{cd}_{ab} (\Sigma_R)^{ef}_{gh} (\Sigma_R)^{gh}_{ef} \, , \\[5pt]
    & V(H_{L,R}, \Phi, \Sigma_{L,R}) =  \mu_1 H_R^\dagger \Phi H_L  + \lambda_9 H_L^\dagger \Phi \Phi^\dagger  H_L  + \lambda_{10} H_L^\dagger H_L {\rm Tr} [\Phi^\dagger \Phi]  \notag \\
    &~~~~~~~~~~~~~~~~~~~~~~~~~~ +  \lambda_{11} H_{L}^{i*} H_{Li} (\Sigma_L)^{ab}_{cd} (\Sigma_L)^{cd}_{ab} + \lambda_{12} H_{L}^{a*} H_{Lc} (\Sigma_L)^{de}_{ab} (\Sigma_L)^{cb}_{de}   \notag \\
    &~~~~~~~~~~~~~~~~~~~~~~~~~~ +  \lambda_{13} H_{L}^{i*} H_{Li} (\Sigma_R)^{ab}_{cd} (\Sigma_R)^{cd}_{ab} + \lambda_{14} {\rm Tr} [\Phi^\dagger \Phi] (\Sigma_L)^{ab}_{cd} (\Sigma_L)^{cd}_{ab} \notag \\ 
    &~~~~~~~~~~~~~~~~~~~~~~~~~~ + \lambda_{15} \Phi^{*a}_i \Phi^{i}_{c} (\Sigma_L)^{de}_{ab} (\Sigma_L)^{cb}_{de} + (L \leftrightarrow R, \Phi \leftrightarrow \Phi^\dagger) \, .
\end{align}
Here we have shown the indices for some of the non-trivial contractions. Note that all the Higgs potential parameters are real, upon imposing parity symmetry, and thus a CP conserving vacuum is admitted, where all the VEVs are real.  By inserting the VEVs of Eqs.~\eqref{eq:VEV0}, \eqref{eq:VEV1}, and \eqref{eq:VEVeta} in Eq.~\eqref{eq:fullpot}, we obtain the following conditions for the potential to be extremum:
\begin{align}
     \kappa_L \left[-m_H^2 + 2 \lambda_1 \kappa_L^2 + \lambda_2 \kappa_R^2 + 3 \lambda_{10} v_\Phi^2 + 24 (3 \lambda_{11} + \lambda_{12}) \la\Sigma_L\ra^2 + 72 \lambda_{13} \la\Sigma_R\ra^2\right] &= 0 \, , \label{eq:minkL}\\
    \kappa_R \left[-m_H^2 + 2 \lambda_1 \kappa_R^2 + \lambda_2 \kappa_L^2 + 3 \lambda_{10} v_\Phi^2 + 24 (3 \lambda_{11} + \lambda_{12}) \la\Sigma_R\ra^2 + 72 \lambda_{13} \la\Sigma_L\ra^2\right] &= 0 \, , \label{eq:minkR} \\
    v_\Phi \left[ -m_\Phi^2 + 2 (3 \lambda_3 + \lambda_4) v_\Phi^2 + \lambda_{10} (\kappa_L^2 + \kappa_R^2) + 8 (9 \lambda_{14} + \lambda_{15}) (\la\Sigma_L\ra^2+ \la\Sigma_R\ra^2) \right] &= 0 \, , \label{eq:minivPhi}\\
    \la\Sigma_L\ra \left[ -3 m_\Sigma^2 - 12 \mu_2 \la\Sigma_L\ra + 4 (30 \lambda_5 + 28 \lambda_6 + 7 \lambda_7) \la\Sigma_L\ra^2 + 216 \lambda_8 \la\Sigma_R\ra^2   \right. & \notag \\
 \left. + (3\lambda_{11}+\lambda_{12}) \kappa_L^2 + 3 \lambda_{13} \kappa_R^2 + (9 \lambda_{14}+ \lambda_{15}) v_\Phi^2 \right] & = 0 \, , \label{eq:minSL} \\
    \la\Sigma_R\ra \left[ -3 m_\Sigma^2 - 12 \mu_2 \la\Sigma_R\ra + 4 (30 \lambda_5 + 28 \lambda_6 + 7 \lambda_7) \la\Sigma_R\ra^2 + 216 \lambda_8 \la\Sigma_L\ra^2   \right. & \notag \\
\left.  + (3\lambda_{11}+\lambda_{12}) \kappa_R^2 + 3 \lambda_{13} \kappa_L^2 + (9 \lambda_{14}+ \lambda_{15}) v_\Phi^2 \right] & = 0 \, . \label{eq:minSR}
\end{align}
From here it is straightforward, although tedious, to obtain the full spectrum of Higgs boson masses.  However, this is not needed for our purpose.  It is clear form these minimization conditions that it admits a solution with $\la\Sigma_R\ra =0$ and $\la\Sigma_L\ra \neq 0$, with the latter VEV obtained from Eq.~\eqref{eq:minSL} as
\begin{equation}
    \la \Sigma_L\ra = \frac{1}{\widetilde{\lambda}} \left( 2 \mu_2 \pm \sqrt{9 \mu^2 + 3/2 \widetilde{\lambda} m_{\Sigma}^2}  \right) \, ,
\end{equation}
where $\widetilde{\lambda} = 60 \lambda_5 + 14 (4 \lambda_6 + \lambda_7) $. 
This corresponds to a minimum of the potential for a large range of the parameters in the potential. 

While the VEV $\langle \Sigma_L \rangle \sim m_\Sigma$ takes its natural value from Eq. (\ref{eq:minSL}), two mini fine-tunings are needed to realize $v_\Phi \sim \kappa_R \ll \langle \Sigma_L \rangle$, as given in Eqs.~\eqref{eq:minkR}, \eqref{eq:minivPhi}.  Furthermore, one strong fine-tuning is needed to realize $\kappa_L \ll \langle \Sigma_L \rangle$ from Eq. (\ref{eq:minkL}). This last condition is the tuning associated with the usual Higgs mass hierarchy problem, while the fine-tuning to realize intermediate scale VEVs are present in all non-SUSY GUTs with an intermediate scale. 

For completeness we show the masses of the $X_L^\mu$ and $Y_L^\mu$  gauge bosons which are of order $\langle\Sigma_L\rangle$. The covariant derivative for the Higgs field $\Sigma_L$ field  reads as
\begin{align}
    D^\mu (\Sigma_L)^{ab}_{cd} &= \partial^\mu (\Sigma_L)^{ab}_{cd} - \frac{i g_{5L}}{2} \left\{ (\Vec{T}.\Vec{V}^\mu_L)^e_c (\Sigma_L)^{ab}_{ed} + (\Vec{T}.\Vec{V}^\mu_L)^e_d (\Sigma_L)^{ab}_{ce} \right. \notag \\
    & ~~~~ \left. - (\Vec{T}.\Vec{V}^\mu_L)^a_f (\Sigma_L)^{fb}_{cd} -(\Vec{T}.\Vec{V}^\mu_L)^b_f (\Sigma_L)^{af}_{cd}\right\} \, .
\end{align}
The $X$ and $Y$ gauge boson mass then given by
\begin{equation}
    M_{X^{\mu}_{L},Y^{\mu}_{L}}  = 4\sqrt{6} g_{5L} \la \Sigma_L \ra \, .
\end{equation}
These are the only twelve gauge bosons that pick up mass at the GUT scale.

\subsection{Decomposition of scalar fields under SM gauge group}
\label{sec:A2}
We define the following fields under SM gauge group $SU(3)_c\times SU(2)_L\times U(1)_Y$:
\begin{align}
    H_L ({\bf 5},1) &= H_{L}^c ({\bf 3},1,-1/3) + H_L^d  (1,{\bf 2},1/2) \notag \\
    H_R (1,{\bf 5}) &= H_R^c  ({\bf 3},1,-1/3) + H_{R}^+(1,1,1) + H_R^0  (1,1,0) \notag \\
    \Phi ({\bf \overline{5}},{\bf 5}) &= \phi_{DD}^o({\bf 8},1,0) +  \Phi_{DD}^s(1,1,0) +  \Phi_{D \nu}({\bf \oline{3}},1,1/3)+  \Phi_{De}({\bf \overline{3}},1,4/3)     \notag \\
    &~~~ +  \Phi_{LD}({\bf 3},{\bf 2},-5/6) + 
    \Phi_{L\nu} (1,{\bf 2},-1/2)  
     + \Phi_{Le}(1,{\bf 2},1/2)     \, .
     \label{eq:qn}
\end{align}

\subsection{Analytical solution of one loop RGE to gauge couplings including threshold corrections from vector-like fermions} \label{app:analytical}
The solutions to the one-loop RGE, $\alpha_i(\mu)$, can be written down as functions of momentum starting from the scale $M_G$ where the couplings are unified.  At the scale $M_I$, we impose the boundary conditions given in Eq. (\ref{eq:matchingcond}).   
For the three gauge couplings at $\mu = m_t$ we obtain
\begin{eqnarray}
\alpha_{1Y}^{-1}(m_t) &=& \alpha_G^{-1} - \left(\frac{5}{13}b_1' + \frac{8}{13}b_5'  \right) \frac{1}{2\pi}{\rm log}\left(\frac{M_I}{M_G} \right) - \frac{b_1}{2\pi}{\rm log} \left(\frac{m_t}{M_I}  \right) - \Delta_{\alpha_{1Y}}  \nonumber \\
\alpha_{2L}^{-1}(M_t) &=& \alpha_G^{-1} - \frac{b_2'}{2\pi}{\rm log}\left(\frac{M_I}{M_G} \right)
- \frac{b_2}{2\pi}{\rm log}\left(\frac{m_t}{M_I} \right) - \Delta_{\alpha_{2L}} \nonumber \\
\alpha_{3c}^{-1}(Mm_t) &=& 2 \alpha_G^{-1} - \frac{b_3'+b_5'}{2\pi}{\rm log}\left(\frac{M_I}{M_G} \right) - \frac{b_3}{2\pi} {\rm log}\left(\frac{m_t}{M_I} \right)- \Delta_{\alpha_{3c}}~.
\label{eq:soln}
\end{eqnarray}
Here $\Delta_{\alpha_i}$ are threshold corrections at $M_I$ arising from the spread in the masses of the $U,\,D,\,E$ fermions. These masses are not all degenerate, as they obey the relations
\begin{equation}
M_{U_1}:M_{U_2}:M_{U_3} = m_u:m_c:m_t,~~~M_{E_1}:M_{E_2}:M_{E_3} = m_e:m_\mu:m_\tau
\label{eq:relations}
\end{equation}
which needs to be incorporated via $\Delta_{\alpha_i}$ in Eq. (\ref{eq:soln}). In addition, the masses of the $D_i$ quarks are also hierarchical, as shown in Eq. (\ref{eq:massesdown}) from our numerical fit to fermion masses.  
The threshold corrections $\Delta_{\alpha_i}$ from these spreads in fermion masses at $\mu = M_I$ are given by
\begin{equation}
\Delta_{\alpha_i} = 
\sum_{F=U,D,E}\left[\frac{\Delta b_i^F}{2\pi} {\rm log}\left(\frac{M_{F_1}}{M_{F_2}}  \right)+
\frac{2 \Delta b_i^F}{2\pi} {\rm log}\left(\frac{M_{F_2}}{M_{F_3}}  \right) +
\frac{3 \Delta b_i^F}{2\pi} {\rm log}\left(\frac{M_{F_3}}{M_{I}}  \right)\right] ~.
\end{equation}
The contributions to $\Delta b_i^F$ from the vector-like fermions $U,\,D,\,E$ are:
\begin{eqnarray}
(\Delta b_1^U, \,\Delta b_1^D, \,\Delta b_1^E) &=& \left(\frac{16}{39}, \,\frac{4}{9}, \,\frac{4}{13}\right) \nonumber \\
(\Delta b_2^U,\, \Delta b_2^D, \,\Delta b_2^E) &=& (0,\,0,\,0) \nonumber \\ 
(\Delta b_3^U, \,\Delta b_3^D,\, \Delta b_3^E) &=& \left(\frac{2}{3}, \,\frac{2}{3},\, 0\right)~.
\end{eqnarray}
When the threshold corrections are ignored, one would obtain from Eq. (\ref{eq:soln}) the result given in Eq. (\ref{eq:weak}) for the weak mixing angle $\sin^2\theta_W(m_t)$.  

Now, let us consider the one-loop solutions to the RGE including the threshold effects shown in Eq.(\ref{eq:soln}). If we use the fermion and scalar spectrum at $M_I$ shown in Eqs. (\ref{eq:MIfermion})-(\ref{eq:MIscalar}), but remove the $\eta_K$ scalar, the one-loop $\beta$-function coefficients will be
\begin{equation}
(b_{1Y}, b_{2L}, b_{3c}) = \left(\frac{41}{26}, -\frac{19}{6}, -7\right),~~(b'_{1L}, b'_{2L}, b'_{3cL}, b'_{5R}) = \left(\frac{133}{30}, -\frac{19}{6}, -\frac{37}{6}, -\frac{41}{3}\right)~.
\label{eq:bi}
\end{equation}
Taking $M_{U_3} = M_I$, $M_{E_3} = (m_\tau/m_t) M_I$, with the relations given in Eq. (\ref{eq:relations}), and with $M_{D_i}$ values given in Eq. (\ref{eq:massesdown}), one obtains for $M_I$, $M_G$ and $\alpha_G^{-1}$ the following values:
\begin{equation}
M_I = 6.3 \times 10^5~{\rm GeV},~~~M_G = 3.3 \times 10^{15}~{\rm GeV},~~~\alpha_G^{-1} = 45.0~.
\label{eq:wrong}
\end{equation}
This choice is however inconsistent with LHC data~\cite{CMS:2016edj}, since $M_{U_3} =  6.3 \times 10^5~{\rm GeV}$ would imply that $M_{U_1} \sim 4$ GeV.  The value of $M_I$ should be raised to $M_I \ge 10^8$ GeV for consistency with LHC limit of $M_{U_1} \geq 1$ TeV.  This is indeed what is achieved with the inclusion of the $\eta_K$ scalar with a mass around $M_I$. We should also note that without the $\eta_K$ field at $M_I$, the $SU(5)_R$ gauge coupling $\alpha_{5R}$ tends to take non-perturbative values at $\mu = M_I$, although this issue may be ameliorated by including  a few percent threshold effect at the GUT scale that lowers $\alpha_{5R}(M_G)$ from its unified value.

\subsection{Renormalization group equations for Yukawa couplings for \texorpdfstring{\boldmath{$M_I \leq \mu \leq M_G$}}{MIMG} }\label{sec:RGE}
Here we give the full two-loop beta functions for the Yukawa couplings of the model in the momentum range $M_I \leq \mu \leq M_G$.  These were generated with {\tt PyR@TE} package \cite{Sartore:2020gou} and cross-checked with other known results. We have used these two-loop RGEs for the numerical results obtained for gauge coupling unification in Sec. \ref{sec:unification}, as well as for determining the running factors $\eta_i'$ precisely for the fermion mass parameters of Eq. (\ref{eq:etaMG}) of Sec. \ref{sec:fermionfitting}.

The most general renormalizable Yukawa interaction involving the fermion and scalar fields listed in Eqs.~(\ref{eq:MIfermion})-(\ref{eq:MIscalar}), valid for the momentum range $M_I \leq \mu \leq M_G$  when the intermediate scale gauge symmetry is $SU(3)_{cL} \times SU(2)_L \times U(1)_L \times SU(5)_R$,  can be written down as
\begin{align}
\begin{autobreak}
-\mathcal{L}_Y = 

 Y_{uL}  {\overline{Q}_L} \widetilde{H}_L^d {U_R}

+ \frac{1}{4} Y_{uR}^\dagger {\chi_R} {H_R} {\chi_R}

+ Y_{dL} {\overline{Q}_L} {H}_L^d {D_R}

+ Y_{lL} {\overline{L}_L} {H}_L^d {E_R}

+\sqrt{2} Y_{dR}^\dagger \psi_R {H_R^{\dagger}} {\chi_R}

+  Y_D {D_R} {\Phi}_D^\dagger {\psi_R}

 + \text{h.c.}
\end{autobreak}
\label{eq:YukI}
\end{align}
Here the $SU(5)_R$ contractions are not shown explicitly, but these are identical to those given in Eq. (\ref{eq:YukLag}). The resulting quark and lepton mass matrices take the form:
\begin{align}
\mathcal{M}_u=\left(\begin{array}{cc}
0 & Y_{uL}^T \kappa_L \\
Y_{uR}^{*} \kappa_R & 0
\end{array}\right), ~~
\mathcal{M}_{\ell}=\left(\begin{array}{cc}
0 & Y_{\ell L}^T \kappa_L \\
Y_{dR}^{*} \kappa_R & 0
\end{array}\right), 
~~\mathcal{M}_d=\left(\begin{array}{cc}
0 & Y_{dL} \kappa_L \\
Y_{d R}^\dagger \kappa_R & Y_D^* v_\Phi
\end{array}\right)   \, .
\label{eq:Fmass}
\end{align}
We have the relation $Y_{uR} = Y_{uR}^T$ in this momentum regime owing to $SU(5)_R$ symmetry.  At $\mu = M_G$, these matrices obey the following boundary conditions (see Eq. (\ref{eq:fermionmass})):
\begin{equation}
Y_{uL} = Y_{uR} = Y_u,~~~Y_{dR} = Y_{dL} = Y_{\ell L}^T= Y_\ell^T~.
\label{eq:boundary}
\end{equation}

Following the definition of Ref.~\cite{Sartore:2020gou}, the RGEs for the Yukawa couplings canbe written as
\begin{equation*}
\beta\left(X\right) \equiv \mu \frac{d X}{d \mu}\equiv\cdot \frac{1}{\left(4 \pi\right)^{2}}\beta^{(1)}(X)+\cdot \frac{1}{\left(4 \pi\right)^{4}}\beta^{(2)}(X) \, ,
\end{equation*}
where $\beta^{(1)}(X)$ and $\beta^{(2)}(X)$ are the one-loop  and two-loop beta functions.
These beta functions, valid in the momentum range $M_I \leq \mu \leq M_G$ are given by
\input{RGE}


 \bibliographystyle{style}
 \bibliography{BIB.bib}

\end{document}

%% file: RGE.tex
{\allowdisplaybreaks

\begin{align}
 \begin{autobreak}
 \beta^{(1)}(Y_{uL}) =

+ \frac{3}{2} Y_{uL} Y_{uL}^{\dagger} Y_{uL}

-  \frac{3}{2} Y_{dL} Y_{dL}^{\dagger} Y_{uL}

+ 3 \tr\left(Y_{uL}^{\dagger} Y_{uL} \right) Y_{uL}

+ 3 \tr\left(Y_{dL}^{\dagger} Y_{dL} \right) Y_{uL}

+ \tr\left(Y_{lL}^{\dagger} Y_{lL} \right) Y_{uL}

-  \frac{17}{20} g_{1L}^{2} Y_{uL}

-  \frac{9}{4} g_{2L}^{2} Y_{uL}

- 8 g_{3L}^{2} Y_{uL} \, .
\end{autobreak} \\[5pt]
 \begin{autobreak}
 \beta^{(2)}(Y_{uL}) =

+ \frac{3}{2} Y_{uL} Y_{uL}^{\dagger} Y_{uL} Y_{uL}^{\dagger} Y_{uL}

-  \frac{1}{4} Y_{uL} Y_{uL}^{\dagger} Y_{dL} Y_{dL}^{\dagger} Y_{uL}

-  Y_{dL} Y_{dL}^{\dagger} Y_{uL} Y_{uL}^{\dagger} Y_{uL}

+ \frac{11}{4} Y_{dL} Y_{dL}^{\dagger} Y_{dL} Y_{dL}^{\dagger} Y_{uL}

+ \frac{35}{8} Y_{dL} Y_D^{*} Y_D^{\trans} Y_{dL}^{\dagger} Y_{uL}

-  \frac{27}{4} \tr\left(Y_{uL}^{\dagger} Y_{uL} Y_{uL}^{\dagger} Y_{uL} \right) Y_{uL}

-  \frac{27}{4} \tr\left(Y_{uL}^{\dagger} Y_{uL} \right) Y_{uL} Y_{uL}^{\dagger} Y_{uL}

+ \frac{15}{4} \tr\left(Y_{uL}^{\dagger} Y_{uL} \right) Y_{dL} Y_{dL}^{\dagger} Y_{uL}

+ \frac{3}{2} \tr\left(Y_{uL}^{\dagger} Y_{dL} Y_{dL}^{\dagger} Y_{uL} \right) Y_{uL}

-  \frac{27}{4} \tr\left(Y_{dL}^{\dagger} Y_{dL} \right) Y_{uL} Y_{uL}^{\dagger} Y_{uL}

-  \frac{27}{4} \tr\left(Y_{dL}^{\dagger} Y_{dL} Y_{dL}^{\dagger} Y_{dL} \right) Y_{uL}

+ \frac{15}{4} \tr\left(Y_{dL}^{\dagger} Y_{dL} \right) Y_{dL} Y_{dL}^{\dagger} Y_{uL}

-  \frac{45}{4} \tr\left(Y_{dL}^{\dagger} Y_{dL} Y_D^{*} Y_D^{\trans} \right) Y_{uL}

-  \frac{9}{4} \tr\left(Y_{lL}^{\dagger} Y_{lL} \right) Y_{uL} Y_{uL}^{\dagger} Y_{uL}

+ \frac{5}{4} \tr\left(Y_{lL}^{\dagger} Y_{lL} \right) Y_{dL} Y_{dL}^{\dagger} Y_{uL}

-  \frac{9}{4} \tr\left(Y_{lL}^{\dagger} Y_{lL} Y_{lL}^{\dagger} Y_{lL} \right) Y_{uL}

+ \frac{223}{80} g_{1L}^{2} Y_{uL} Y_{uL}^{\dagger} Y_{uL}

+ \frac{135}{16} g_{2L}^{2} Y_{uL} Y_{uL}^{\dagger} Y_{uL}

+ 16 g_{3L}^{2} Y_{uL} Y_{uL}^{\dagger} Y_{uL}

-  \frac{43}{80} g_{1L}^{2} Y_{dL} Y_{dL}^{\dagger} Y_{uL}

+ \frac{9}{16} g_{2L}^{2} Y_{dL} Y_{dL}^{\dagger} Y_{uL}

- 16 g_{3L}^{2} Y_{dL} Y_{dL}^{\dagger} Y_{uL}

+ \frac{17}{8} g_{1L}^{2} \tr\left(Y_{uL}^{\dagger} Y_{uL} \right) Y_{uL}

+ \frac{45}{8} g_{2L}^{2} \tr\left(Y_{uL}^{\dagger} Y_{uL} \right) Y_{uL}

+ 20 g_{3L}^{2} \tr\left(Y_{uL}^{\dagger} Y_{uL} \right) Y_{uL}

+ \frac{5}{8} g_{1L}^{2} \tr\left(Y_{dL}^{\dagger} Y_{dL} \right) Y_{uL}

+ \frac{45}{8} g_{2L}^{2} \tr\left(Y_{dL}^{\dagger} Y_{dL} \right) Y_{uL}

+ 20 g_{3L}^{2} \tr\left(Y_{dL}^{\dagger} Y_{dL} \right) Y_{uL}

+ \frac{15}{8} g_{1L}^{2} \tr\left(Y_{lL}^{\dagger} Y_{lL} \right) Y_{uL}

+ \frac{15}{8} g_{2L}^{2} \tr\left(Y_{lL}^{\dagger} Y_{lL} \right) Y_{uL}

+ \frac{5561}{1800} g_{1L}^{4} Y_{uL}

-  \frac{9}{20} g_{1L}^{2} g_{2L}^{2} Y_{uL}

+ \frac{19}{15} g_{1L}^{2} g_{3L}^{2} Y_{uL}

+ \frac{67}{4} g_{2L}^{4} Y_{uL}

+ 9 g_{2L}^{2} g_{3L}^{2} Y_{uL}

-  \frac{202}{9} g_{3L}^{4} Y_{uL} \, .
\end{autobreak}\\[5pt]
\begin{autobreak}
\beta^{(1)}(Y_{uR}) =

+ 3 Y_{uR} Y_{uR}^{*} Y_{uR}

- 3 Y_{uR} Y_{dR}^{\dagger} Y_{dR}

- 3 Y_{dR}^{\trans} Y_{dR}^{*} Y_{uR}

+ 3 \tr\left(Y_{uR}^{*} Y_{uR} \right) Y_{uR}

+ 4 \tr\left(Y_{dR}^{\dagger} Y_{dR} \right) Y_{uR}

-  \frac{108}{5} g_{5R}^{2} Y_{uR} \, .
\end{autobreak}\\[5pt]
\begin{autobreak}
\beta^{(2)}(Y_{uR}) =

-  \frac{33}{4} Y_{uR} Y_{uR}^{*} Y_{uR} Y_{uR}^{*} Y_{uR}

- 6 Y_{uR} Y_{uR}^{*} Y_{uR} Y_{dR}^{\dagger} Y_{dR}

-  \frac{3}{4} Y_{uR} Y_{uR}^{*} Y_{dR}^{\trans} Y_{dR}^{*} Y_{uR}

-  \frac{3}{4} Y_{uR} Y_{dR}^{\dagger} Y_{dR} Y_{uR}^{*} Y_{uR}

+ 11 Y_{uR} Y_{dR}^{\dagger} Y_{dR} Y_{dR}^{\dagger} Y_{dR}

+ \frac{21}{4} Y_{uR} Y_{dR}^{\dagger} Y_D^{\dagger} Y_D Y_{dR}

- 6 Y_{dR}^{\trans} Y_{dR}^{*} Y_{uR} Y_{uR}^{*} Y_{uR}

- 8 Y_{dR}^{\trans} Y_{dR}^{*} Y_{uR} Y_{dR}^{\dagger} Y_{dR}

+ 11 Y_{dR}^{\trans} Y_{dR}^{*} Y_{dR}^{\trans} Y_{dR}^{*} Y_{uR}

+ \frac{21}{4} Y_{dR}^{\trans} Y_D^{\trans} Y_D^{*} Y_{dR}^{*} Y_{uR}

-  \frac{27}{2} \tr\left(Y_{uR}^{*} Y_{uR} Y_{uR}^{*} Y_{uR} \right) Y_{uR}

-  \frac{27}{2} \tr\left(Y_{uR}^{*} Y_{uR} \right) Y_{uR} Y_{uR}^{*} Y_{uR}

+ \frac{15}{2} \tr\left(Y_{uR}^{*} Y_{uR} \right) Y_{uR} Y_{dR}^{\dagger} Y_{dR}

+ 6 \tr\left(Y_{uR}^{*} Y_{uR} Y_{dR}^{\dagger} Y_{dR} \right) Y_{uR}

+ \frac{15}{2} \tr\left(Y_{uR}^{*} Y_{uR} \right) Y_{dR}^{\trans} Y_{dR}^{*} Y_{uR}

- 18 \tr\left(Y_{dR}^{\dagger} Y_{dR} \right) Y_{uR} Y_{uR}^{*} Y_{uR}

+ 10 \tr\left(Y_{dR}^{\dagger} Y_{dR} \right) Y_{uR} Y_{dR}^{\dagger} Y_{dR}

- 18 \tr\left(Y_{dR}^{\dagger} Y_{dR} Y_{dR}^{\dagger} Y_{dR} \right) Y_{uR}

+ 10 \tr\left(Y_{dR}^{\dagger} Y_{dR} \right) Y_{dR}^{\trans} Y_{dR}^{*} Y_{uR}

- 9 \tr\left(Y_{dR}^{\dagger} Y_D^{\dagger} Y_D Y_{dR} \right) Y_{uR}

+ \frac{396}{5} g_{5R}^{2} Y_{uR} Y_{uR}^{*} Y_{uR}

-  \frac{261}{5} g_{5R}^{2} Y_{uR} Y_{dR}^{\dagger} Y_{dR}

-  \frac{261}{5} g_{5R}^{2} Y_{dR}^{\trans} Y_{dR}^{*} Y_{uR}

+ 54 g_{5R}^{2} \tr\left(Y_{uR}^{*} Y_{uR} \right) Y_{uR}

+ 60 g_{5R}^{2} \tr\left(Y_{dR}^{\dagger} Y_{dR} \right) Y_{uR}

-  \frac{2427}{25} g_{5R}^{4} Y_{uR} \, .
\end{autobreak}
\label{eq:RGEYuR}\\[5pt]
\begin{autobreak}
\beta^{(1)}(Y_{dL}) =

-  \frac{3}{2} Y_{uL} Y_{uL}^{\dagger} Y_{dL}

+ \frac{3}{2} Y_{dL} Y_{dL}^{\dagger} Y_{dL}

+ \frac{5}{2} Y_{dL} Y_D^{*} Y_D^{\trans}

+ 3 \tr\left(Y_{uL}^{\dagger} Y_{uL} \right) Y_{dL}

+ 3 \tr\left(Y_{dL}^{\dagger} Y_{dL} \right) Y_{dL}

+ \tr\left(Y_{lL}^{\dagger} Y_{lL} \right) Y_{dL}

-  \frac{1}{4} g_{1L}^{2} Y_{dL}

-  \frac{9}{4} g_{2L}^{2} Y_{dL}

- 8 g_{3L}^{2} Y_{dL} \, .
\end{autobreak}\\[5pt]
\begin{autobreak}
\beta^{(2)}(Y_{dL}) =

+ \frac{11}{4} Y_{uL} Y_{uL}^{\dagger} Y_{uL} Y_{uL}^{\dagger} Y_{dL}

-  Y_{uL} Y_{uL}^{\dagger} Y_{dL} Y_{dL}^{\dagger} Y_{dL}

-  \frac{1}{4} Y_{dL} Y_{dL}^{\dagger} Y_{uL} Y_{uL}^{\dagger} Y_{dL}

+ \frac{3}{2} Y_{dL} Y_{dL}^{\dagger} Y_{dL} Y_{dL}^{\dagger} Y_{dL}

-  \frac{5}{2} Y_{dL} Y_D^{*} Y_{dR}^{*} Y_{dR}^{\trans} Y_D^{\trans}

-  \frac{5}{8} Y_{dL} Y_D^{*} Y_D^{\trans} Y_{dL}^{\dagger} Y_{dL}

-  \frac{15}{8} Y_{dL} Y_D^{*} Y_D^{\trans} Y_D^{*} Y_D^{\trans}

-  \frac{27}{4} \tr\left(Y_{uL}^{\dagger} Y_{uL} Y_{uL}^{\dagger} Y_{uL} \right) Y_{dL}

+ \frac{15}{4} \tr\left(Y_{uL}^{\dagger} Y_{uL} \right) Y_{uL} Y_{uL}^{\dagger} Y_{dL}

-  \frac{27}{4} \tr\left(Y_{uL}^{\dagger} Y_{uL} \right) Y_{dL} Y_{dL}^{\dagger} Y_{dL}

+ \frac{3}{2} \tr\left(Y_{uL}^{\dagger} Y_{dL} Y_{dL}^{\dagger} Y_{uL} \right) Y_{dL}

+ \frac{15}{4} \tr\left(Y_{dL}^{\dagger} Y_{dL} \right) Y_{uL} Y_{uL}^{\dagger} Y_{dL}

-  \frac{27}{4} \tr\left(Y_{dL}^{\dagger} Y_{dL} Y_{dL}^{\dagger} Y_{dL} \right) Y_{dL}

-  \frac{27}{4} \tr\left(Y_{dL}^{\dagger} Y_{dL} \right) Y_{dL} Y_{dL}^{\dagger} Y_{dL}

-  \frac{45}{4} \tr\left(Y_{dL}^{\dagger} Y_{dL} Y_D^{*} Y_D^{\trans} \right) Y_{dL}

+ \frac{5}{4} \tr\left(Y_{lL}^{\dagger} Y_{lL} \right) Y_{uL} Y_{uL}^{\dagger} Y_{dL}

-  \frac{9}{4} \tr\left(Y_{lL}^{\dagger} Y_{lL} \right) Y_{dL} Y_{dL}^{\dagger} Y_{dL}

-  \frac{9}{4} \tr\left(Y_{lL}^{\dagger} Y_{lL} Y_{lL}^{\dagger} Y_{lL} \right) Y_{dL}

-  \frac{15}{4} \tr\left(Y_D^{\dagger} Y_D \right) Y_{dL} Y_D^{*} Y_D^{\trans}

-  \frac{79}{80} g_{1L}^{2} Y_{uL} Y_{uL}^{\dagger} Y_{dL}

+ \frac{9}{16} g_{2L}^{2} Y_{uL} Y_{uL}^{\dagger} Y_{dL}

- 16 g_{3L}^{2} Y_{uL} Y_{uL}^{\dagger} Y_{dL}

+ \frac{187}{80} g_{1L}^{2} Y_{dL} Y_{dL}^{\dagger} Y_{dL}

+ \frac{135}{16} g_{2L}^{2} Y_{dL} Y_{dL}^{\dagger} Y_{dL}

+ 16 g_{3L}^{2} Y_{dL} Y_{dL}^{\dagger} Y_{dL}

+ \frac{11}{12} g_{1L}^{2} Y_{dL} Y_D^{*} Y_D^{\trans}

+ \frac{55}{3} g_{3L}^{2} Y_{dL} Y_D^{*} Y_D^{\trans}

+ 51 g_{5R}^{2} Y_{dL} Y_D^{*} Y_D^{\trans}

+ \frac{17}{8} g_{1L}^{2} \tr\left(Y_{uL}^{\dagger} Y_{uL} \right) Y_{dL}

+ \frac{45}{8} g_{2L}^{2} \tr\left(Y_{uL}^{\dagger} Y_{uL} \right) Y_{dL}

+ 20 g_{3L}^{2} \tr\left(Y_{uL}^{\dagger} Y_{uL} \right) Y_{dL}

+ \frac{5}{8} g_{1L}^{2} \tr\left(Y_{dL}^{\dagger} Y_{dL} \right) Y_{dL}

+ \frac{45}{8} g_{2L}^{2} \tr\left(Y_{dL}^{\dagger} Y_{dL} \right) Y_{dL}

+ 20 g_{3L}^{2} \tr\left(Y_{dL}^{\dagger} Y_{dL} \right) Y_{dL}

+ \frac{15}{8} g_{1L}^{2} \tr\left(Y_{lL}^{\dagger} Y_{lL} \right) Y_{dL}

+ \frac{15}{8} g_{2L}^{2} \tr\left(Y_{lL}^{\dagger} Y_{lL} \right) Y_{dL}

-  \frac{31}{1800} g_{1L}^{4} Y_{dL}

-  \frac{27}{20} g_{1L}^{2} g_{2L}^{2} Y_{dL}

+ \frac{31}{15} g_{1L}^{2} g_{3L}^{2} Y_{dL}

+ \frac{67}{4} g_{2L}^{4} Y_{dL}

+ 9 g_{2L}^{2} g_{3L}^{2} Y_{dL}

-  \frac{202}{9} g_{3L}^{4} Y_{dL} \, .
\end{autobreak}\\[5pt]
\begin{autobreak}
\beta^{(1)}(Y_{lL}) =

+ \frac{3}{2} Y_{lL} Y_{lL}^{\dagger} Y_{lL}

+ 3 \tr\left(Y_{uL}^{\dagger} Y_{uL} \right) Y_{lL}

+ 3 \tr\left(Y_{dL}^{\dagger} Y_{dL} \right) Y_{lL}

+ \tr\left(Y_{lL}^{\dagger} Y_{lL} \right) Y_{lL}

-  \frac{9}{4} g_{1L}^{2} Y_{lL}

-  \frac{9}{4} g_{2L}^{2} Y_{lL} \, .
\end{autobreak}\\[5pt]
\begin{autobreak}
\beta^{(2)}(Y_{lL}) =

+ \frac{3}{2} Y_{lL} Y_{lL}^{\dagger} Y_{lL} Y_{lL}^{\dagger} Y_{lL}

-  \frac{27}{4} \tr\left(Y_{uL}^{\dagger} Y_{uL} Y_{uL}^{\dagger} Y_{uL} \right) Y_{lL}

-  \frac{27}{4} \tr\left(Y_{uL}^{\dagger} Y_{uL} \right) Y_{lL} Y_{lL}^{\dagger} Y_{lL}

+ \frac{3}{2} \tr\left(Y_{uL}^{\dagger} Y_{dL} Y_{dL}^{\dagger} Y_{uL} \right) Y_{lL}

-  \frac{27}{4} \tr\left(Y_{dL}^{\dagger} Y_{dL} Y_{dL}^{\dagger} Y_{dL} \right) Y_{lL}

-  \frac{27}{4} \tr\left(Y_{dL}^{\dagger} Y_{dL} \right) Y_{lL} Y_{lL}^{\dagger} Y_{lL}

-  \frac{45}{4} \tr\left(Y_{dL}^{\dagger} Y_{dL} Y_D^{*} Y_D^{\trans} \right) Y_{lL}

-  \frac{9}{4} \tr\left(Y_{lL}^{\dagger} Y_{lL} Y_{lL}^{\dagger} Y_{lL} \right) Y_{lL}

-  \frac{9}{4} \tr\left(Y_{lL}^{\dagger} Y_{lL} \right) Y_{lL} Y_{lL}^{\dagger} Y_{lL}

+ \frac{387}{80} g_{1L}^{2} Y_{lL} Y_{lL}^{\dagger} Y_{lL}

+ \frac{135}{16} g_{2L}^{2} Y_{lL} Y_{lL}^{\dagger} Y_{lL}

+ \frac{17}{8} g_{1L}^{2} \tr\left(Y_{uL}^{\dagger} Y_{uL} \right) Y_{lL}

+ \frac{45}{8} g_{2L}^{2} \tr\left(Y_{uL}^{\dagger} Y_{uL} \right) Y_{lL}

+ 20 g_{3L}^{2} \tr\left(Y_{uL}^{\dagger} Y_{uL} \right) Y_{lL}

+ \frac{5}{8} g_{1L}^{2} \tr\left(Y_{dL}^{\dagger} Y_{dL} \right) Y_{lL}

+ \frac{45}{8} g_{2L}^{2} \tr\left(Y_{dL}^{\dagger} Y_{dL} \right) Y_{lL}

+ 20 g_{3L}^{2} \tr\left(Y_{dL}^{\dagger} Y_{dL} \right) Y_{lL}

+ \frac{15}{8} g_{1L}^{2} \tr\left(Y_{lL}^{\dagger} Y_{lL} \right) Y_{lL}

+ \frac{15}{8} g_{2L}^{2} \tr\left(Y_{lL}^{\dagger} Y_{lL} \right) Y_{lL}

+ \frac{2021}{200} g_{1L}^{4} Y_{lL}

+ \frac{27}{20} g_{1L}^{2} g_{2L}^{2} Y_{lL}

+ \frac{67}{4} g_{2L}^{4} Y_{lL} \, .
\end{autobreak}\\[5pt]
\begin{autobreak}
\beta^{(1)}(Y_{dR}) =

-  \frac{9}{2} Y_{dR} Y_{uR}^{*} Y_{uR}

+ 3 Y_{dR} Y_{dR}^{\dagger} Y_{dR}

+ \frac{3}{2} Y_D^{\dagger} Y_D Y_{dR}

+ 3 \tr\left(Y_{uR}^{*} Y_{uR} \right) Y_{dR}

+ 4 \tr\left(Y_{dR}^{\dagger} Y_{dR} \right) Y_{dR}

- 18 g_{5R}^{2} Y_{dR} \, .
\end{autobreak}\\[5pt]
\begin{autobreak}
\beta^{(2)}(Y_{dR}) =

+ \frac{135}{8} Y_{dR} Y_{uR}^{*} Y_{uR} Y_{uR}^{*} Y_{uR}

-  \frac{3}{2} Y_{dR} Y_{uR}^{*} Y_{uR} Y_{dR}^{\dagger} Y_{dR}

-  \frac{75}{4} Y_{dR} Y_{uR}^{*} Y_{dR}^{\trans} Y_{dR}^{*} Y_{uR}

- 6 Y_{dR} Y_{dR}^{\dagger} Y_{dR} Y_{uR}^{*} Y_{uR}

- 6 Y_{dR} Y_{dR}^{\dagger} Y_{dR} Y_{dR}^{\dagger} Y_{dR}

-  \frac{3}{4} Y_{dR} Y_{dR}^{\dagger} Y_D^{\dagger} Y_D Y_{dR}

-  \frac{3}{4} Y_D^{\dagger} Y_{dL}^{\trans} Y_{dL}^{*} Y_D Y_{dR}

-  \frac{15}{8} Y_D^{\dagger} Y_D Y_D^{\dagger} Y_D Y_{dR}

-  \frac{27}{2} \tr\left(Y_{uR}^{*} Y_{uR} Y_{uR}^{*} Y_{uR} \right) Y_{dR}

+ \frac{45}{4} \tr\left(Y_{uR}^{*} Y_{uR} \right) Y_{dR} Y_{uR}^{*} Y_{uR}

+ 6 \tr\left(Y_{uR}^{*} Y_{uR} Y_{dR}^{\dagger} Y_{dR} \right) Y_{dR}

-  \frac{27}{2} \tr\left(Y_{uR}^{*} Y_{uR} \right) Y_{dR} Y_{dR}^{\dagger} Y_{dR}

+ 15 \tr\left(Y_{dR}^{\dagger} Y_{dR} \right) Y_{dR} Y_{uR}^{*} Y_{uR}

- 18 \tr\left(Y_{dR}^{\dagger} Y_{dR} Y_{dR}^{\dagger} Y_{dR} \right) Y_{dR}

- 18 \tr\left(Y_{dR}^{\dagger} Y_{dR} \right) Y_{dR} Y_{dR}^{\dagger} Y_{dR}

- 9 \tr\left(Y_{dR}^{\dagger} Y_D^{\dagger} Y_D Y_{dR} \right) Y_{dR}

-  \frac{9}{4} \tr\left(Y_D^{\dagger} Y_D \right) Y_D^{\dagger} Y_D Y_{dR}

-  \frac{432}{5} g_{5R}^{2} Y_{dR} Y_{uR}^{*} Y_{uR}

+ \frac{369}{5} g_{5R}^{2} Y_{dR} Y_{dR}^{\dagger} Y_{dR}

+ \frac{17}{20} g_{1L}^{2} Y_D^{\dagger} Y_D Y_{dR}

+ 17 g_{3L}^{2} Y_D^{\dagger} Y_D Y_{dR}

+ \frac{99}{5} g_{5R}^{2} Y_D^{\dagger} Y_D Y_{dR}

+ 54 g_{5R}^{2} \tr\left(Y_{uR}^{*} Y_{uR} \right) Y_{dR}

+ 60 g_{5R}^{2} \tr\left(Y_{dR}^{\dagger} Y_{dR} \right) Y_{dR}

-  \frac{1729}{25} g_{5R}^{4} Y_{dR} \, .
\end{autobreak}\\[5pt]
\begin{autobreak}
\beta^{(1)}(Y_D) =

+ Y_{dL}^{\trans} Y_{dL}^{*} Y_D

+ 2 Y_D Y_{dR} Y_{dR}^{\dagger}

+ 4 Y_D Y_D^{\dagger} Y_D

+ \tr\left(Y_D^{\dagger} Y_D \right) Y_D

-  \frac{1}{5} g_{1L}^{2} Y_D

- 4 g_{3L}^{2} Y_D

-  \frac{36}{5} g_{5R}^{2} Y_D \, .
\end{autobreak}\\[5pt]
\begin{autobreak}
\beta^{(2)}(Y_D) =

-  \frac{1}{4} Y_{dL}^{\trans} Y_{uL}^{*} Y_{uL}^{\trans} Y_{dL}^{*} Y_D

-  \frac{1}{4} Y_{dL}^{\trans} Y_{dL}^{*} Y_{dL}^{\trans} Y_{dL}^{*} Y_D

-  \frac{3}{2} Y_D Y_{dR} Y_{uR}^{*} Y_{uR} Y_{dR}^{\dagger}

-  Y_D Y_{dR} Y_{dR}^{\dagger} Y_{dR} Y_{dR}^{\dagger}

-  \frac{5}{2} Y_D Y_{dR} Y_{dR}^{\dagger} Y_D^{\dagger} Y_D

-  \frac{3}{4} Y_D Y_D^{\dagger} Y_{dL}^{\trans} Y_{dL}^{*} Y_D

-  \frac{7}{4} Y_D Y_D^{\dagger} Y_D Y_D^{\dagger} Y_D

-  \frac{9}{2} \tr\left(Y_{uL}^{\dagger} Y_{uL} \right) Y_{dL}^{\trans} Y_{dL}^{*} Y_D

- 9 \tr\left(Y_{uR}^{*} Y_{uR} \right) Y_D Y_{dR} Y_{dR}^{\dagger}

-  \frac{9}{2} \tr\left(Y_{dL}^{\dagger} Y_{dL} \right) Y_{dL}^{\trans} Y_{dL}^{*} Y_D

-  \frac{3}{2} \tr\left(Y_{dL}^{\dagger} Y_{dL} Y_D^{*} Y_D^{\trans} \right) Y_D

-  \frac{3}{2} \tr\left(Y_{lL}^{\dagger} Y_{lL} \right) Y_{dL}^{\trans} Y_{dL}^{*} Y_D

- 12 \tr\left(Y_{dR}^{\dagger} Y_{dR} \right) Y_D Y_{dR} Y_{dR}^{\dagger}

- 3 \tr\left(Y_{dR}^{\dagger} Y_D^{\dagger} Y_D Y_{dR} \right) Y_D

- 6 \tr\left(Y_D^{\dagger} Y_D Y_D^{\dagger} Y_D \right) Y_D

- 6 \tr\left(Y_D^{\dagger} Y_D \right) Y_D Y_D^{\dagger} Y_D

+ \frac{133}{120} g_{1L}^{2} Y_{dL}^{\trans} Y_{dL}^{*} Y_D

+ \frac{51}{8} g_{2L}^{2} Y_{dL}^{\trans} Y_{dL}^{*} Y_D

-  \frac{16}{3} g_{3L}^{2} Y_{dL}^{\trans} Y_{dL}^{*} Y_D

+ \frac{114}{5} g_{5R}^{2} Y_D Y_{dR} Y_{dR}^{\dagger}

+ \frac{89}{30} g_{1L}^{2} Y_D Y_D^{\dagger} Y_D

+ \frac{106}{3} g_{3L}^{2} Y_D Y_D^{\dagger} Y_D

+ \frac{354}{5} g_{5R}^{2} Y_D Y_D^{\dagger} Y_D

+ \frac{1}{6} g_{1L}^{2} \tr\left(Y_D^{\dagger} Y_D \right) Y_D

+ \frac{10}{3} g_{3L}^{2} \tr\left(Y_D^{\dagger} Y_D \right) Y_D

+ 6 g_{5R}^{2} \tr\left(Y_D^{\dagger} Y_D \right) Y_D

+ \frac{22}{45} g_{1L}^{4} Y_D

-  \frac{16}{15} g_{1L}^{2} g_{3L}^{2} Y_D

-  \frac{36}{25} g_{1L}^{2} g_{5R}^{2} Y_D

+ \frac{91}{9} g_{3L}^{4} Y_D

-  \frac{144}{5} g_{3L}^{2} g_{5R}^{2} Y_D

+ \frac{41}{25} g_{5R}^{4} Y_D \, .
\end{autobreak}
\end{align}

}